\documentclass[a4paper,fleqn]{mnras}
\usepackage{epsfig}
\usepackage{threeparttable}
\usepackage{multicol}
\usepackage{aas_macros,graphicx,times,multirow,amsmath,amssymb,color,longtable}
\usepackage[T1]{fontenc}
\usepackage{graphicx}
\usepackage{pdflscape}
\usepackage{subfig}

\title[SPH simulations of binary TDEs]{Hydrodynamical simulations of the tidal stripping of binary stars by massive black holes}
\author[Mainetti et al.]{Deborah Mainetti$^{1, 2, 3}$ \thanks{E-mail: d.mainetti1@campus.unimib.it},
Alessandro Lupi$\rm ^{3, 4}$, Sergio Campana$\rm ^{2}$ and Monica Colpi$\rm ^{1, 3}$
\smallskip\\
$\rm ^1$Dipartimento di Fisica G. Occhialini, Universit\`a degli Studi di Milano-Bicocca, Piazza della Scienza 3, I-20126 Milano, Italy\\
$\rm ^2$INAF, Osservatorio Astronomico di Brera, Via E. Bianchi 46, I-23807, Merate (LC), Italy\\
$\rm ^3$INFN, Sezione di Milano-Bicocca, Piazza della Scienza 3, I-20126 Milano, Italy\\
$\rm ^4$Institut d'Astrophysique de Paris, Sorbonne Universit\`{e}s, UPMC Univ Paris 6 et CNRS, UMR 7095, 98 bis bd Arago, F-75014 Paris, France
}

\date{Accepted 2016 January 20. Received 2016 January 19; in original form 2015 December 10}
\pagerange{\pageref{firstpage}--\pageref{lastpage}} 
\pubyear{2016}

\def\ltsima{$\; \buildrel < \over \sim \;$}
\def\lsim{\lower.5ex\hbox{\ltsima}}
\def\gtsima{$\; \buildrel $\geq$ \over \sim \;$}
\def\gsim{\lower.5ex\hbox{\gtsima}}

\newcommand{\be}{\begin{equation}}
\newcommand{\en}{\end{equation}}

\def\j1023{\mbox{PSR\,J1023$+$0038}}

\def\LaTeX{L\kern-.36em\raise.3ex\hbox{a}\kern-.15em
    T\kern-.1667em\lower.7ex\hbox{E}\kern-.125emX}

\begin{document}

\label{firstpage}
\pagerange{\pageref{firstpage}--\pageref{lastpage}}
\maketitle

\begin{abstract}
In a galactic nucleus, a star on a low angular momentum orbit around the central massive black hole can be fully or partially disrupted by the black hole tidal field, lighting up the compact object via gas accretion. This phenomenon can repeat if the star, not fully disrupted, is on a closed orbit. Because of the multiplicity of stars in binary systems, also binary stars may experience in pairs such a fate, immediately after being tidally separated. The consumption of both the binary components by the black hole is expected to power a double-peaked flare. In this paper, we perform for the first time, with \textsc{\small GADGET}2, a suite of smoothed particle hydrodynamics simulations of binary stars around a galactic central black hole in the Newtonian regime. We show that accretion luminosity light curves from double tidal disruptions reveal a more prominent knee, rather than a double peak, when decreasing the impact parameter of the  encounter and when elevating the difference between the mass of the star which leaves the system after binary separation and the mass of the companion. The detection of a knee can anticipate the onset of periodic accretion luminosity flares if one of the stars, only partially disrupted, remains bound to the black hole after binary separation. Thus knees could be precursors of periodic flares, which can then be predicted, followed up and better modelled. Analytical estimates in the black hole mass range $10^5-10^8 \rm M_{\rm \odot}$ show that the knee signature is enhanced in the case of black holes of mass $10^{6} - 10^{7} \rm M_{\rm \odot}$.
\end{abstract}

\begin{keywords}
hydrodynamics - methods: numerical - binaries: close - galaxies: kinematics and dynamics - galaxies: nuclei  
\end{keywords}

%%%%%%

\section{Introduction}
Supermassive black holes (BHs) are ubiquitous in the centre of massive galaxies. For most of the time they are in a quiescent state, but sometimes they can accrete matter from the surroundings and power an active galactic nucleus (AGN; Ho 2008). 
Stars orbiting around the central BH of a galaxy interact with each other, increasing the probability for one of them to be scattered on a low angular momentum orbit (Alexander 2012). A tidal disruption event (TDE) could thus occur contributing to the BH flaring on time-scales of months or years (e.g. Rees 1988; Phinney 1989). For solar mass stars, this occurs when the (non-spinning) BH mass is less than about $10^{8} \rm M_{\rm \odot}$. For heavier BHs, these stars cross the horizon and are fully swallowed before being tidally disrupted (Macleod, Ramirez-Ruiz \& Guillochon 2012). As a consequence, TDEs contribute to the detection of otherwise quiescent BHs in inactive (or weakly active) galaxies in a mass interval somewhat complementary to that probed in surveys of bright AGNs and QSOs (Vestergaard \& Osmer 2009).

The total tidal disruption of a single star of mass $M_{\rm*}$ and radius $R_{\rm*}$ moving on a parabolic orbit around the central BH of a galaxy of mass $M_{\rm BH}$ occurs when its pericentre radius $r_{\rm p}$ is less than about the so-called BH tidal radius
\begin{equation} 
r_{\rm t}=R_{\rm*}\Bigl(\frac{M_{\rm BH}}{M_{\rm*}}\Bigr)^{1/3}\sim 10^{2} \rm R_{\rm \odot} \Bigl(\frac{\it R_{\rm *}}{1 \rm R_{\rm \odot}}\Bigr) \Bigl(\frac{\it M_{\rm BH}}{10^{6}\rm M_{\rm \odot}}\frac{1 \rm M_{\rm \odot}}{\it M_{\rm *}}\Bigr)^{1/3}  \label{eq1},
\end{equation}
corresponding to the distance where the BH tidal force overcomes the star self-gravity at its surface (Hills 1975; Frank \& Rees 1976). On the contrary, if $r_{\rm p} \gtrsim r_{\rm t}$, the star undergoes less distortion and suffers only partial disruption. The value of the impact parameter 
$\beta=r_{\rm t}/r_{\rm p}$
defines how deep the disruption is (Guillochon \& Ramirez-Ruiz 2013, 2015a). Roughly, only about half of the produced stellar debris remains bound to the BH and accretes on to it, powering the emission of a characteristic flare (e.g. Rees 1988; Phinney 1989). In the regime of partial TDEs, the star, if on a bound orbit, could transfer a fraction of its mass to the BH every time it passes through the pericentre of its orbit, thus powering one flare for every orbital period and maybe `spoon-feeding' the quiescent luminosity of weakly active galaxies (MacLeod et al. 2013).

TDEs are quite rare events, with estimated rates of $\sim 10^{-5}$ $\rm galaxy^{-1}$ $\rm yr^{-1}$ (e.g. Donley et al. 2002). Despite this and sparse observations, a few TDEs have been observed mainly in the optical-UV (Renzini et al. 1995; Gezari et al. 2006, 2008, 2009, 2012; Komossa et al. 2008; van Velzen et al. 2011; Wang et al. 2011, 2012; Cenko et al. 2012a; Gezari 2012; Arcavi et al. 2014; Chornock et al. 2014; Holoien et al. 2014; Vinko et al. 2015) and soft X-ray bands (Bade, Komossa \& Dahlem 1996; Komossa \& Bade 1999; Komossa \& Greiner 1999; Grupe, Thomas \& Leighly 1999; Greiner et al. 2000; Li, Ramesh \& Kristen 2002; Halpern, Gezari \& Komossa 2004; Komossa 2004, 2012, 2015; Komossa et al. 2004; Esquej et al. 2007, 2008; Cappelluti et al. 2009; Maksym, Ulmer \& Eracleous 2010; Lin et al. 2011, 2015; Saxton et al. 2012, 2015; Maksym et al. 2013; Donato et al. 2014; Khabibullin \& Sazonov 2014; Maksym, Lin \& Irwin 2014), but also in the radio and hard X-ray bands (Bloom et al. 2011; Burrows et al. 2011; Levan et al. 2011; Zauderer et al. 2011; Cenko et al. 2012b; Hryniewicz \& Walter 2016; Lei et al. 2016). Many theoretical studies have been carried out to understand the physics of TDEs and model their accretion luminosity light curves (hereafter just light curves or flares), considering stars approaching the BH on a variety of orbits, from parabolic to bound (Nolthenius \& Katz 1982; Bicknell \& Gingold 1983; Carter \& Luminet 1985; Luminet \& Marck 1985; Luminet \& Carter 1986; Rees 1988; Evans \& Kochanek 1989; Phinney 1989;  Khokhlov, Novikov \& Pethick 1993a, b; Laguna et al. 1993; Diener et al. 1995, 1997; Ivanov \& Novikov 2001; Kobayashi et al. 2004; Rosswog, Ramirez-Ruiz \& His 2008, 2009; Guillochon et al. 2009; Lodato, King \& Pringle 2009; Ramirez-Ruiz \& Rosswog 2009; Strubbe \& Quataert 2009; Kasen \& Ramirez-Ruiz 2010; Lodato \& Rossi 2010; Amaro-Seoane, Miller \& Kennedy 2012; MacLeod et al. 2012, 2013; Guillochon \& Ramirez-Ruiz 2013, 2015a; Hayasaki, Stone \& Loeb 2013). 

So far, only single-star TDEs have been taken into account. However, most of the stars in the field are in binaries (Duquennoy \& Mayor 1991b; Fischer \& Marcy 1992); hence, it is worth also studying close encounters between binaries and galactic central BHs which can lead to the disruption of both the binary members. The topic was first discussed by Mandel \& Levin (2015), suggesting that in a binary-BH encounter under certain conditions both binary components may undergo tidal disruption in sequence immediately after the tidal binary break-up. A double-peaked flare is expected to occur, signature of such a peculiar event. 

In this paper, we present for the first time the results of a series of smoothed particle hydrodynamics (SPH) simulations performed using the \textsc{\small GADGET}2 code (Springel 2005; the code can be freely downloaded from http://wwwmpa.mpa-garching.mpg.de/gadget/) in the aim at studying the physics of double tidal disruptions and at characterizing the expected  light curves. As a first exploratory study, we consider parabolic encounters of binaries with galactic central BHs in the Newtonian regime, in order to explore which are the most favourable conditions for the occurrence of double-peaked flares. In particular, we address the following questions. Are all simulated encounters leading to double-peaked light curves or are there cases of single-peaked light curves? How can we disentangle the different outcomes? How prominent are the double peaks?

The paper is organized as follows. In Section 2, we resume the conditions required for double TDEs and the associated space of binary parameters (Mandel \& Levin 2015). In Section 3, we initialize low-resolution SPH simulations of binary-BH encounters with different $r_{\rm p}$ values of the centre of mass (CM) of the binaries around the BH. Not all encounters can lead to double TDEs, and in Section 4 we introduce a classification of the obtained outcomes. In Section 5, we show the results of a selected sample of high-resolution simulations and the light curves directly inferred from them. Section 6 sums up results and conclusions.

\section{Basics for double tidal disruptions} \label{basics}
We are here interested in identifying the set conditions for the sequential tidal disruption of binary stars around galactic central BHs, 
following Mandel \& Levin (2015). 

Tidal break-up of a binary on a parabolic orbit around a BH occurs if the binary CM around the BH enters a sphere  of radius 
\begin{equation}
r_{\rm tb}=a_{\rm bin}\Bigl(\frac{M_{\rm BH}}{M_{\rm bin}}\Bigr)^{1/3}\sim 10^{3} \rm R_{\rm \odot} \Bigl(\frac{\it a_{\rm bin}}{10 \rm R_{\rm \odot}}\Bigr) \Bigl(\frac{\it M_{\rm BH}}{10^{6}\rm M_{\rm \odot}}\frac{1 \rm M_{\rm \odot}}{\it M_{\rm *}}\Bigr)^{1/3}, \label{eq2}
\end{equation}
where $a_{\rm bin}$ and $M_{\rm bin}$ are the binary semimajor axis and total mass (Miller et al. 2005; Sesana, Madau \& Haardt 2009). We notice that binary break-up comes before single-star tidal disruptions, given that $r_{\rm tb} \gtrsim r_{\rm t}$ (see equations \ref{eq1} and \ref{eq2}). Tidal break-up  occurs when the specific angular momentum (in modulus) of the binary CM at pericentre   becomes less than
\begin{equation}
l_{\rm CM}(r_{\rm tb})\sim \sqrt{G M_{\rm BH} a_{\rm bin} \Bigl(\frac{M_{\rm BH}}{M_{\rm bin}}\Bigr)^{1/3}}.
\end{equation}
Orbits which allow tidal binary break-up are called loss cone orbits (Merritt 2013). A binary on a loss cone orbit is broken up after one pericentre passage, over a  time-scale $T \sim 2\pi \sqrt{r^{3}/GM_{\rm BH}}$, corresponding to the orbital period of a binary on a circular orbit at the same distance from the BH. 

Both stars of a binary can undergo a sequential tidal disruption immediately after the tidal binary break-up only if the specific angular momentum of the binary CM around the BH at the closest approach,
defined as $l_{\rm CM}(r_{\rm p})\sim \sqrt{G M_{\rm BH} r_{\rm p}},$ instantly changes from being greater than $l_{\rm CM}(r_{\rm tb})$ to becoming less than $l_{\rm CM}(r_{\rm t})$, where
\begin{equation}
l_{\rm CM}(r_{\rm t})\sim \sqrt{G M_{\rm BH} R_{\rm*} \Bigl(\frac{M_{\rm BH}}{M_{\rm*}}\Bigr)^{1/3}}.
\end{equation}
In this way, the binary enters intact the region of single-star TDEs. 
This occurs if the binary experiences a large enough change $\Delta L$, at least of the order of $l_{\rm CM} (r_{\rm tb})$, in the specific circular angular momentum $l_{\rm circ}(r)\sim \sqrt{G M_{\rm BH}r}$, over a time-scale $T$. 
Interactions with surrounding stars and/or massive perturbers can promote such a change (Perets, Hopman \& Alexander 2007; Alexander 2012). 
We consider empty the portion of the loss cone, in phase space, corresponding to binaries that break up before entering the region of single-star TDEs, and full the portion of the loss cone corresponding to binaries which can enter intact the region of single-star TDEs (Merritt 2013). 

In order to evaluate the distribution of the binary parameters associated with double disruptions, it is useful to 
determine  $r_{\rm min}$, defined as the distance of the binary from the BH
before experiencing the change $\Delta L$ in $l_{\rm circ}$,  separating the two regimes. Considering two-body relaxation over a time-scale $t_{\rm r}=0.065(G M_{\rm BH}/r)^{3/2}/[G^{2}M_{\rm*}^{2}n(r) \ln\Lambda]$ (Spitzer \& Hart 1971) as the main mechanism which drives changes in angular momentum, it is known that the change in specific circular angular momentum $l_{\rm circ}(r)$ over a period $T$ is of the order of
\begin{equation}
\Delta L\sim (T/t_{\rm r})^{1/2}l_{\rm circ}
\end{equation}
(Merritt 2013). Thus, the critical condition $\Delta L\sim l_{\rm CM}(r_{\rm tb})$ enables us to infer
$r_{\rm min}$. If the binary is orbiting inside a Bahcall-Wolf density profile $n(r)=n_{\rm 0}\bigl(r/r_{\rm 0}\bigr)^{-7/4}$ (Bahcall \& Wolf 1976), $r_{\rm min}$  reads
\begin{multline}
r_{\rm min}\sim \Biggl[\frac{0.065}{2\pi n_{\rm 0}r_{\rm 0}^{7/4} \ln \Lambda}\Biggr]^{4/9} \Bigl(\frac{M_{\rm BH}}{M_{\rm *}}\Bigr)^{28/27} a_{\rm bin}^{4/9}\\ \sim10^{7} \rm R_{\rm \odot} \Bigl(\frac{1.3\times 10^6 \rm pc^{-3}}{\it n_{\rm 0}}\Bigr)^{4/9} \Bigl(\frac{0.3 \rm pc}{\it r_{\rm 0}}\Bigr)^{7/9} \Bigl(\frac{10}{\ln \Lambda}\Bigr)^{4/9} \\ \times \Bigl(\frac{\it M_{\rm BH}}{10^{6} \rm M_{\rm \odot}}\frac{1 \rm M_{\rm \odot}}{\it M_{\rm *}}\Bigr)^{28/27} 
 \Bigl(\frac{\it a_{\rm bin}}{10 \rm R_{\rm \odot}}\Bigr)^{4/9}, \label{eq9}
\end{multline}
taking $n_{\rm 0}$ and $r_{\rm 0}$ as for the Milky Way (Merritt 2010). We note that $r_{\rm min}$ is comparable to the radius of gravitational  influence of a BH 
\begin{equation}
r_{\rm h}=\frac{G M_{\rm BH}}{\sigma^2}\sim 5\times 10^7 \rm R_{\rm \odot} 
\left ({M_{\rm BH}\over 10^6\, {\rm M_\odot}}\right )
\Bigl(\frac{65 \rm km/s}{\it \sigma}\Bigr)^{2}
\end{equation}
(Peebles 1972; Merritt 2000).
 
A binary carries internal degrees of freedom, and in particular the relative velocity of the two binary components,  $\sqrt{G M_{\rm bin}/a_{\rm bin}}$,  is clearly smaller than the orbital velocity of the binary CM relative to the BH, $\sqrt{G M_{\rm BH}/r}$. The velocity of the two stars relative to the centre of mass of the stellar binary gives then a small contribution to the specific angular momentum of each binary star relative to the BH at $r_{\rm tb}$ that approximately is
\begin{equation}
\delta l \sim \sqrt{G M_{\rm bin} a_{\rm bin}}\Bigl(\frac{M_{\rm BH}}{M_{\rm bin}}\Bigr)^{1/3}.
\end{equation}
Sequential disruptions are expected to be favoured when $\delta l$ is small. Indeed, the smaller $\delta l$ is, the more each binary component has an orbit around the BH similar to the one of the binary CM, i.e. a similar pericentre passage. Thus, we require
\begin{equation}
\frac{\delta l}{l_{\rm CM}(r_{\rm t})}\sim \sqrt{\frac{a_{\rm bin}}{R_{\rm*}}}\Bigl(\frac{M_{\rm*}}{M_{\rm BH}}\Bigr)^{1/3}\ll 1,
\end{equation}
where we approximated $M_{\rm bin}\sim M_{\rm *}$. For $M_{\rm *}=1\rm M_{\rm \odot}$, $R_{\rm *}=1\rm R_{\rm \odot}$, $M_{\rm BH}=10^{6} \rm M_{\rm \odot}$ we need $a_{\rm bin}\ll 10^{4} \rm R_{\rm \odot}$. Hence, the second condition for double TDEs, which joins the condition on $\Delta L$, is the involvement of close binaries. Furthermore, very close binaries are required in order to avoid their evaporation due to interactions with field stars before tidal binary break-up (Merritt 2013).

In the full loss cone regime, the parameter space of binaries that can undergo double TDEs can be inferred from the rate of binary entrance in the region of stellar TDEs per unit of $r$ and $a_{\rm bin}$ as found in Mandel \& Levin (2015):
\begin{equation}
\frac{d^{3}N (a_{\rm bin}, r)}{dr da_{\rm bin} dt} \sim \Bigl(\frac{l_{\rm CM}(r_{\rm t})}{l_{\rm CM} (r_{\rm tb})}\Bigr)^{2}\frac{4\pi r^{2}n(r)\xi(a_{\rm bin})}{T}, \label{eq10}
\end{equation}
where $\bigl(l_{\rm CM}(r_{\rm t})/l_{\rm CM}(r_{\rm tb})\bigr)^{2}$
is the probability for a binary to enter directly the single TDE region (Merritt 2013) and 
$\xi(a_{\rm bin})=\bigl[ \ln\bigl(a_{\rm max}/a_{\rm min}\bigr)\bigr]^{-1} a_{\rm bin}^{-1}$ is the distribution function for $a_{\rm bin}$ given in $\rm \ddot Opik$ (1924), with $a_{\rm max}$ and $a_{\rm min}$ being the maximum and the minimum semimajor axes of stellar binaries in a generic galactic field.

Integration of equation \ref{eq10} over $r$, between $r_{\rm min}$ and $+\infty$, enables us to evaluate the number of binaries that may undergo sequential tidal disruption of their components per unit of time and unit of $a_{\rm bin}$. The resulting integral scales as 
\begin{equation}
\frac{d^{2}N (a_{\rm bin})}{da_{\rm bin} dt} \propto \Bigl[\ln\Bigl(\frac{a_{\rm max}}{a_{\rm min}}\Bigr)\Bigr]^{-1} R_{\rm *} a_{\rm bin}^{-19/9}.  \label{eqrate}
\end{equation}

From Kepler's law, we can connect $a_{\rm bin}$ with the internal orbital period of the stellar binaries $P_{\rm bin}$ to infer the number of events per unit of time and unit of $P_{\rm bin}$. The resulting rate is
\begin{equation}
\frac{d^{2}N (P_{\rm bin})}{dP_{\rm bin}dt} \propto \Bigl[\ln\Bigl(\frac{a_{\rm max}}{a_{\rm min}}\Bigr)\Bigr]^{-1}  R_{\rm *} P_{\rm bin}^{-47/27}. 
\end{equation}
We use this scaling to extract the initial conditions of our SPH simulations. 

Thus, in the case of solar mass stars (i.e. $R_{\rm *}=1 \rm R_{\rm \odot}$), the contribution of double TDEs to all TDEs could be approximately estimated by integrating
equation \ref{eqrate} over $a_{\rm bin}$ between $1$ and $10^{4} \rm R_{\rm \odot}$ and dividing it by the corresponding integral obtained after integration over $r$ of equation \ref{eq10}, with $R_{\rm *}$ in place of $a_{\rm bin}$ (also in equation \ref{eq9}) \footnote{Note that substituting $R_{\rm *}$ to $a_{\rm bin}$ in equation \ref{eq10}, $d^3N(R_{\rm *}, r)/drdR_{\rm *}dt \sim 4\pi r^2n(r)/T$.}.
This ratio scales as $(19/9)\bigl[\ln\bigl(a_{\rm max}/a_{\rm min}\bigr)\bigr]^{-1}$, which gives a maximum of $\sim 20$ per cent assuming $a_{\rm max}=10^{4} \rm R_{\rm \odot}$ and $a_{\rm min}=1 \rm R_{\rm \odot}$ and considering that the multiplicity of stars is single:double $\sim$ 50:50 for 100 solar-type stars (Duquennoy \& Mayor 1991b), disregarding uncertainties in the number of very close binaries.

The definition of the parameter space of binaries that may be double tidally disrupted is fundamental to guide us to sensibly define the initial conditions of a small number of
representative low-resolution simulations aimed at checking different outcomes from different initial parameters, and particularly from different pericentre radii of the binary CM.     

\section{General parameter definition for low-resolution SPH simulations} \label{low}
The simulations in this paper are performed using the TreeSPH code \textsc{\small GADGET}2 (Springel 2005). In SPH codes, a star is represented by a set of gas particles. Each particle is characterized by a spatial distance, the smoothing length, over which its properties are `smoothed' by its kernel function, i.e. evaluated by summing the properties of particles in the range of the kernel according to the kernel itself (Price 2005). In particular, in \textsc{\small GADGET}2 the smoothing length of each particle is defined so that its kernel volume contains a constant mass, and is allowed to vary with time, thus adapting to the local conditions. The kernel adopted here is the one used most commonly and is based on cubic splines (Monaghan \& Lattanzio 1985). On the other hand, gravitational interactions between particles are computed through a hierarchical oct-tree algorithm, which significantly reduces the number of pair interactions needed to be computed. The definition of a gravitational softening length $\epsilon \sim 0.1 R_{\rm *}/(N_{\rm part})^{1/3}$, where $N_{\rm part}$ is the total number of particles, prevents particle overlapping. 
\textsc{\small GADGET}2 enables us to follow the temporal evolution of single particle properties and to infer from them TDE light curves (see Section \ref{subsection}). 

We run 14 low-resolution simulations of parabolic encounters between equal-mass binaries and BHs (LE runs) to test the nature of the outcomes for 
different initial conditions, varying binary parameters, $M_{\rm BH}$ and $r_{\rm p}$.
The stellar binaries are first evolved in isolation for several dynamical times to ensure their stability.
The BH force is implemented in the code analytically, as a Newtonian potential, and particles which fall below the innermost stable circular orbit radius $R_{\rm ISCO}$ are excised from simulations.
We consider equal solar mass stars modelled as polytropes of index 5/3 and we sample each of them with $10^3$ particles. Some correspondent high-resolution simulations are presented in Section \ref{subsection}.
The initial binary internal orbital periods $P_{\rm bin}$ and semimajor axes $a_{\rm bin}$ are extracted according to the distributions described in Section \ref{basics}. 
Based on the work of  Duquennoy \& Mayor (1991a), we consider binaries with $0.1$d ($a_{\rm bin}\sim 1 \rm R_{\rm \odot}$) $< P_{\rm bin} < 10$d ($a_{\rm bin}\sim 10 \rm R_{\rm \odot}$) to be circular, binaries with $10$d $\leq P_{\rm bin} \leq 1000$d ($a_{\rm bin}\sim 500 \rm R_{\rm \odot}$) to have internal eccentricities distributed according to a Gaussian of mean 0.3 and standard deviation 0.15 and binaries with $1000$d $< P_{\rm bin} < 1000$yr ($a_{\rm bin}\sim 10^{4} \rm R_{\rm \odot}$) to have internal eccentricities which follow a thermal distribution $p(e_{\rm bin})\sim 2e_{\rm bin}$. 
In order to avoid immediate collisions between the binary components, the initial pericentre radius of the internal binaries (i.e. $a_{\rm bin}(1-e_{\rm bin})$) is set greater than twice the sum of the stellar radii, which are 
\begin{equation}
R_{\rm*}=\Bigl(\frac{M_{\rm*}}{\rm M_{\rm \odot}}\Bigr)^{k} \rm R_{\rm \odot},
\end{equation}
with $k=0.8$ for $M_{\rm*}< 1\rm M_{\rm \odot}$ and $k=0.6$ for $M_{\rm*}> 1 \rm M_{\rm \odot}$, according to Kippenhahn \& Weigert (1994), $R_{\rm *}=1 \rm R_{\rm \odot}$ for $M_{\rm *}=1 \rm M_{\rm \odot}$. 
Binaries are then placed on parabolic orbits around the BH at an initial distance 10 times greater than the tidal binary break-up radius $r_{\rm tb}$, thus preventing initial tidal distortions from the BH.  BHs of masses $10^{5}$ and $10^{6} \rm M_{\rm \odot}$ are considered.  The nominal pericentre distances $r_{\rm p}$ are generated between $1$ and $300 \rm R_{\rm \odot}$ (Mandel \& Levin 2015). 
Stars are placed on Keplerian orbits, and their positions and velocities relative to their binary centre of mass and to the BH are assigned accordingly. The initial internal binary plane is set, arbitrarily, perpendicular to the orbital plane around the BH. The results of these simulations are shown in Section \ref{following}.

\section{Outcomes of  low-resolution SPH simulations}  \label{following}
Tables \ref{1} and \ref{2} summarize the results of our low-resolution simulations as a function of $M_{\rm BH}$, $a_{\rm bin}$ and $r_{\rm p}$. 
\begin{table}
\centering
\caption{Outcomes of our low-resolution SPH simulations of parabolic binary-BH encounters ($M_{\rm *}=1 \rm M_{\rm \odot}$, $R_{\rm *}=1 \rm R_{\rm \odot}$) as a function of $a_{\rm bin}$ and $r_{\rm p}$. Here $M_{\rm BH}=10^{6}\rm M_{\rm \odot}$, $r_{\rm t}=100.0 \rm R_{\rm \odot}$, $r_{\rm tb}({a_{\rm bin}=4.9 \rm R_{\rm \odot})}=390.0 \rm R_{\rm \odot}$, $r_{\rm tb}({a_{\rm bin}=9.8 \rm R_{\rm \odot})}=780.0 \rm R_{\rm \odot}$. TD-TDE stands for total double TDE, ATD-TDE for almost total double TDE (i.e. more than $\sim 70\%$ of stellar mass lost), PD-TDE for partial double TDE,  MG for merger, BBK for binary break-up without stellar disruptions.\label{1}}
\begin{tabular}{c c c c c c c}
\hline
\footnotesize{$a_{\rm bin}$$\backslash$$r_{\rm p}$} &  \footnotesize{$50.0$} & \footnotesize{$100.0$} & \footnotesize{$142.6$} & \footnotesize{$200.0$} & \footnotesize{$420.0$} & \footnotesize{$780.0$} \\
$(\rm R_{\rm \odot})$ & & & & & & \\
\hline
\small{$4.9$} & \footnotesize{LE1:} & \footnotesize{LE2:} & \footnotesize{LE3:} & \footnotesize{LE4:} & \footnotesize{LE5:} &  \\
 & \footnotesize{TD-} &\footnotesize{ATD-} & \footnotesize{PD-} & \footnotesize{PD-} & \footnotesize{MG} &  \\
 &  \footnotesize{TDE} & \footnotesize{TDE} & \footnotesize{TDE} & \footnotesize{TDE} & \\
& & & & & & \\
\small{$9.8$} & & \footnotesize{LE9:} &  & \footnotesize{LE10:} & & \footnotesize{LE11:} \\
 &  & \footnotesize{ATD-} & & \footnotesize{PD-} & & \footnotesize{BBK} \\
 & & \footnotesize{TDE} & & \footnotesize{TDE} & & \\
\hline
\end{tabular}
\end{table}
\begin{table}
\centering
\caption{Same as Table \ref{1}, with $M_{\rm BH}=10^{5}\rm M_{\rm \odot}$, $r_{\rm t}=50.0 \rm R_{\rm \odot}$, $r_{\rm tb}({a_{\rm bin}=4.9 \rm R_{\rm \odot})}=180.0 \rm R_{\rm \odot}$, $r_{\rm tb}({a_{\rm bin}=9.8 \rm R_{\rm \odot})}=360.0 \rm R_{\rm \odot}$. PD-TDE stands for partial double TDE, MG for merger, BBK for binary break-up without stellar disruptions, UN for undisturbed binary. \label{2}}
\begin{tabular}{c c c c c c c}
\hline
\footnotesize{$a_{\rm bin}$$\backslash$$r_{\rm p}$} & \footnotesize{$50.0$} & \footnotesize{$100.0$} & \footnotesize{$142.6 $} & \footnotesize{$200.0$} & \footnotesize{$420.0$} & \footnotesize{$780.0$} \\
$(\rm R_{\rm \odot})$ & & & & & & \\
\hline\small{$4.9$} & & \footnotesize{LE6:} & & \footnotesize{LE7:} & \footnotesize{LE8:} & \\
 & &  \footnotesize{PD-} & & \footnotesize{MG} & \footnotesize{UN} & \\
 & & \footnotesize{TDE} & & & & \\
& & & & & & \\
\small{$9.8$} & &  \footnotesize{LE12:}&  & \footnotesize{LE13:} & & \footnotesize{LE14:} \\
 & &  \footnotesize{PD-} & & \footnotesize{BBK} & & \footnotesize{UN} \\
 & &  \footnotesize{TDE} & & & & \\ 
\hline
\end{tabular}
\end{table}
Several outcomes from binary-BH encounters are possible, including the results of our simulations:\\
\begin{enumerate}
\item PD-TDE: partial double TDE, 
\item ATD-TDE: almost total double TDE, i.e. more than $\sim 70\%$ of stellar mass is lost, 
\item P\&T-TDE: single partial plus single total TDE, 
\item TD-TDE: total double TDE, 
\item MG: merger of the binary components, 
\item BBK: tidal binary break-up without stellar tidal disruptions,
\item UN: undisturbed binary.
\end{enumerate}
The intensity of the disruptions, i.e. the morphology of the resulting objects, is estimated from our simulation results based on
the tidal deformation, the extent of stellar mass loss and possible orbital changes of the binary stars with pericentre passage. 
After closest approach, the orbital evolution of the binary stars around the BH is computed using an $N$-body Hermite code (e.g. Hut \& Makino 1995; the code can be freely downloaded from https://www.ids.ias.edu/$\sim$piet/act/comp/algorithms/starter/), knowing the current position and velocity of the centre of mass of each binary component from SPH simulations (see Section \ref{subsection} for the recipe used to infer the position and velocity of the centres of mass). The use of the Hermite code enables us to overcome the high computational time required by SPH simulations to track the dynamics of stars when the bulk of the hydrodynamical processes have subsided. 

Appendix \ref{appA} contains an inventory of representative orbits according to the classification highlighted above. Tables \ref{app1} and \ref{app2}, respectively, refer to the simulations described in Tables \ref{1} and \ref{2}. There, we show the orbital evolution of the binary components around the BH, for each simulation, in the ($x,y$) and ($y,z$) planes, starting from (0,0), (0,0). Units are in $\rm R_{\rm \odot}$. Blue curves represent the initial parabolic orbits of the binary CM around the BH, each inferred from the position of the BH and the pericentre radius $r_{\rm p}$. Red curves trace the orbital evolution of the binary components as inferred from SPH simulations, while green curves trace the orbital evolution of the stars as computed using the Hermite code. Black dots indicate the position of the BH. Mergers (MGs; LE5, LE7) are found when the two binary components progressively reduce their relative separation starting from just before the pericentre passage around the BH, without being tidally separated. The MG product, which is represented by stars at a fixed minimum distance in simulations performed using the Hermite code, follows an orbit which overlaps the initial parabolic one of the binary CM. In the UN case (LE8, LE14), 
the binary keeps its internal and external orbits unchanged, even after pericentre passage. D-TDEs (LE1, LE2, LE3, LE4, LE6, LE9, LE10, LE12) are preceded by tidal binary separation, which can also occur
without stellar disruptions (BBK; LE11, LE13). Binary break-up leads one star to get bound to the BH and the other to remain unbound. In the case of BBKs or partial disruptions, the latter may leave the system at a high velocity, becoming a hypervelocity star (Hills 1988; Antonini, Lombardi \& Merritt 2011). 

\section{High-resolution SPH simulations} \label{3.1}
\subsection{A glimpse to simulated double TDEs} \label{glimpse}
The low-resolution simulations described in Sections \ref{low} and \ref{following} serve as guide for the selection of three higher resolution SPH simulations, with an increased number of particles per star equal to $10^5$. A number of particles per star of $10^6$ would require too much computational time. Indeed, the computational cost in \textsc{\small GADGET}2 scales as $N_{\rm part}\log (N_{\rm part})$, which is a factor of 12 higher in the case of $N_{\rm part}=10^6$ with respect to $N_{\rm part}=10^5$.

\begin{figure}
\includegraphics[width=3.cm, angle=-90]{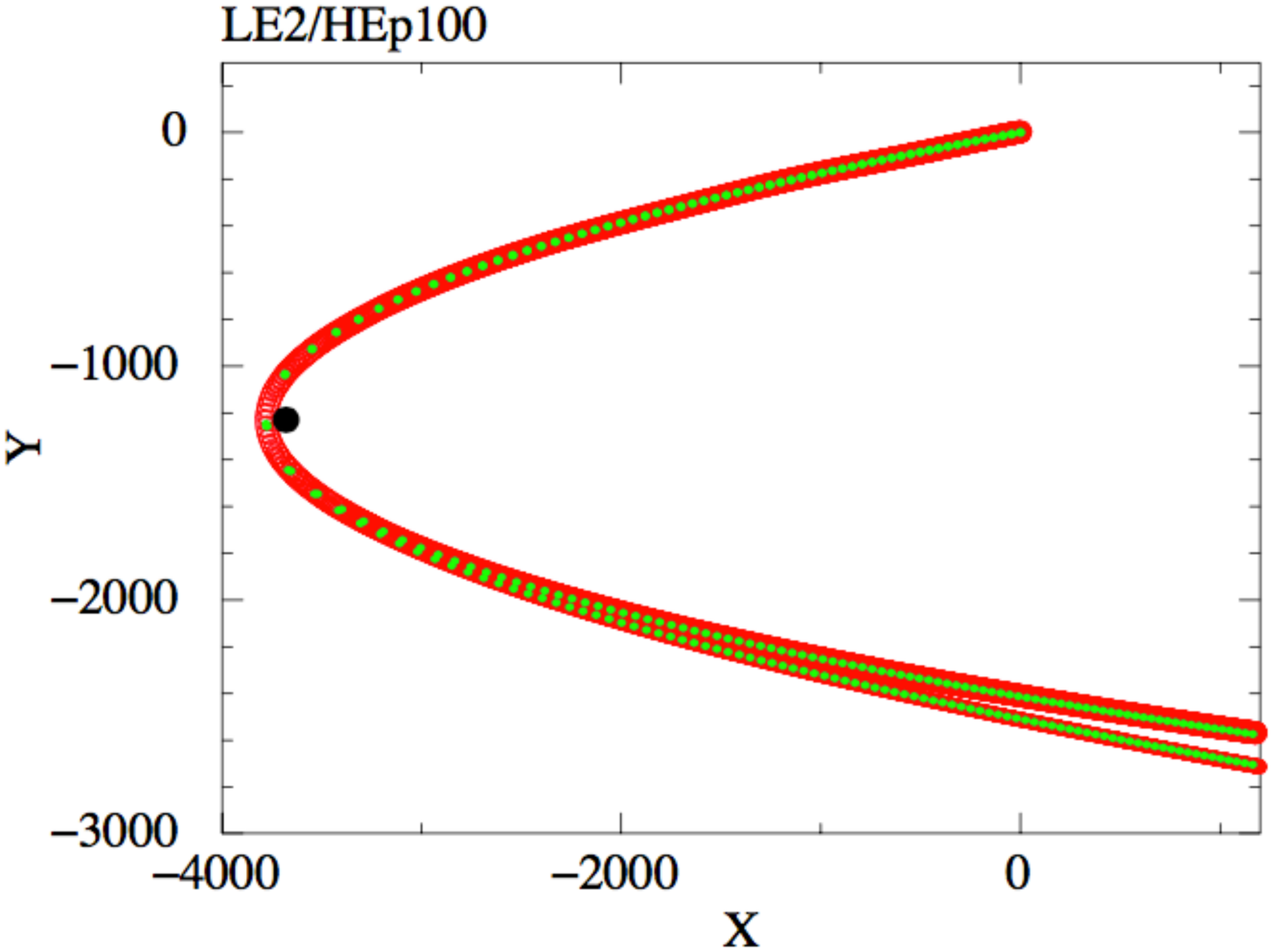} \includegraphics[width=3.cm, angle=-90]{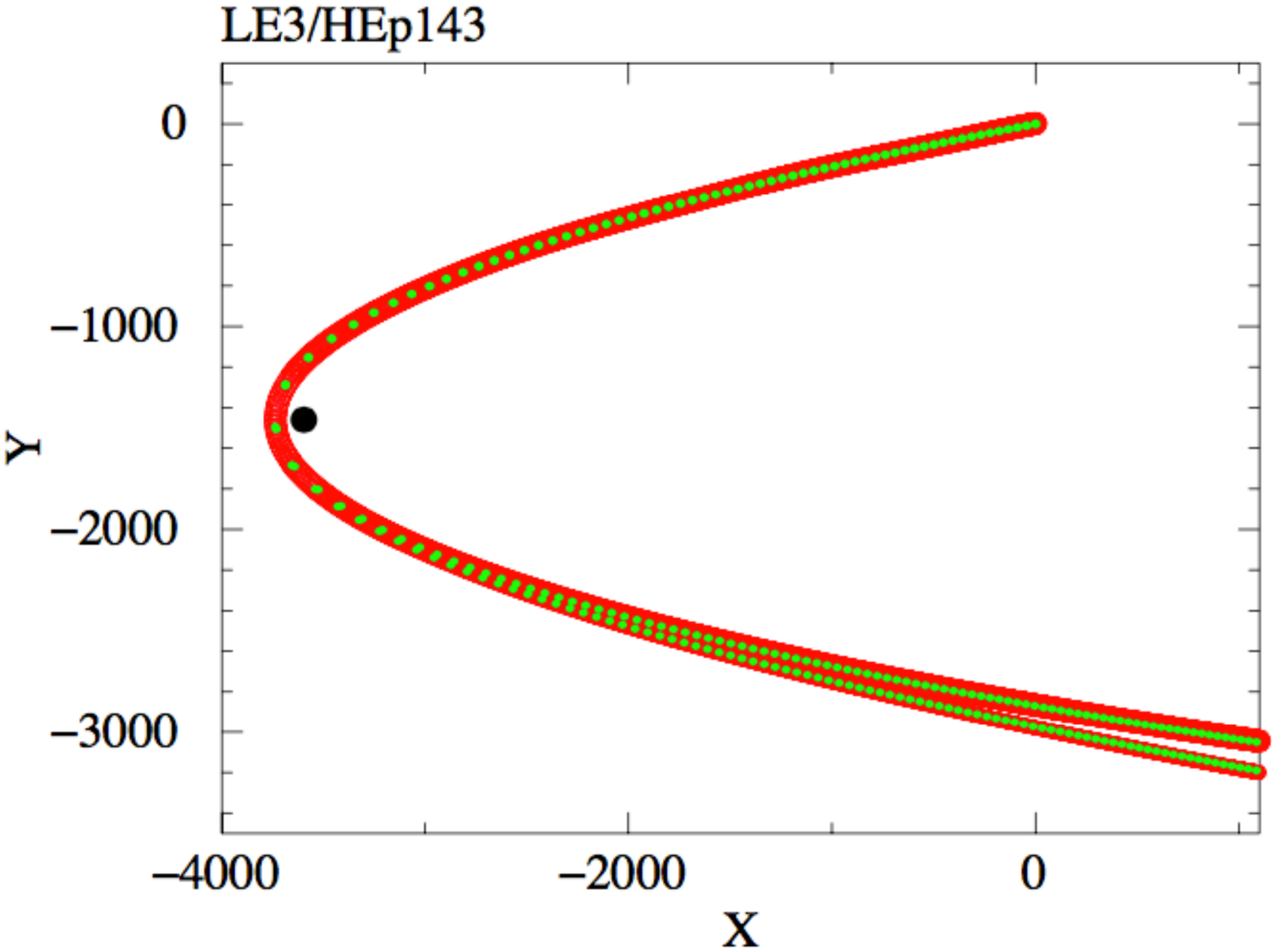} \\
\includegraphics[width=3.cm, angle=-90]{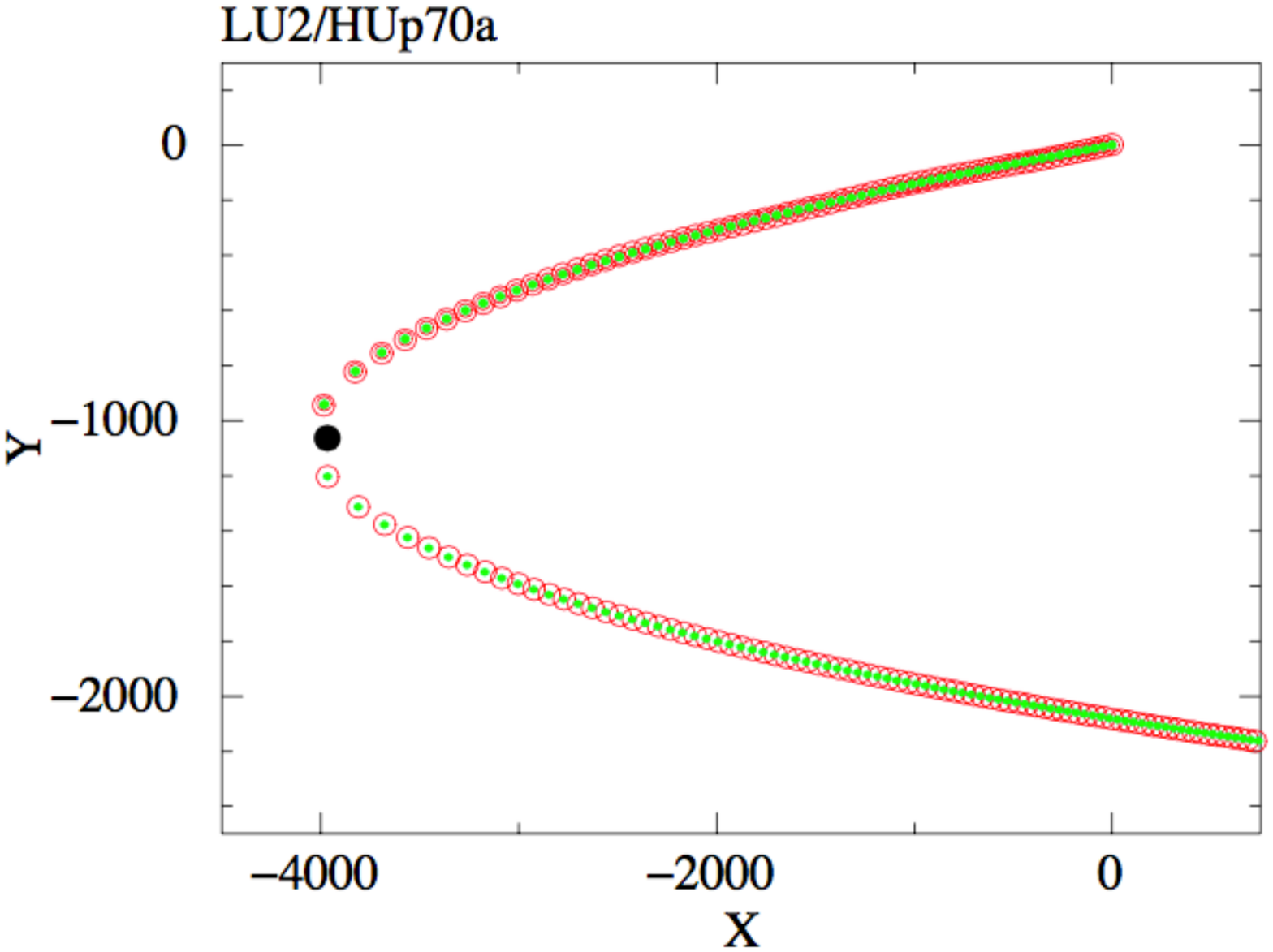} \includegraphics[width=3.cm, angle=-90]{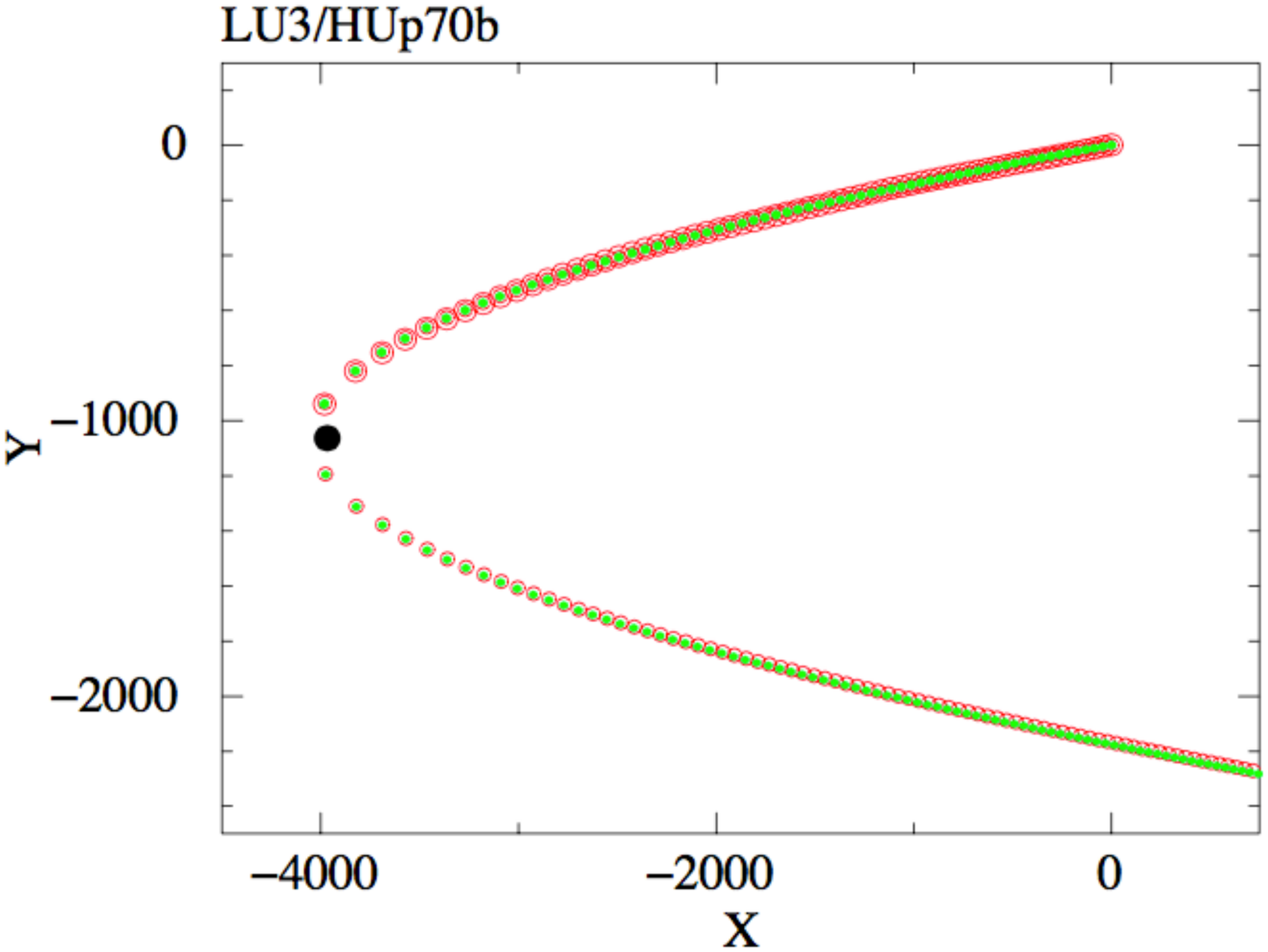}
\caption{Orbital evolution of the binary stars, starting from (0,0) in the ($x,y$) plane, as inferred from the corresponding low- (red curves) and high-resolution (green curves) SPH simulations for  LE2/HEp100 and LE3/HEp143 (upper panels) and LU2/HUp70a and LU3/HUp70b (bottom panels). We do not consider simulations LE1/HEp50 and LU1/HUp42 given that both the binary stars are totally disrupted when approaching the BH. Black dots indicate the position of the BH. Units are in $\rm R_{\rm \odot}$.
\label{new}}
\end{figure}

Our goal is to infer directly from simulations the light curves associated with double TDEs of different intensities. For this reason, we set the initial conditions for an almost total, a partial and a total double disruption event, following simulations LE2 and LE3 for the not fully disruptive events and simulation LE1 in order to obtain a total double disruption. Table \ref{referee3} summarizes the outcomes which come out from these three high-resolution simulations (HE runs) as a function of $a_{\rm bin}$ and $r_{\rm p}$. These results are the same as expected from the corresponding low-resolution simulations  (see Table \ref{1}). Furthermore, Fig. \ref{new} (upper panels) points out that the orbits of the binary stars follow the same evolution in corresponding low- (red curves) and high-resolution (green curves) SPH simulations after pericentre passage, assuring numerical convergence.
\begin{table}
\centering
\caption{Same as Table \ref{1} for our high resolution simulations involving equal-mass binaries. \label{referee3}}
\begin{tabular}{c c c c}
\hline
\footnotesize{$a_{\rm bin}$$\backslash$$r_{\rm p}$} & \footnotesize{$50.0$} & \footnotesize{$100.0$} & \small{$142.6$} \\
$(\rm R_{\rm \odot})$ & & & \\
\hline
\small{$4.9$} & \footnotesize{HEp50:} & \footnotesize{HEp100:} & \footnotesize{HEp143:} \\
 &  \footnotesize{TD-TDE} & \footnotesize{ATD-TDE} & \footnotesize{PD-TDE} \\
\hline
\end{tabular}
\end{table}

\begin{figure*} 
\includegraphics[width=4.05cm, angle=90]{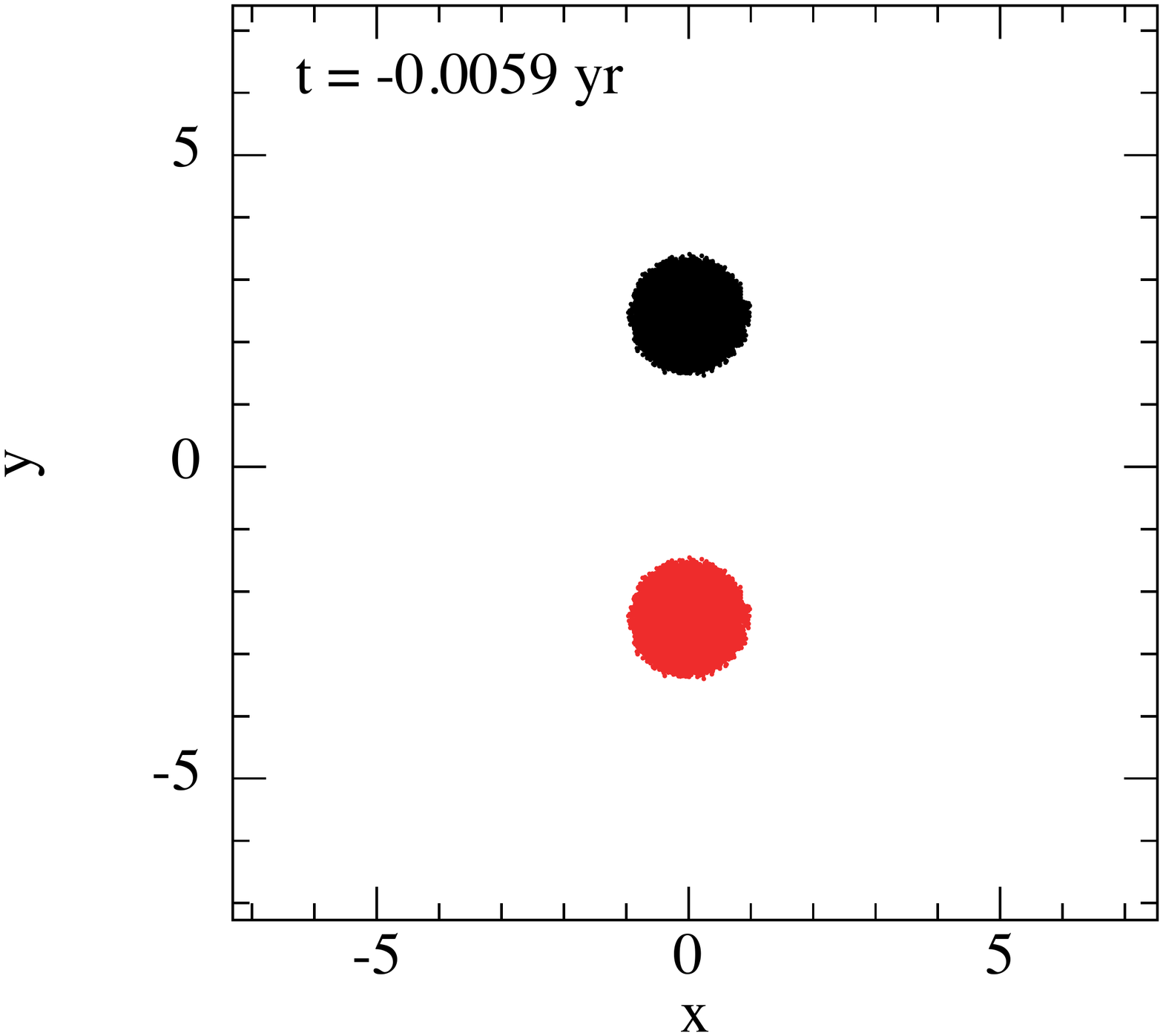} \includegraphics[width=4.05cm, angle=90]{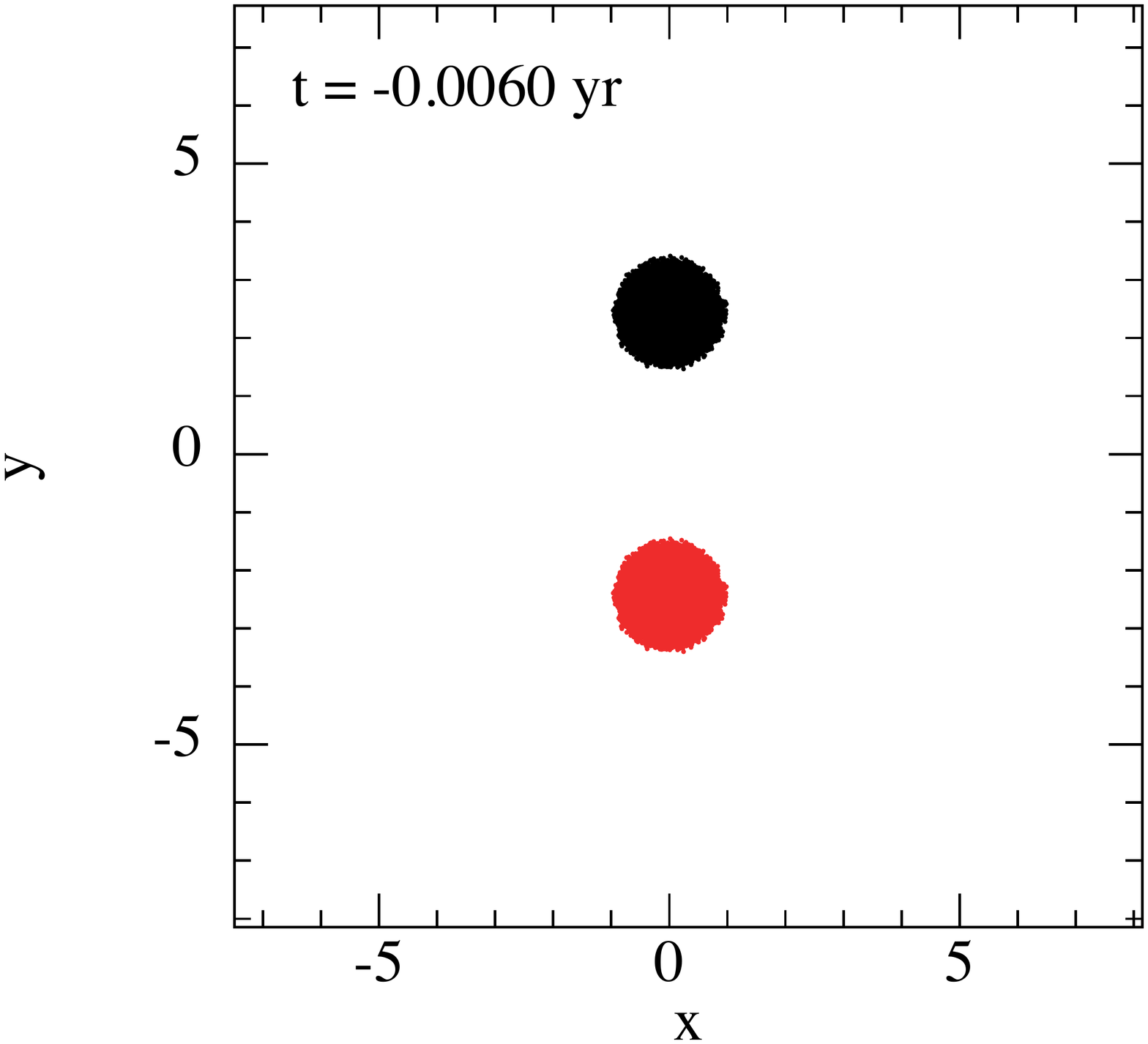}   \includegraphics[width=4.05cm, angle=90]{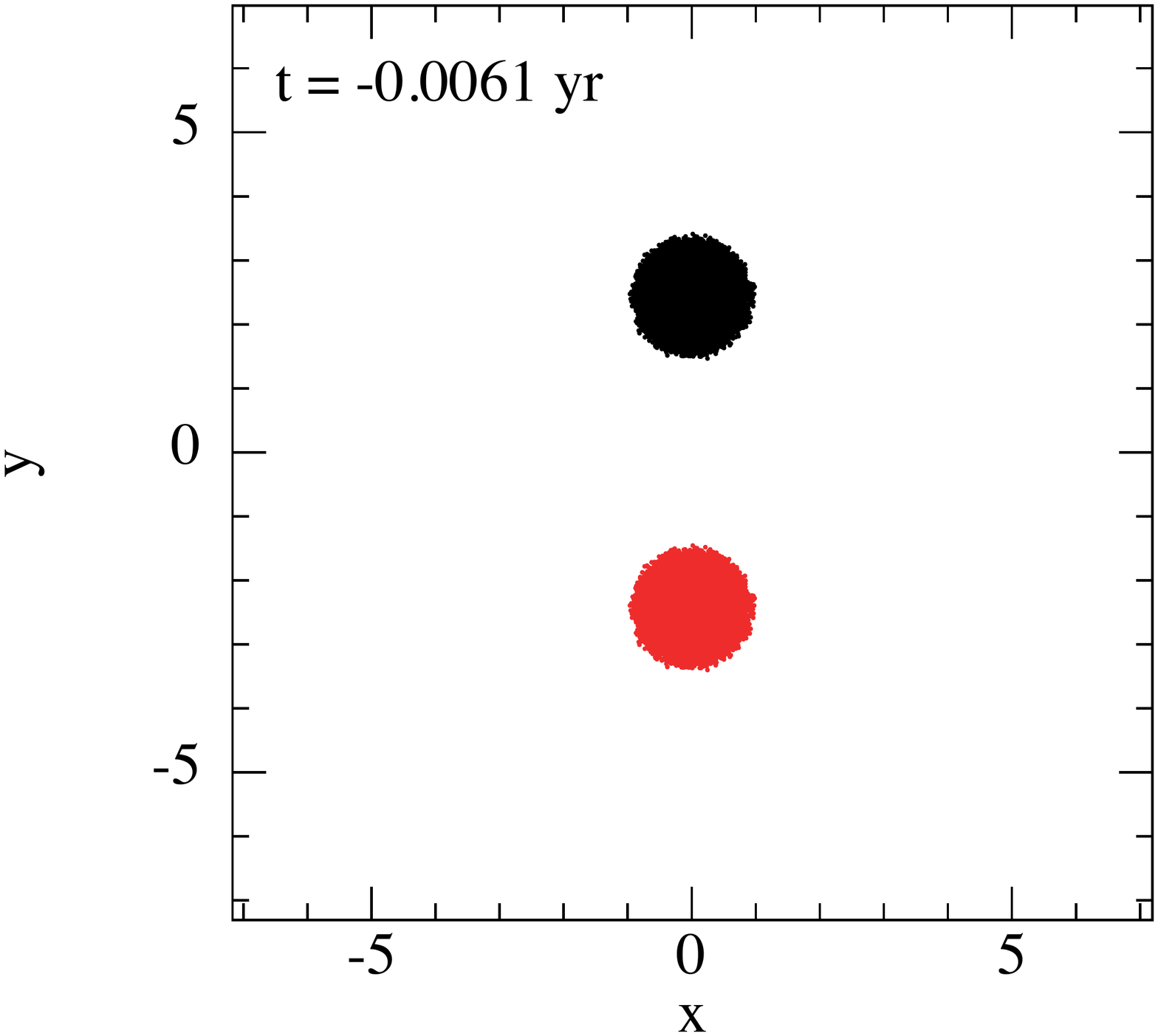} \\   
\includegraphics[width=4.05cm, angle=90]{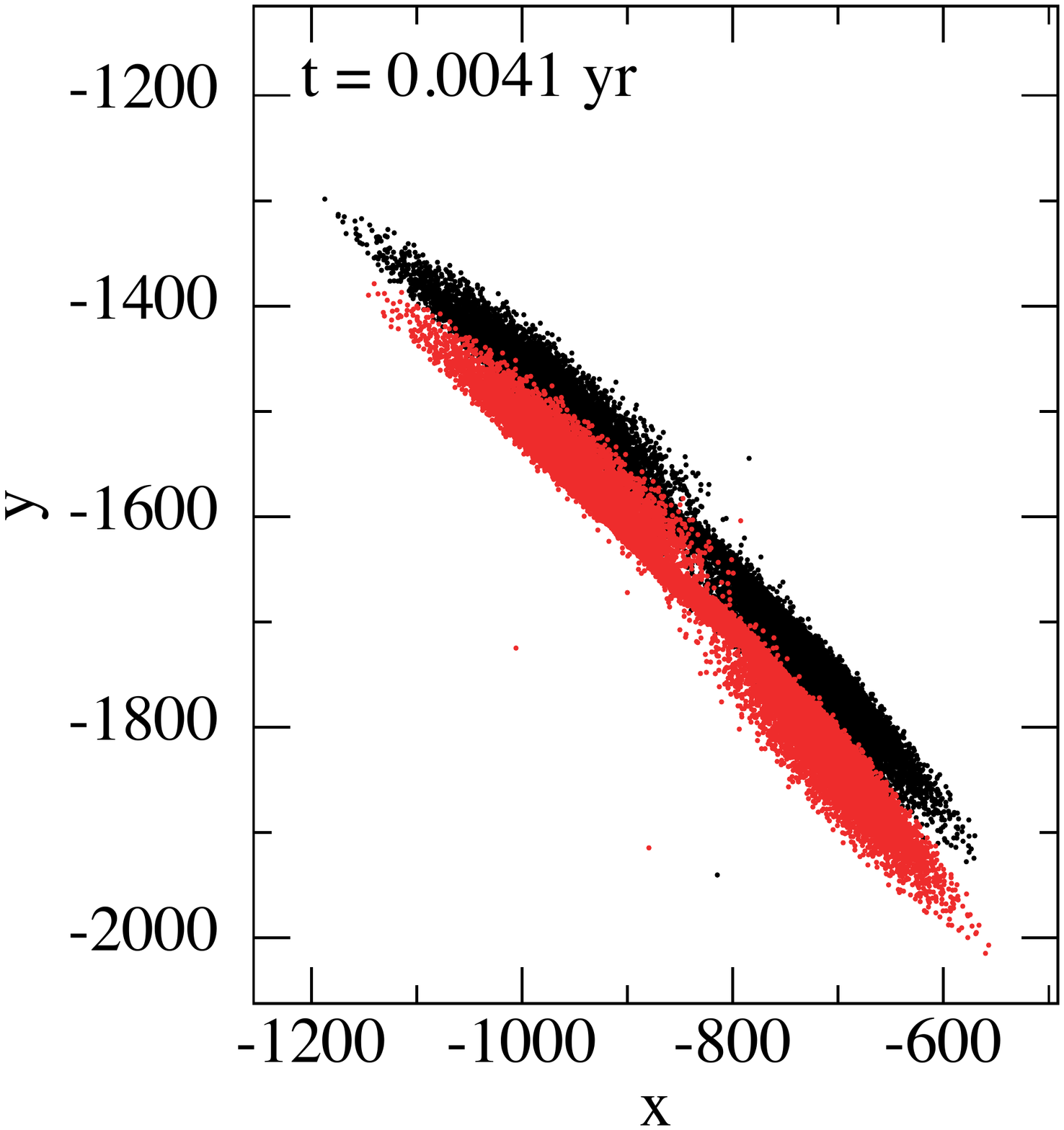} \includegraphics[width=4.05cm, angle=90]{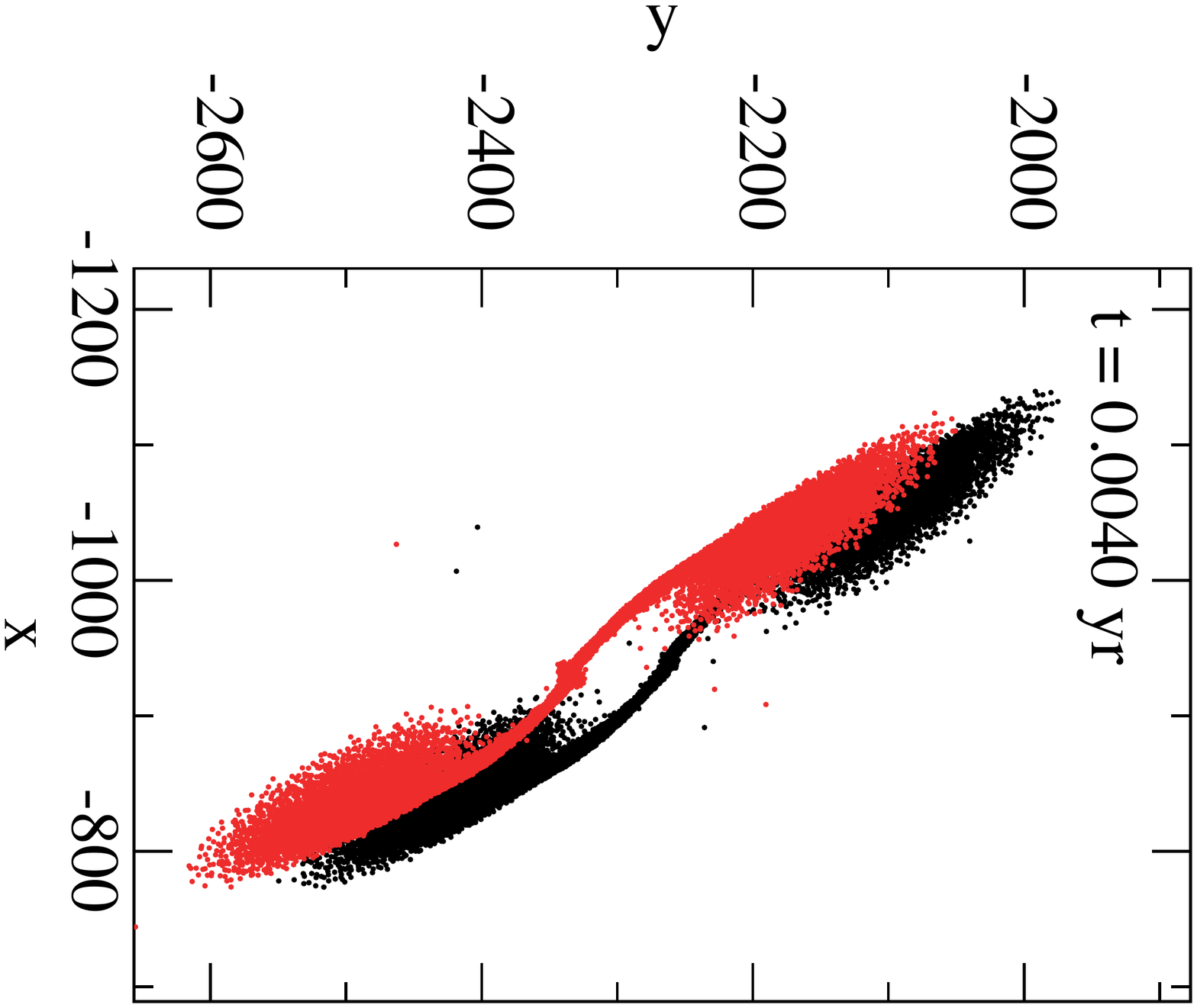}       \includegraphics[width=4.05cm, angle=90]{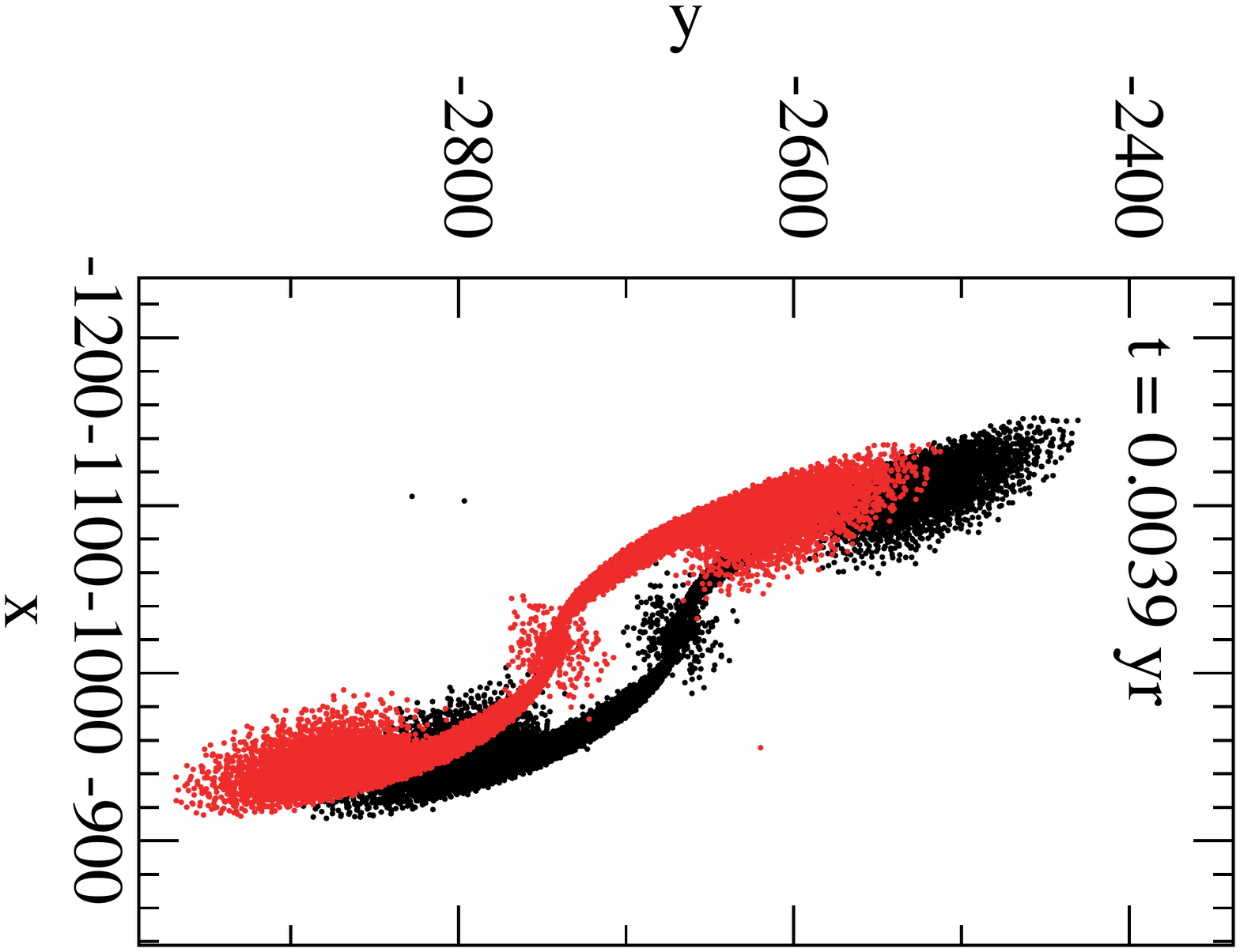}      \\ 
\includegraphics[width=4.05cm, angle=90]{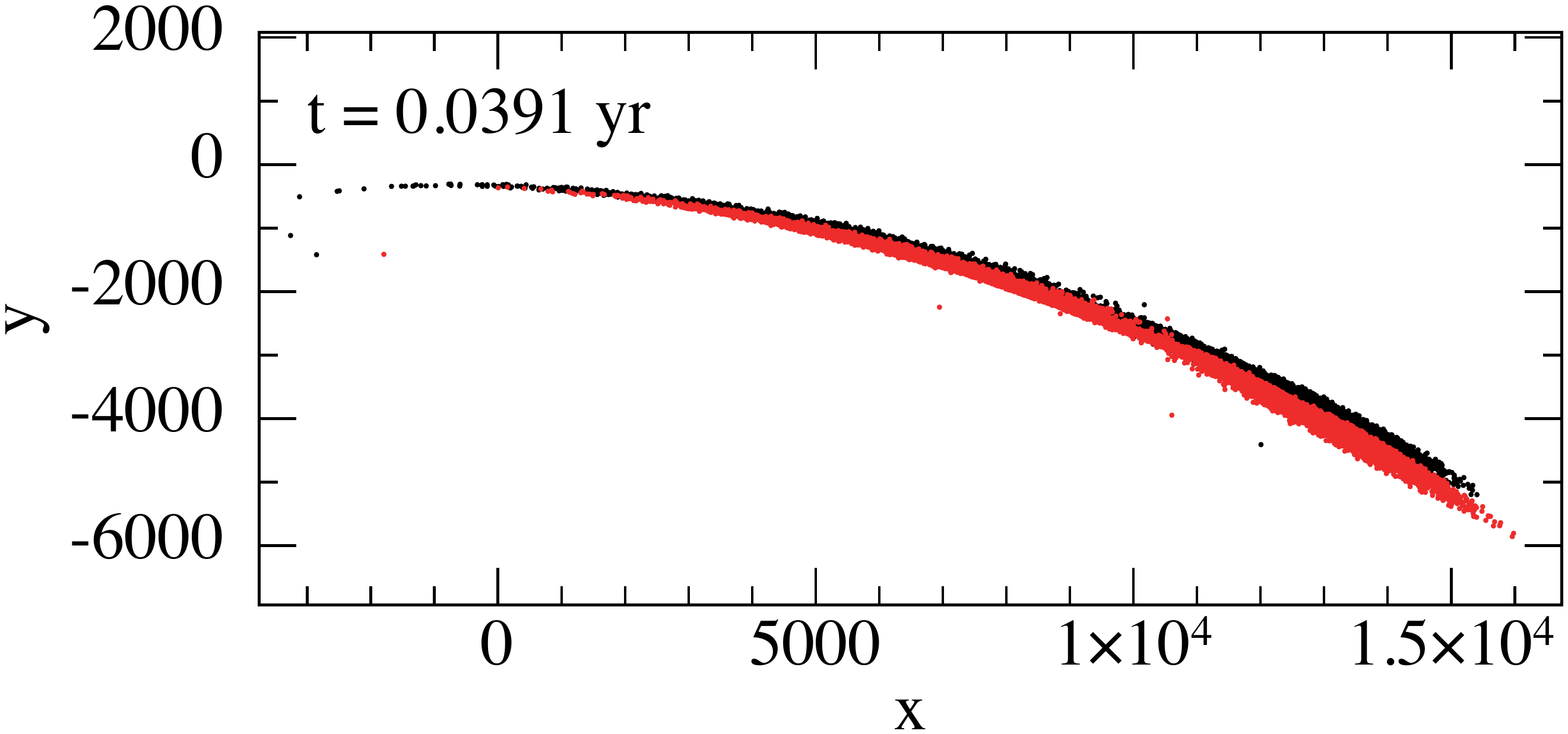} \includegraphics[width=4.05cm, angle=90]{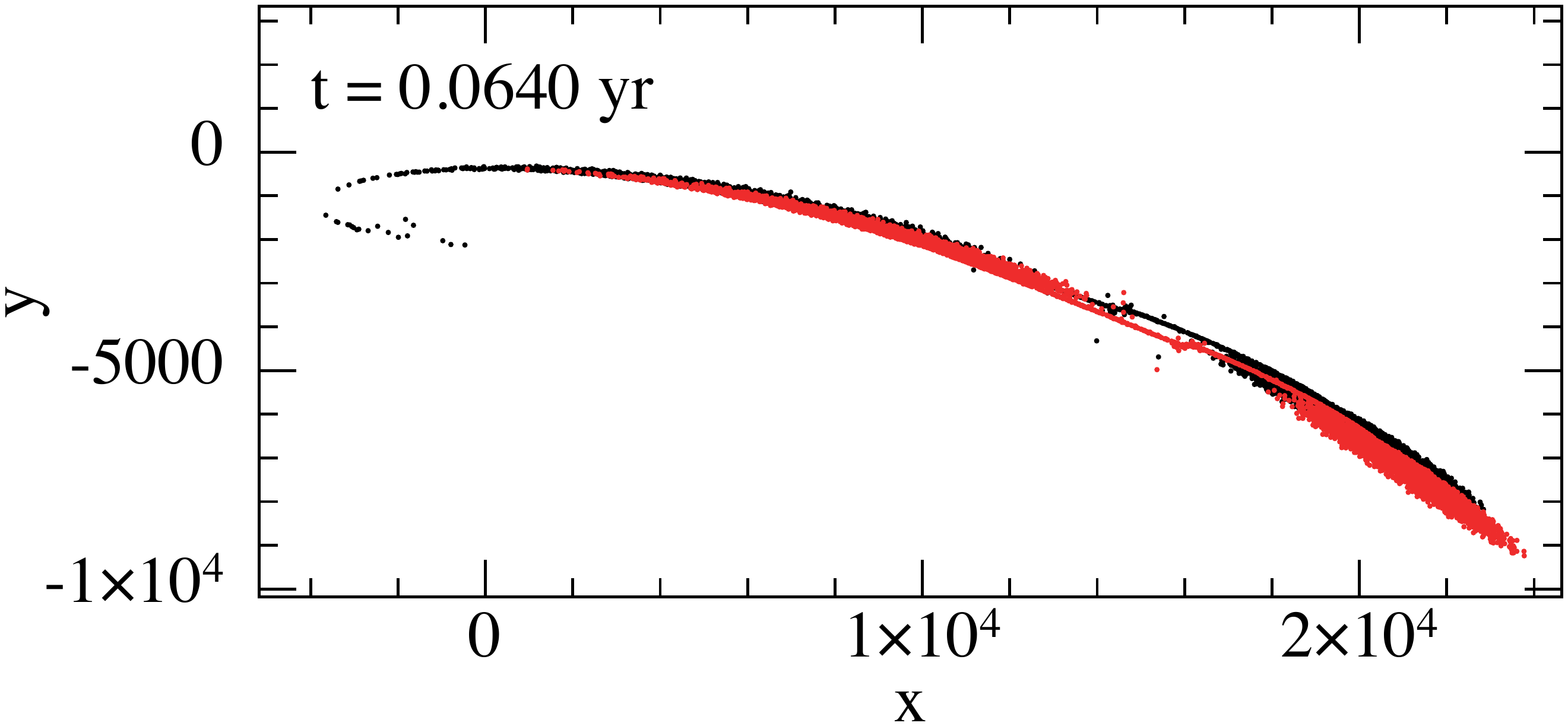}      \includegraphics[width=4.05cm, angle=90]{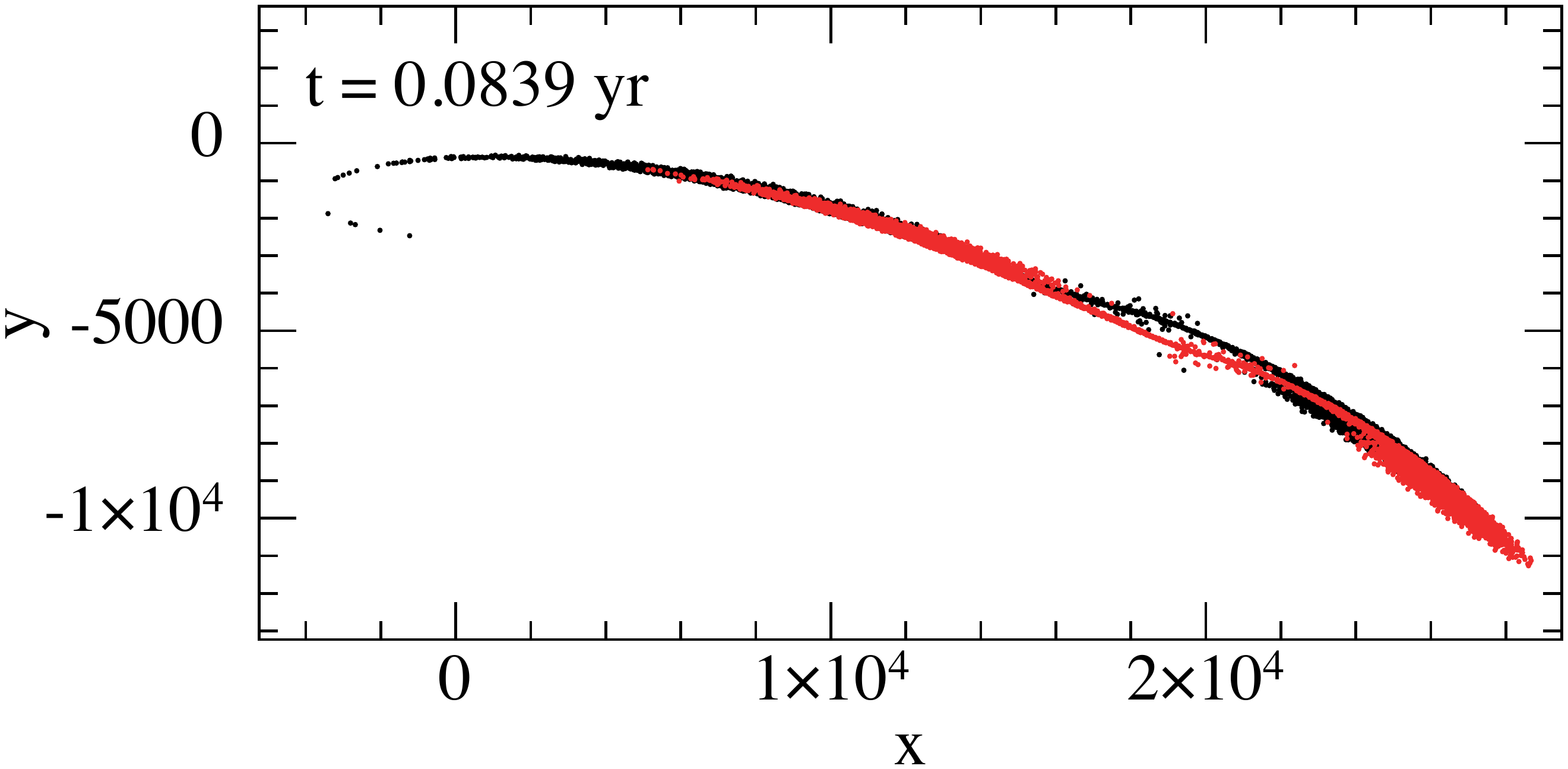}      \\  
\caption{Representative high-resolution snapshots of the SPH particle distribution, respectively, in simulations HEp50 (left column), HEp100 (central column) and HEp143 (right column), projected in the ($x,y$) plane. Positional units are in $\rm R_{\rm \odot}$ and times are in fractions of pericentre time. Black particles originally belong to the star which will get bound to the BH after tidal binary break-up and red particles depict its companion. The BH is at position ($x,y$)=(-3779.62,-875.17) (simulation HEp50), ($x,y$)=(-3679.62,-1229.57) (simulation HEp100), ($x,y$)=(-3594.48,-1459.84) (simulation HEp143). 
The survived binary components are clearly visible in the almost total (HEp100) and partial (HEp143) TDE cases, whereas stars are fully disrupted after pericentre passage in the total TDE case (HEp50).
 \label{snapshots}}
\end{figure*}

\begin{figure}
\includegraphics[width=5.5cm, angle=90]{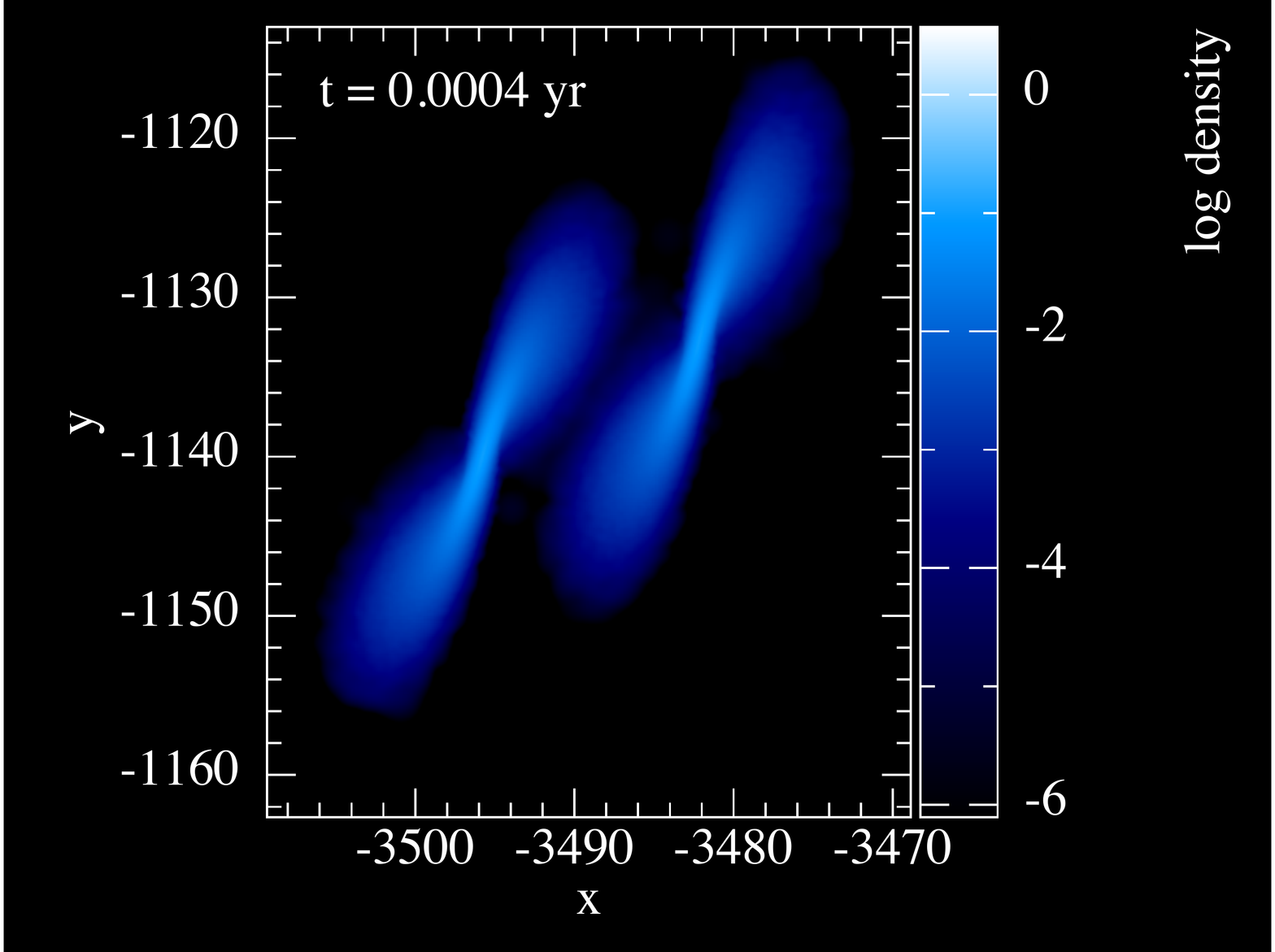}\\
\includegraphics[width=5.5cm, angle=90]{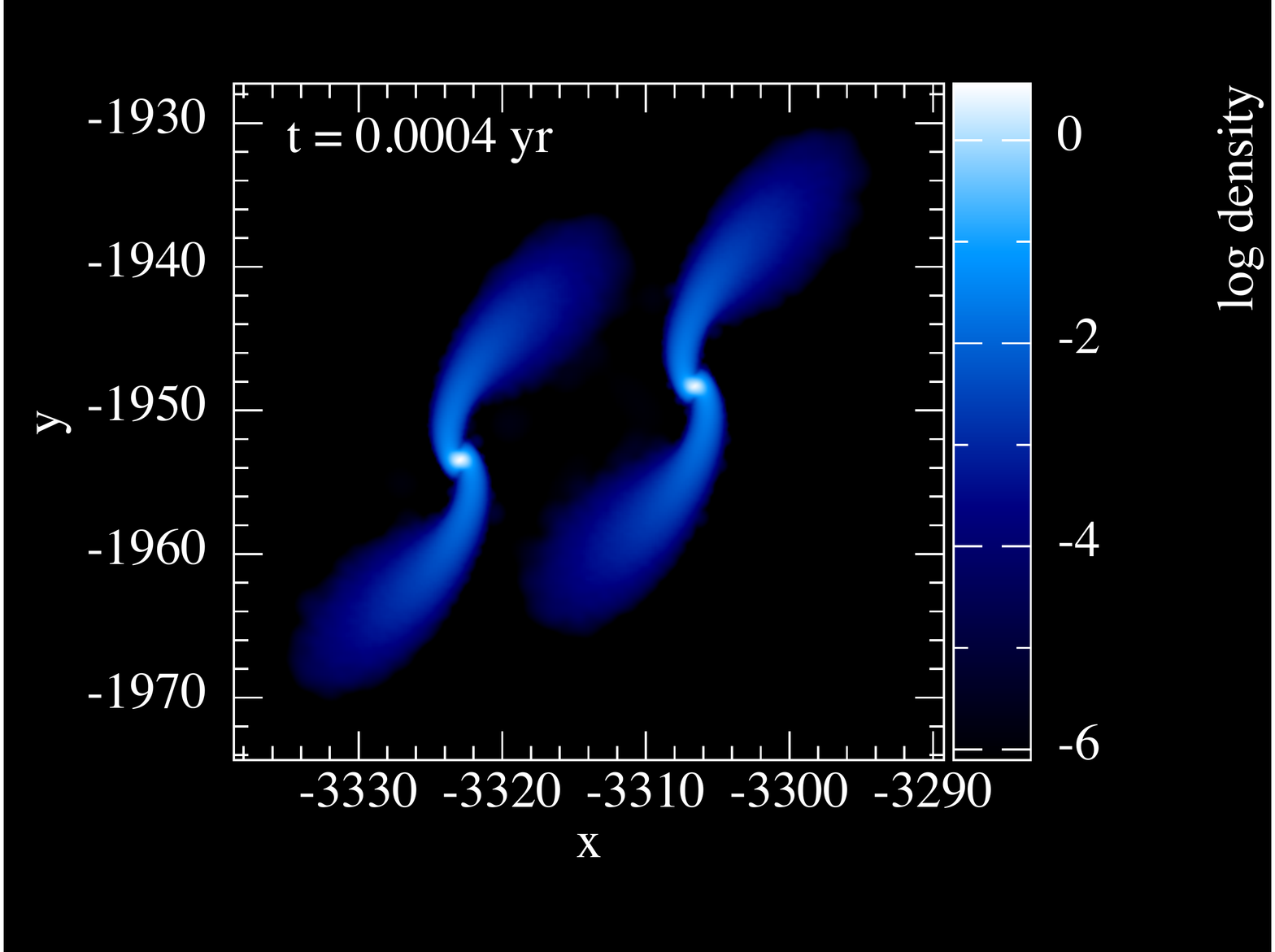} \\
\caption{Snapshots of the SPH particle density (in log scale) for the simulated total double (HEp50; upper panel) and
partial double (HEp143; bottom panel) TDE at $t=0.0004 \rm yr$ ($\sim 0.15 \rm d$) after pericentre passage, projected in the ($x,y$) plane. The remnant binary components are clearly visible in the partial double disruption case. \label{blue}}
\end{figure}

Fig \ref{snapshots} shows representative snapshots of the SPH particle distribution, projected in the ($x,y$) plane and in fractions of pericentre time, depicting the dynamics of simulations HEp50  (left column), HEp100 (central column) and HEp143 (right column).
Panels are in $\rm R_{\rm \odot}$. In each simulation, black particles originally shape the star which will get bound to the BH after binary separation, whereas red particles initially belong to the one which will unbind. The remnant of the binary components after disruption is clearly visible in the almost total (HEp100) and partial double (HEp143) TDE cases. Forward in time, the distribution of the particles which leave the stars once tidally disrupted visibly spreads, and particles originally associated with the two different stars tend to mix, preventing their by-eye distinction.
For this reason, snapshots of the SPH particle distribution are introduced in place of snapshots of the SPH particle density, which are shown for the first time in Fig. \ref{blue} (in log scale), projected in the ($x,y$) plane, only at $0.0004 \rm yr$ ($\sim 0.15 \rm d$) after pericentre passage for the simulated total double (HEp50) and partial double (HEp143) TDE. Again, the remnant of the binary components is clearly visible in the partially disruptive encounter. 

The selection of the stellar debris associated with a specific star is possible thanks to a detailed analysis of the snapshots. This enables us to extract the light curves associated with each single-star disruption and then to infer the composite light curves associated with double disruptions. 
We discuss this in Section \ref{subsection}.      

\subsection{Double TDE light curves: the case of equal-mass binaries}  \label{subsection}
The basic (simplifying) assumption when inferring the light curves associated with TDEs is that the accretion rate on to the BH has close correspondence to the rate of stellar debris which returns to pericentre after disruption.  
Indeed, if the viscous time (Li et al. 2002) driving the fallback of stellar debris on to the BH is negligible compared to the returning time at pericentre of the most bound material since the time of stellar disruption (which is generally the case in our simulations), 
then the rate of debris returning at pericentre  
\begin{equation}
\dot M(t)=\frac{\bigl(2 \pi G M_{\rm BH}\bigr)^{2/3}}{3}\frac{dM}{dE}t^{-5/3},     \label{mdoteq}
\end{equation}
coincides to first approximation to the rate of accretion on to the BH. Inferring $\dot M(t)$ is thus equivalent to computing the luminosity $L(t)$ associated with a TDE 
\begin{equation}
L(t)=\eta \dot M(t) c^2,  \label{LM}
\end{equation}
assuming an appropriate efficiency $\eta$.

In equation \ref{mdoteq}, $dM/dE$ is the distribution of the stellar debris per unit energy as a function of $E$, the specific binding energy relative to the BH. Generally, such a distribution is neither flat nor constant in time (e.g. Lodato et al. 2009; Guillochon \& Ramirez-Ruiz 2013), allowing $\dot M(t)$ to deviate from the classically assumed $t^{-5/3}$ trend, inferred from equation \ref{mdoteq} when taking a uniform distribution in $E$ (e.g. Rees 1988; Phinney 1989). 

Here we compute $dM/dE$ as a function of time for each binary component directly from our simulations, following the recipe from Guillochon \& Ramirez-Ruiz (2013).
The position and velocity of the centre of mass of each star around the BH are computed through an iterative approach. The initial reference point is the particle with the highest local density. Particles within $2 \rm R_{\rm \odot}$ from it (a bit more than $R_{\rm *}$) are considered to be still bound to the star and their total mass is denoted as $M_{\rm B}$. The specific binding energy of the $i$-th particle relative to the star is calculated as 
\begin{equation}
E_{*_{i}}=\frac{1}{2}|\mathbfit{v}_{i}-\mathbfit{v}_{\rm peak}|^{2}-\frac{GM_{\rm B}}{|{\mathbfit{r}_{i} -\mathbfit{r}_{\rm peak}}|}, \label{energy}
\end{equation}
where $\mathbfit{v}_{i}-\mathbfit{v}_{\rm peak}$ and $\mathbfit{r}_{i}-\mathbfit{r}_{\rm peak}$ are the velocity and position of the $i$-th particle relative to the reference particle. Velocity and position of the temporary centre of mass are determined through the standard formulae by considering only particles with $E_{*_{i}}< 0$. Equation \ref{energy} is then re-evaluated with the new velocity and position of the centre of mass in place of $\mathbfit{v}_{\rm peak}$ and $\mathbfit{r}_{\rm peak}$. This process is re-iterated until the velocity of the centre of mass converges to a constant value, to less than $10^{-5} \rm R_{\rm \odot} \rm yr^{-1}$. Particles with $E_{*_{i}}>0$, i.e. unbound from the star, are then selected in the aim at evaluating their specific binding energy relative to the BH
\begin{equation}
E_{i}=\frac{1}{2}|\mathbfit{v}_{i}|^2-\frac{GM_{\rm BH}}{|\mathbfit{r}_{i}-\mathbfit{r}_{\rm BH}|},
\end{equation}
where $\mathbfit{v}_{i}$ and $\mathbfit{r}_{i}-\mathbfit{r}_{\rm BH}$ are the velocity and position of the $i$-th particle relative to the BH. Particles with $E_{i}>0$ are unbound from the BH, whereas particles with $E_{i}<0$ form the stream of debris bound to the BH. Data are then binned in $E$, i.e. the specific binding energies $E_{i}<0$ are grouped in bins and the correspondent particles fill this histogram. $dM/dE$ as a function of $E$ (i.e. time) is obtained dividing the total mass of particles in each bin by the bin amplitude. 

The time $t$ in equation \ref{mdoteq} is the time since disruption, which is coincident with the first pericentre passage for our purposes. Thus, only material with orbital periods $P_{\rm orb}=2 \pi G M_{\rm BH}/(2 E)^{3/2}$  around the BH less than $t$ contributes to the accretion till that time. 

\begin{figure}
\includegraphics[width=6.5cm, angle=-90]{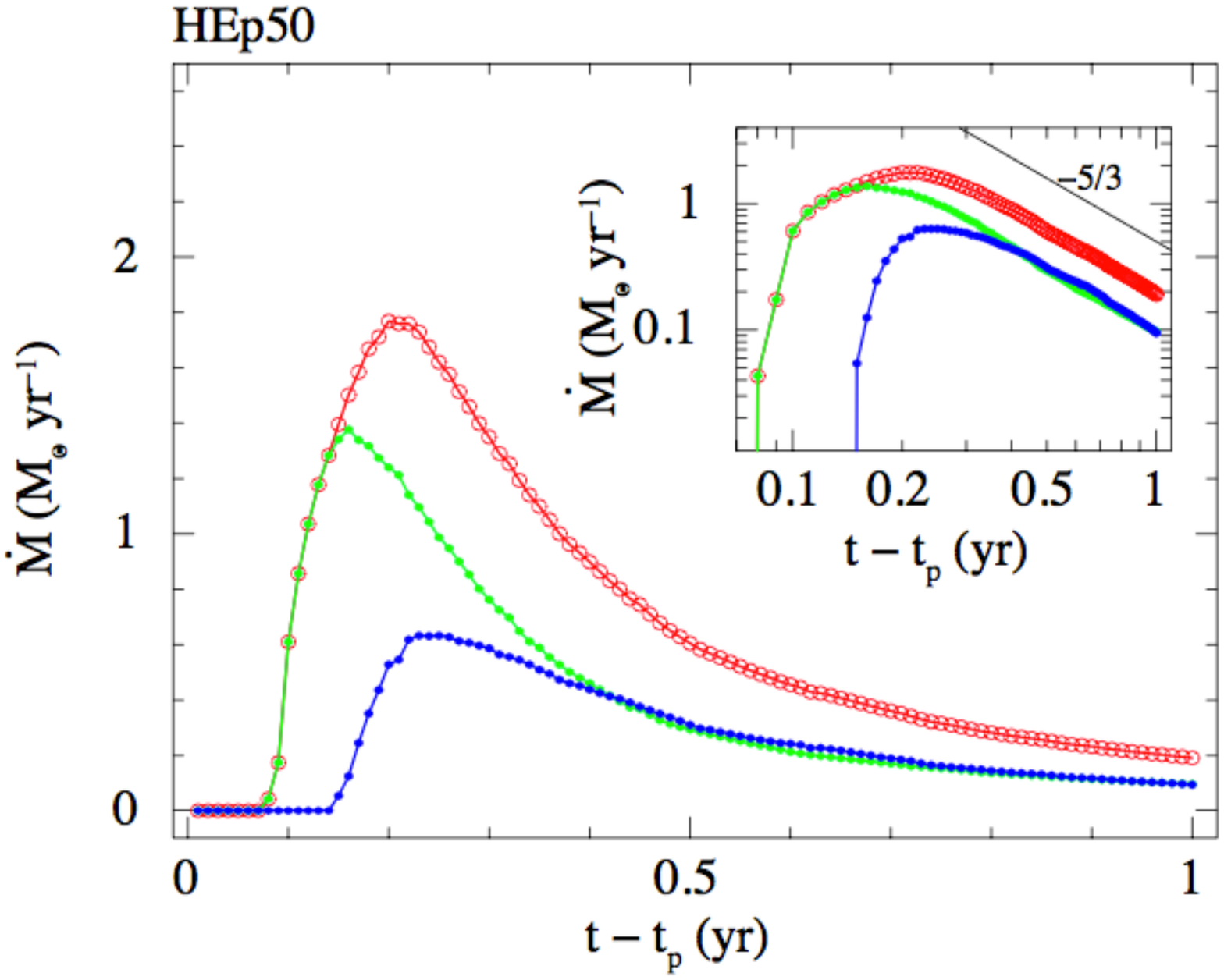}\\
\includegraphics[width=6.5cm, angle=-90]{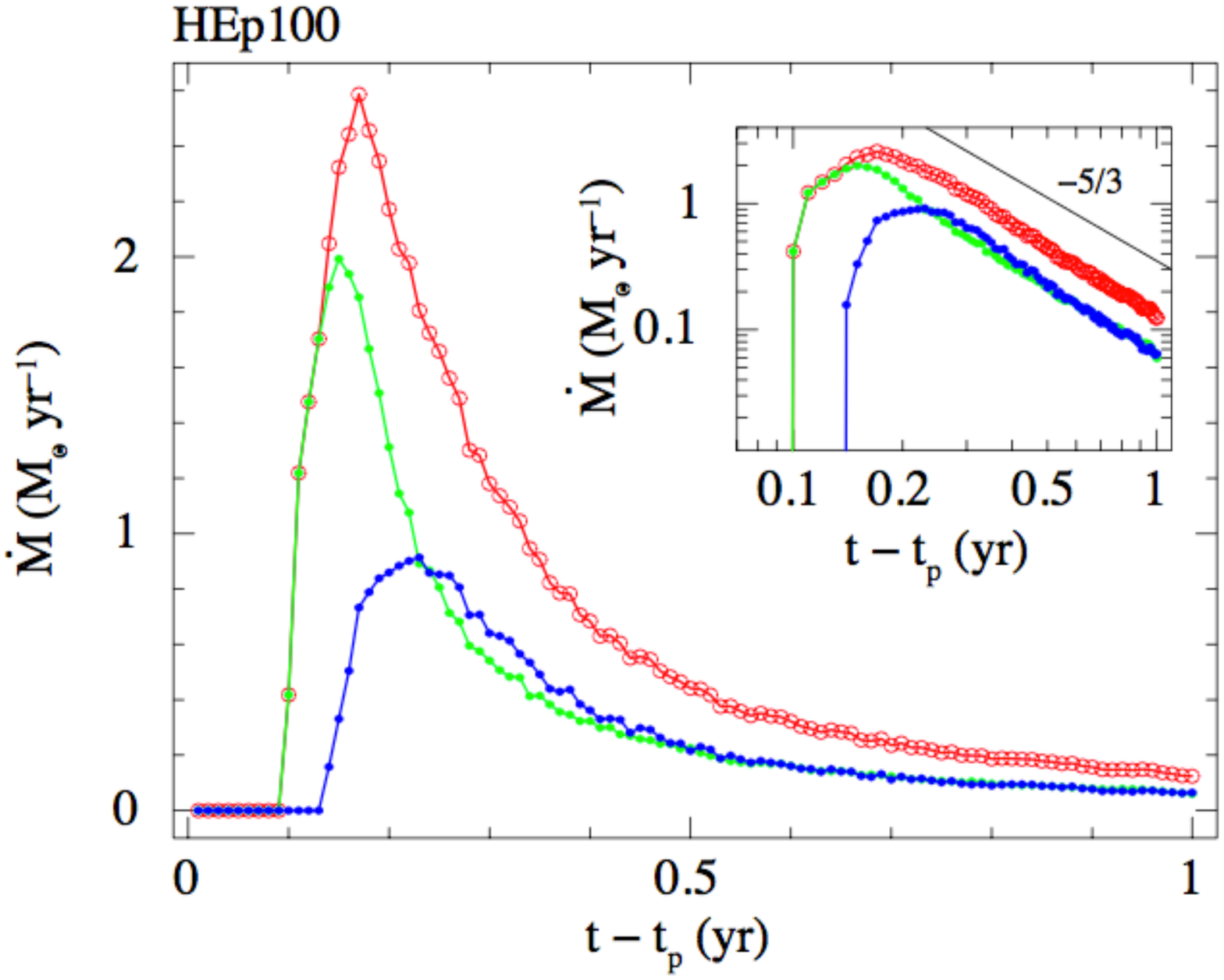}\\
 \includegraphics[width=6.5cm, angle=-90]{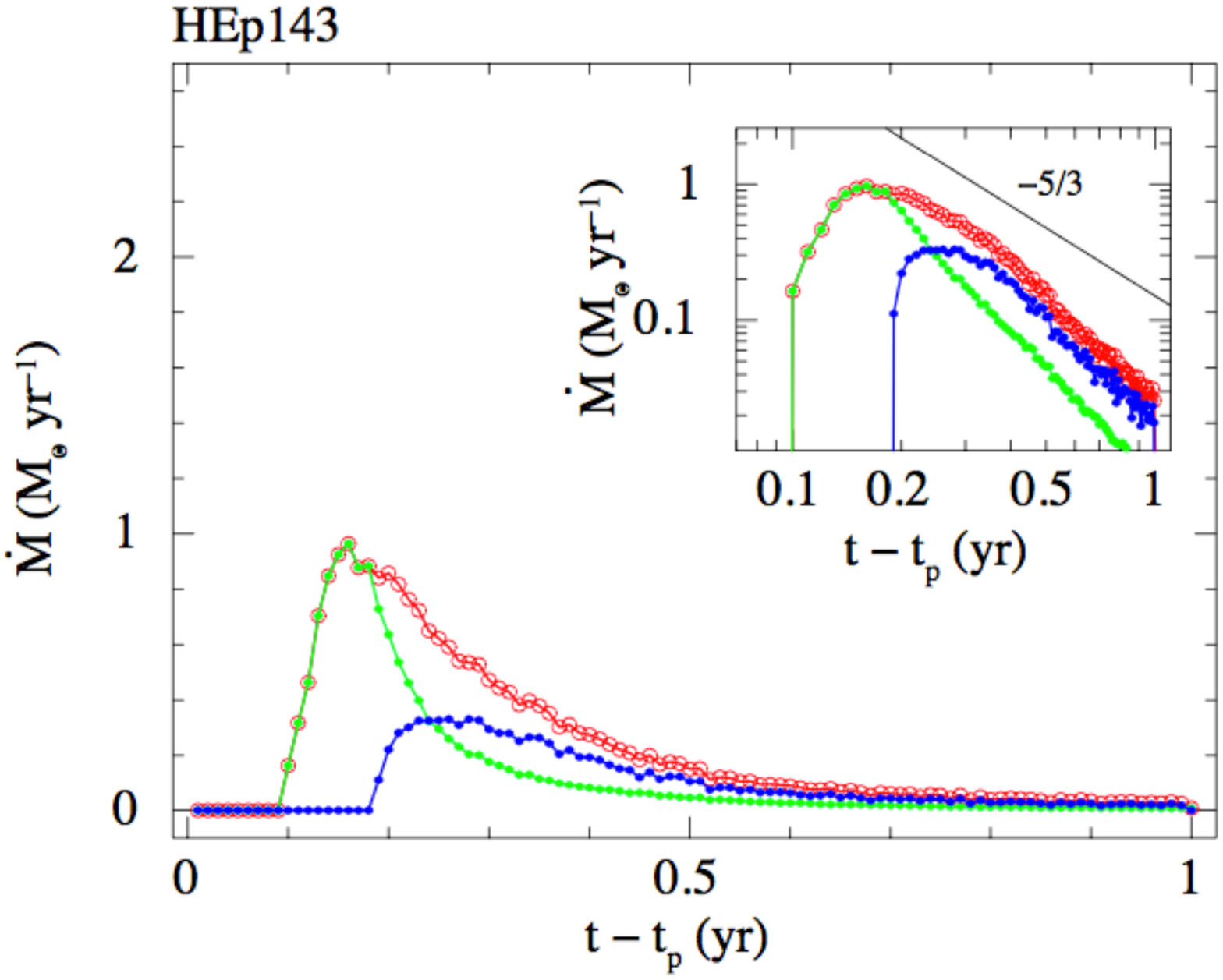}
 \caption{Light curves [$\dot M$ versus time; see equation \ref{LM} in Section \ref{subsection} to convert accretion rates into luminosities] inferred from our high-resolution simulations of parabolic equal-mass binary-BH encounters, depicting a fully disruptive encounter (simulation HEp50), 
an almost total double disruption (simulation HEp100) and a partial double TDE (simulation HEp143). Green and blue curves are associated with the disruption of the binary components; red curves reproduce the point-wise sum of the green and blue curves. On the top-right corners, we show the same plots in logarithmic scale. A knee in the red curve is somehow visible in simulation HEp143, especially in the logarithmic plot, and it decays more steeply than the classically assumed power law of index -5/3. \label{figure1}}
\end{figure}

To build the composite light curves, we need to compute the light curve for each star by interpolating the data coming from different snapshots, and then we sum the results of interpolations, point to point. Green and blue curves in Fig. \ref{figure1} are associated with the disruption of the single binary components, while red curves represent the point-wise sum of the green and blue curves.
Panels on the top-right corners show logarithmic plots. 

The light curves associated with TDEs are described by characteristic parameters, which can be assessed  directly from the light curves and also analytically, in order to check the reliability of the recipe that we have followed. The first characteristic parameter is $t_{\rm most}$, the returning time at pericentre of the most bound stellar debris since disruption. For a star on a parabolic orbit around a BH, it can be evaluated as
\begin{equation}
\Tilde t_{\rm most}=\frac{\pi}{\sqrt{2}}\frac{G M_{\rm BH}}{E^{3/2}}\sim \frac{\pi}{\sqrt{2}}\frac{1}{\sqrt{G}}\frac{M_{\rm BH}^{1/2}}{M_{\rm*}}R_{\rm*}^{3/2},
\end{equation}
where $E$ is the specific energy spread caused by the disruption $\Delta E\sim G M_{\rm BH} R_{\rm*}/r_{\rm t}^{2}$, given that the orbital energy associated with a parabolic orbit is zero. In our simulations, the binary CM is set on a parabolic orbit around the BH but the binary components are a bit out of it. Moreover, after the tidal binary separation, they follow new orbits: an ellipse for the bound star and hyperbola for the unbound star.
Thus, the returning time associated with each binary component is not simply $\Tilde t_{\rm most}$, as it requires knowledge of the new orbits of the separated stars. 
Hereafter, we denote with subscript 1 (2) the bound (unbound) binary component. 

For the bound star, the returning time can be evaluated as
\begin{multline}
t_{\rm most_{\rm1}}=\frac{\pi}{\sqrt{2}}\frac{G M_{\rm BH}}{E^{3/2}} \sim \Tilde t_{\rm most}\Bigl(\frac{M_{\rm *}}{M_{\rm BH}}\Bigr)^{1/2} \\ \times \frac{1}{(\beta_{\rm 1}(1-e_{\rm 1}))^{3/2}}\Bigl(\frac{1}{2}+\frac{\bigl(M_{\rm *}/M_{\rm BH}\bigr)^{1/3}}{\beta_{\rm 1}(1-e_{\rm 1})}\Bigr)^{-3/2},
\end{multline}
where $e_{\rm 1}$ is the eccentricity of its new orbit (computed through the Hermite code), $\beta_{\rm 1}$ the impact parameter of its centre of mass and $E\sim E_{\rm orb}+ \Delta E$, with $E_{\rm orb}\sim G M_{\rm BH}\beta_{\rm 1}(1-e_{\rm 1})/(2 r_{\rm t}) \neq 0$. 
We infer this time also from our simulations, considering as `mostly bound' the first returned particles after disruption associated with the bound star. 
As minimum of significance we assume 
10 particles out of the set of particles, associated with the bound star, bound to the BH. 
If the impact parameters of both the binary components, $\beta_{\rm 1}$ and $\beta_{\rm 2}$, are close to unity, the two estimates of $t_{\rm most_{\rm 1}}$ are in good agreement. In this case, we infer the returning time for the unbound star, $t_{\rm most_{\rm 2}}$, directly from our simulations. On the contrary, the more $\beta_{\rm 1}$ and $\beta_{\rm 2}$ depart from unity, the worse the agreement is. In this case, we introduce a correction factor between the two estimates of $t_{\rm most_{\rm 1}}$, and we use it to correct $t_{\rm most_{\rm 2}}$ as inferred from simulations. $\Tilde t_{\rm most}$, $t_{\rm most_{\rm 1}}$ and $t_{\rm most_{\rm 2}}$ are reported in Table \ref{3} for our three high-resolution simulations.

\begin{table}
\centering
\caption{Characteristic parameters of the light curves inferred from our high-resolution simulations of equal-mass binary-BH encounters, as analytically estimated (see Section \ref{subsection}). Simulations HEp50, HEp100 and HEp143, respectively, correspond to the ones in Fig. \ref{figure1}. $t_{\rm most}$ is the returning time at pericentre of the most bound stellar debris since disruption and $t_{\rm peak}$ the rise time from stellar disruption to accretion rate peak, $\dot M_{\rm peak}$. Tilded values are evaluated setting the binary components on parabolic orbits corresponding to the initial one of the binary CM; untilded values consider the effective orbits of the binary stars. The 1 (2) subscript denote the BH bound (unbound) star. $\Delta t_{\rm peak}$ and $\Delta \dot M_{\rm peak}$ are the differences in rise times and accretion rate peaks between the two `humps' expected in the composite light curves associated with double TDEs, actually visible only in the partially disruptive encounter (simulation HEp143; see Fig. \ref{figure1}).  \label{3}}
\begin{tabular}{c  c  c  c}
\hline
& HEp50: & HEp100: & HEp143: \\
& TD-TDE & ATD-TDE & PD-TDE \\
\hline
\small{$\Tilde t_{\rm most} (\rm yr)$} & \small{0.1126} & \small{0.1126} & \small{0.1126} \\
\small{$t_{\rm most_{\rm 1}} (\rm yr)$} & \small{0.0987} & \small{0.0963} & \small{0.0946} \\
\small{$t_{\rm most_{\rm 2}} (\rm yr)$} &  \small{ 0.1777} & \small{0.1681} & \small{0.1873} \\
\small{$\Tilde t_{\rm peak_{\rm 1}} (\rm yr)$} &  \small{0.1807} & \small{0.1618} & \small{0.1738}\\
\small{$t_{\rm peak_{\rm 1}} (\rm yr)$}& \small{0.1585}  & \small{0.1384} & \small{0.1460} \\
\small{$\Tilde t_{\rm peak_{\rm 2}} (\rm yr)$} &\small{0.1779} &  \small{0.1617} & \small{0.1751} \\
\small{$t_{\rm peak_{\rm 2}} (\rm yr)$} &  \small{0.2809} & \small{0.2415} & \small{0.2915} \\
\small{$\Tilde{\dot M}_{\rm peak_{\rm 1}} (\rm M_{\odot} yr^{-1})$} &\small{1.254} &  \small{1.672} & \small{0.595} \\
\small{$\dot M_{\rm peak_{\rm 1}} (\rm M_{\odot} yr^{-1})$} &\small{1.566} &  \small{2.088} & \small{0.743}\\
\small{$\Tilde{\dot M}_{\rm peak_{\rm 2}} (\rm M_{\odot} yr^{-1})$} &  \small{1.266} & \small{1.563} & \small{0.519} \\
\small{$\dot M_{\rm peak_{\rm 2}} (\rm M_{\odot} yr^{-1})$} & \small{0.792} & \small{0.978} & \small{0.325}  \\
\small{$\Delta t_{\rm peak} (\rm d)$} & - & - & 50 \\
\small{$\Delta \dot M_{\rm peak} (\rm M_{\odot} d^{-1})$} & - &  - & $10^{-3}$  \\
\hline
\end{tabular}
\end{table}

Corrections for the new orbits of the separated stars also involve the second characteristic parameter of TDE light curves, $t_{\rm peak}$, that is the rise time between the time of stellar disruption and the time at which the accretion rate peaks. If the two binary components were on parabolic orbits corresponding to the initial one of their binary CM, the rise time for each star would be denoted as $\Tilde t_{\rm peak}$ (1,2) and could be evaluated following Guillochon \& Ramirez-Ruiz (2013, 2015a). Corrected values come out to be
\begin{equation}
t_{\rm peak}\sim \Tilde t_{\rm peak}\frac{t_{\rm most}}{\Tilde t_{\rm most}}, \label{tpeak}
\end{equation}
assuming that $t_{\rm most}$ and $t_{\rm peak}$ change proportionally. Table \ref{3} collects $\Tilde t_{\rm peak}$ (1,2) and $t_{\rm peak}$ (1,2) for our three high-resolution simulations.

The last characteristic parameter of TDE light curves is the peak of accretion rate, $\dot M_{\rm peak}$. According to MacLeod et al. (2013), this parameter is linked to the mass of the debris which binds to the BH $M_{\rm bound_{\rm BH}}$ and to the rise time $t_{\rm peak}$ through the relation 
\begin{equation}
\dot M_{\rm peak}\sim \frac{2}{3}\frac{M_{\rm bound_{\rm BH}}}{t_{\rm peak}}.
\end{equation}
Values for stars on parabolic orbits, $\Tilde{\dot M}_{\rm peak}$ (1,2), can be evaluated considering $M_{\rm bound_{\rm BH}}$ to be half the mass lost from each star (e.g. Rees 1988) and $t_{\rm peak}\equiv \Tilde t_{\rm peak}$. Corrected values require $M_{\rm bound_{\rm BH}}$ as inferred from our simulations and $t_{\rm peak}$ from equation \ref{tpeak}. Given that standard assumptions work for $\beta\sim 1$, we estimate $\Tilde{\dot M}_{\rm peak}$ (1,2) and $\dot M_{\rm peak}$ (1,2) as just mentioned for simulation HEp100 (see Section \ref{glimpse}), and then we convert them in the corresponding values for the other two simulations, based on the dependence of $\dot M_{\rm peak}$ from the impact parameter $\beta$ reported in Guillochon \& Ramirez-Ruiz (2013, 2015a). Indeed, the only difference among our simulations is the value of the pericentre radius, i.e. $\beta$ \footnote{In the case of unequal-mass binaries, we need to consider also the dependence of $\dot M_{\rm peak}$ from $M_{\rm *}$ and $R_{\rm *}$.}.
However, recall that the relation between
$\dot M_{\rm peak}$ and $\beta$ works for parabolic orbits. Consequently, some differences between the values assessed from the inferred light curves (Fig. \ref{figure1}) and our analytical estimates are to be expected.
Values of $\Tilde{\dot M}_{\rm peak}$ (1,2) and $\dot M_{\rm peak}$ for our three simulations are reported in Table \ref{3}.
Good agreement is found between light-curve parameters inferred from Fig. \ref{figure1} and analytical evaluations, motivating the recipe we have followed in the aim to derive TDE light curves.

As previously said in this section, the composite light curves associated with double TDEs are obtained by summing the light curves associated with the disruption of the single binary components. 
Given that the binary components have different returning and rising times, one should expect to observe a double peak in their composite light curve. In Table \ref{3}, we collect, where possible, the values of 
$\Delta t_{\rm peak}$ and $\Delta \dot M_{\rm peak}$ as inferred from Fig. \ref{figure1}, which are the differences in rise times and accretion rate peaks between the two `humps' in the composite light curves. From Table \ref{3} and Fig. \ref{figure1}, we see that only in simulation HEp143, which corresponds to a grazing encounter, the composite light curve shows not exactly a double peak, as predicted, but anyway a knee. In this case, the single-star light curves are distinguishable enough to be both glimpsed in the composite light curve. As shown in hydrodynamical simulations of single TDEs of Guillochon \& Ramirez-Ruiz (2013), grazing encounters give rise to steep light curves (i.e. steeper than -5/3) immediately after the peak and, in the context of double disruptions, this favours the visibility of the knee in the composite light curves. Therefore, in the case of double TDEs of equal-mass binaries, only grazing encounters can produce a knee in the composite light curve.

\subsection{Double TDE light curves: the case of unequal-mass binaries} \label{unequal}
\begin{figure}
\includegraphics[width=5.5cm, angle=90]{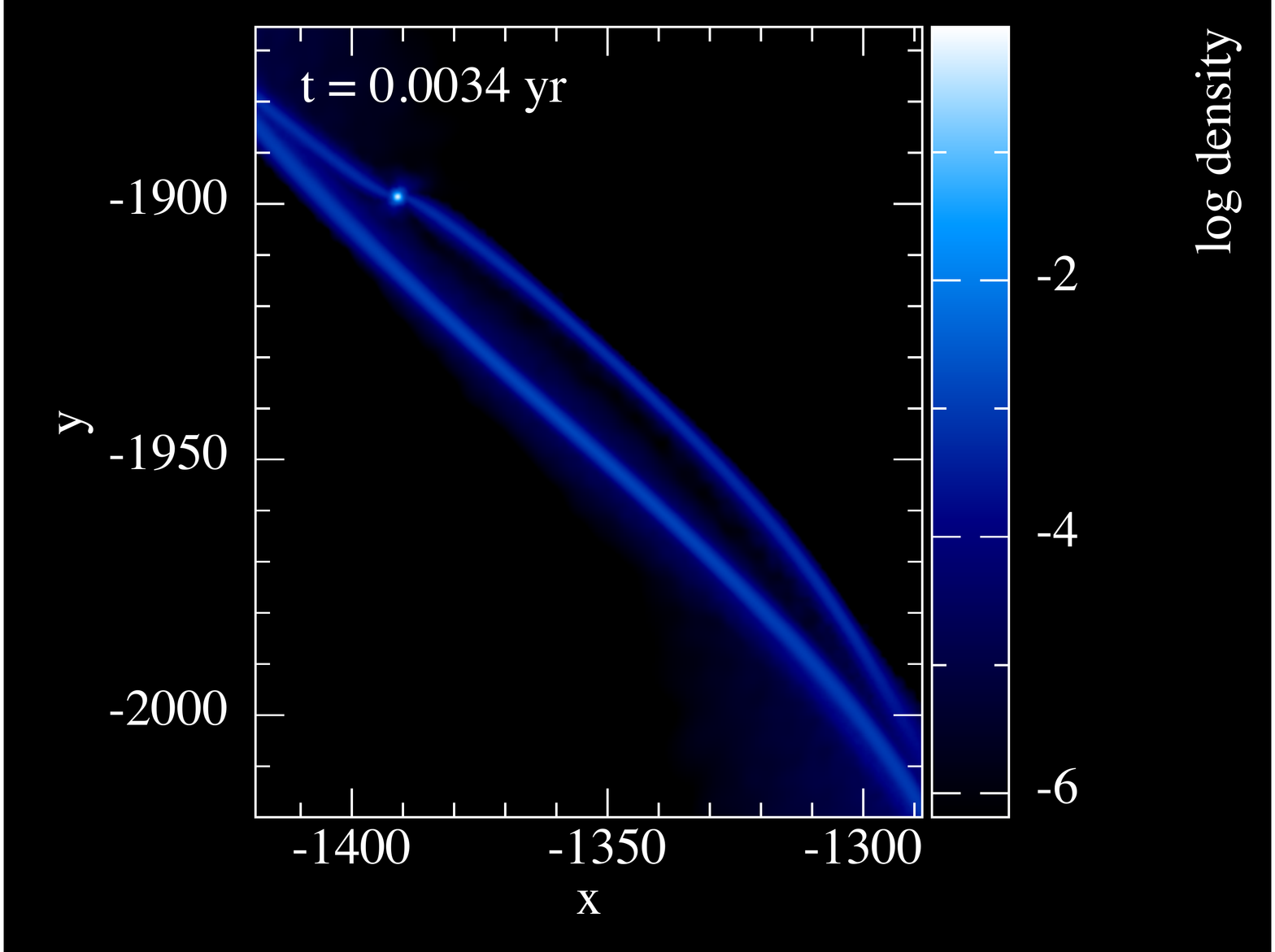} \\
\includegraphics[width=5.5cm, angle=90]{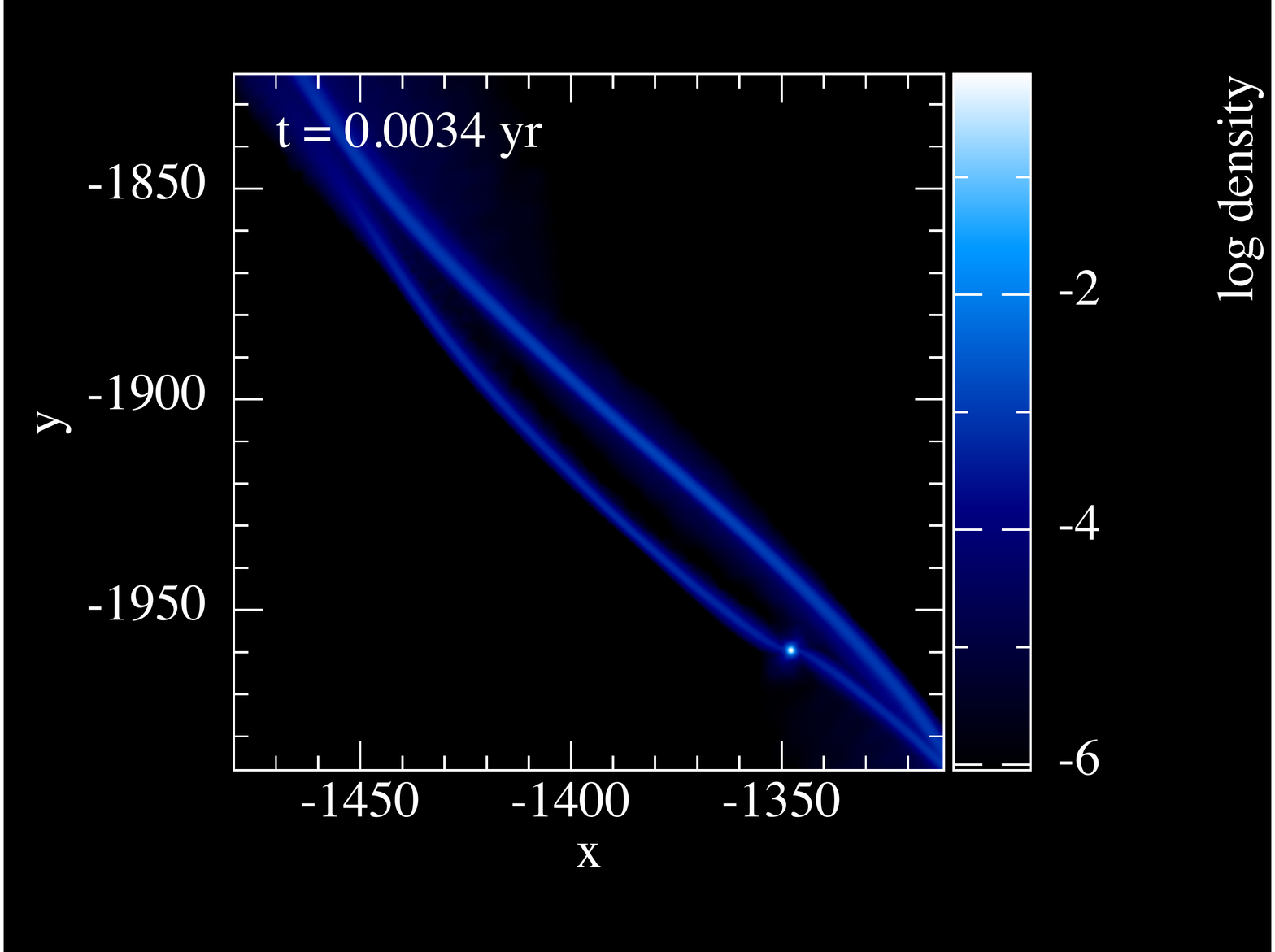} \\
\caption{Zoom in the SPH particle density (in log scale) for simulations HUp70a (upper panel) and HUp70b (bottom panel) in Section \ref{unequal}, projected in the ($x,y$) plane, at $t=0.0034 \rm yr$ ($\sim 1.2 \rm d$) after pericentre passage. The remnant less massive star is clearly visible in both the simulations. \label{de}}
\end{figure}
What happens in the case of deeper encounters if the binary components have unequal masses? Using the same procedure described in Section \ref{subsection}, we carry on and analyse three high-resolution SPH simulations of unequal-mass binaries on parabolic orbits around a BH ($M_{\rm BH}=10^6 \rm M_{\rm \odot}$) (HU runs). 
Table \ref{referee4} collects the outcomes of these simulations as a function of $a_{\rm bin}$ and $r_{\rm p}$. We also perform the correspondent low-resolution SPH simulations (LU runs), respectively, denoted as LU1, LU2 and LU3, finding out the same outcomes and the same orbital evolution of the binary components (Fig. \ref{new}, bottom panels).
\begin{table}
\centering
\caption{Same as Table \ref{1} for our high-resolution simulations involving unequal-mass binaries. \label{referee4}}
\begin{tabular}{c c c}
\hline
\footnotesize{$a_{\rm bin}$$\backslash$$r_{\rm p}$} & \footnotesize{$42.0$} & \footnotesize{$70.0$} \\
$(\rm R_{\rm \odot})$ & & \\
\hline
& & \footnotesize{HUp70a:}\\
\small{$4.9$} & \footnotesize{HUp42:} & \footnotesize{P\&T-TDE} \\
 &  \footnotesize{TD-TDE} & \footnotesize{HUp70b:} \\
 & &  \footnotesize{P\&T-TDE} \\
\hline
\end{tabular}
\end{table}

In particular, simulations LU1/HUp42 consider $M_{\rm 1}=0.4 \rm M_{\rm \odot}$, $r_{\rm t_{\rm 1}}\sim 65.2 \rm R_{\rm \odot}$, $M_{\rm 2}=0.27 \rm M_{\rm \odot}$, $r_{\rm t_{\rm 2}}=54.3 \rm R_{\rm \odot}$, simulations LU2/HUp70a: $M_{\rm 1}=0.5 \rm M_{\rm \odot}$, $r_{\rm t_{\rm 1}}\sim 72.4 \rm R_{\rm \odot}$, $M_{\rm 2}=1 \rm M_{\rm \odot}$, $r_{\rm t_{\rm 2}}=100.0 \rm R_{\rm \odot}$, simulations LU3/HUp70b: $M_{\rm 1}=1. \rm M_{\rm \odot}$, $r_{\rm t_{\rm 1}}\sim 100. \rm R_{\rm \odot}$, $M_{\rm 2}=0.5 \rm M_{\rm \odot}$, $r_{\rm t_{\rm 2}}=72.4 \rm R_{\rm \odot}$. The initial conditions of  simulations LU1/HUp42 are those considered in Mandel \& Levin (2015). With simulations HUp70a and HUp70b, we explore the dependence of the visibility of a double peak on the mass difference between the binary components and on the mass of the captured star, whether it is the less or the more massive of the two. Indeed, simulations HUp70a and HUp70b only differ in that they are out of phase by $180^{\circ}$.
In the high-resolution regime, stars denoted as 1, which remain bound to the BH after binary separation, are modelled respectively with $4\times 10^{4}$, $10^{5}$ and $2\times 10^{5}$ particles and stars 2, which unbind from the BH, with $2.7\times 10^{4}$, $2\times 10^{5}$ and $10^{5}$ particles. Fig. \ref{de} shows a zoom in the SPH particle density (in log scale), projected in the ($x,y$) plane, at $t=0.0034 \rm yr$ ($\sim 1.2 \rm d$) after pericentre passage for simulations HUp70a and HUp70b. The remnant less massive star is clearly visible in both the simulations.

\begin{figure}
\centering
\includegraphics[width=6.5cm, angle=-90]{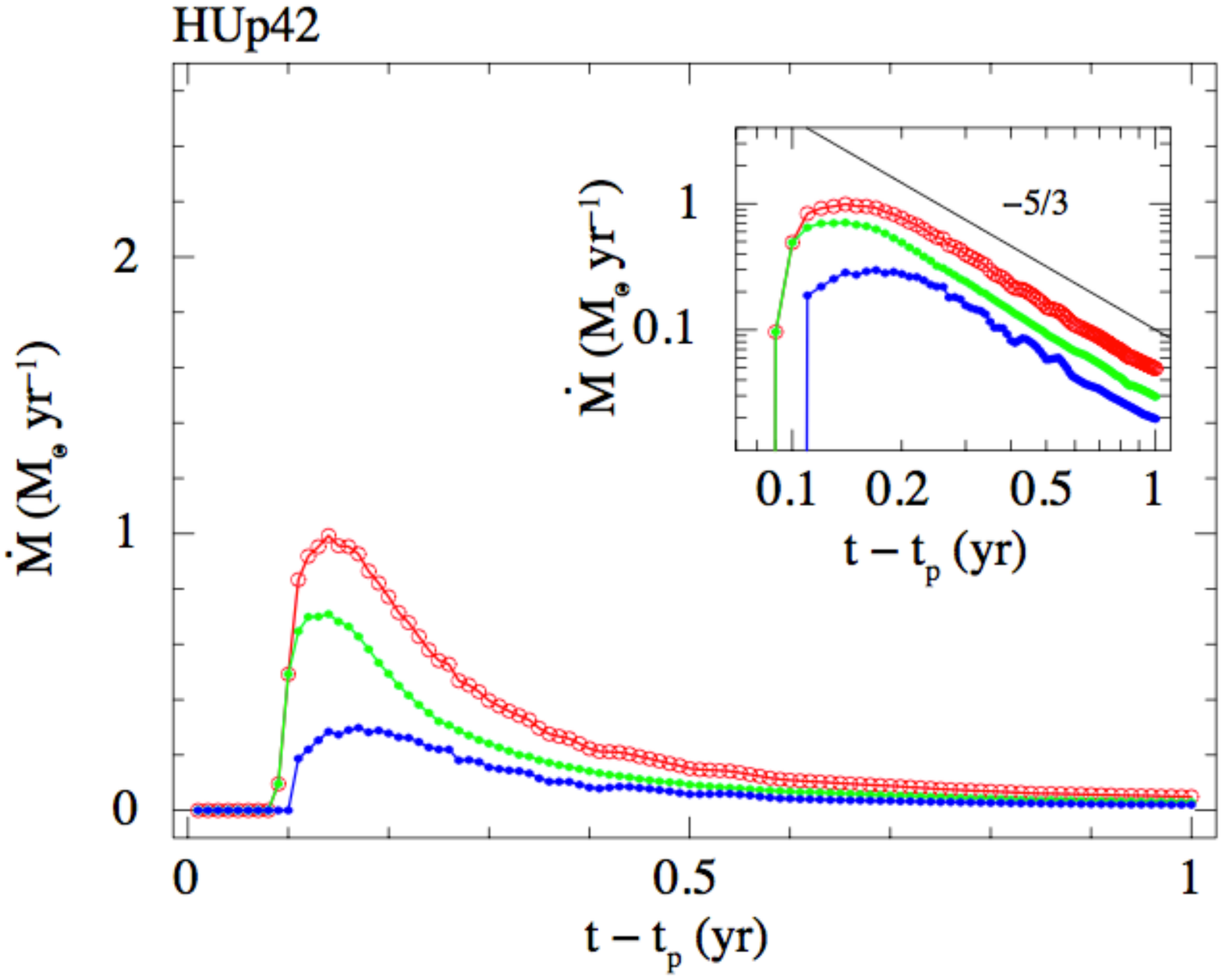}\\
\includegraphics[width=6.5cm, angle=-90]{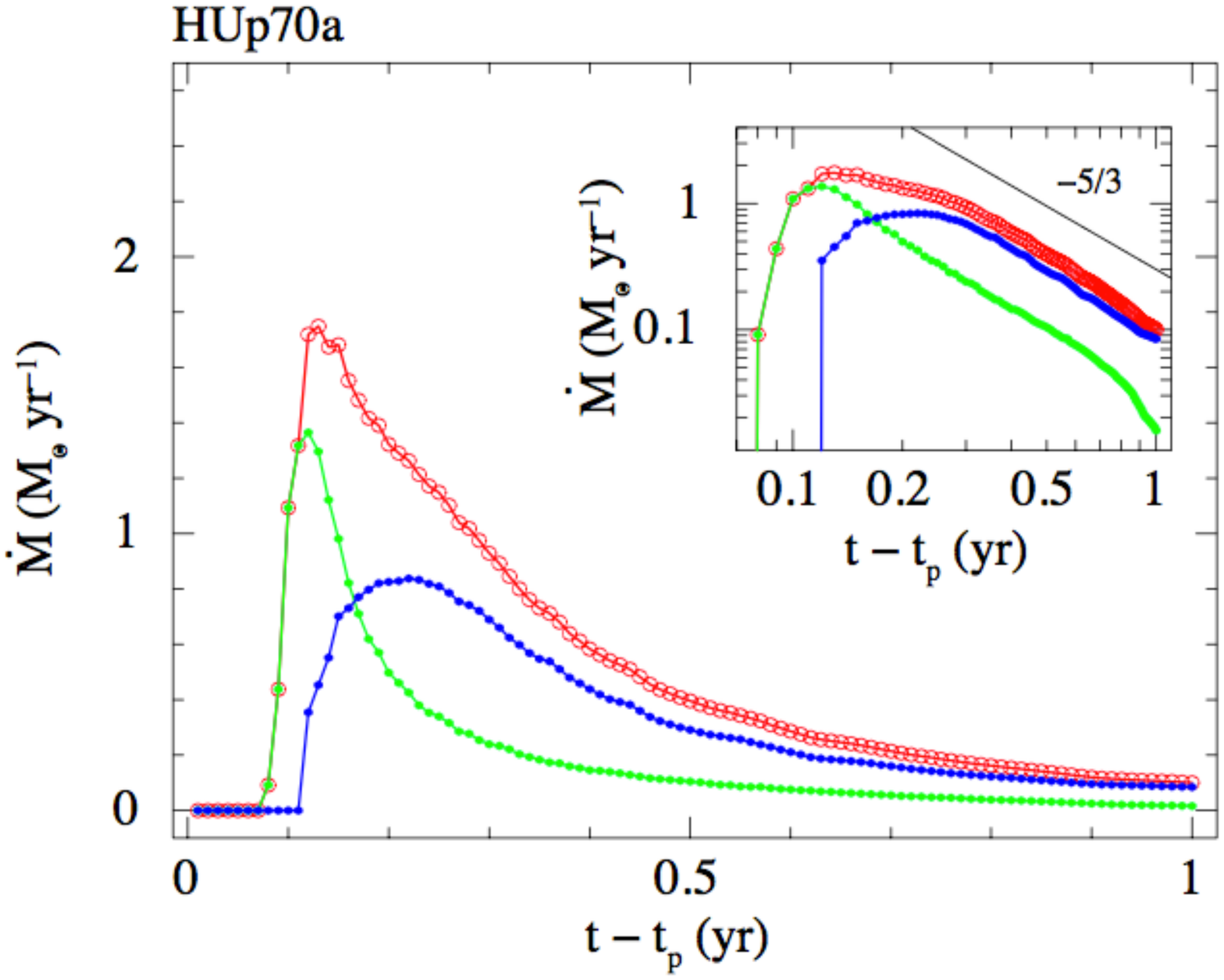}\\
\includegraphics[width=6.5cm, angle=-90]{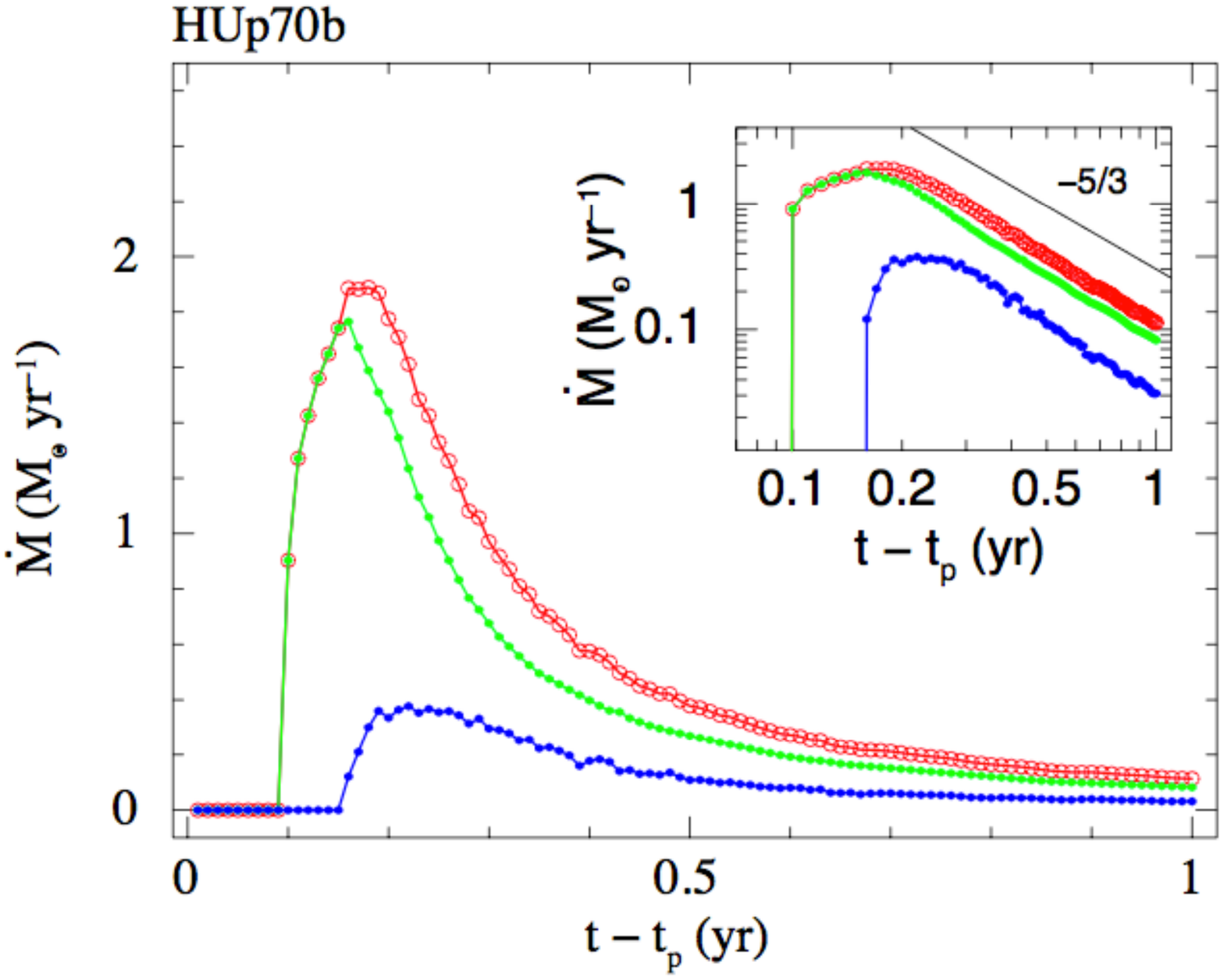}\\
\caption{Light curves [$\dot M$ versus time; see equation \ref{LM} in Section \ref{subsection}] inferred from the high-resolution simulations HUp42, HUp70a and HUp70b in Section \ref{unequal}. Green and blue curves are associated with single-star disruptions; the composite light curves are the red ones. On the top-right corners, we show the same plots in logarithmic scale. A knee in the composite light curve is visible in simulation HUp70a. \label{figure5}}
\end{figure}

Table \ref{5} collects the characteristic parameters of the light curves inferred from simulations HUp42, HUp70a and HUp70b, respectively, as analytically
estimated following Section \ref{subsection}. Fig. \ref{figure5} shows single-star and composite light curves inferred from simulations HUp42, HUp70a and HUp70b following the recipe described in Section \ref{subsection}. Not exactly a double peak, but a knee in the composite light curve is observed when the mass difference between the two stars is increased and when the star which gets bound to the BH is the less massive of the two (simulation HUp70a). This is because a low-mass star is less compact than a higher-mass star (the compactness parameter is $ \propto M_{\rm *}/R_{\rm *}$), and this leads to an increased difference between the narrow peak of the low-mass star light curve and the broader peak of the higher-mass star light curve.

\begin{table}
\centering
\caption{Same as Table \ref{3} for simulations HUp42, HUp70a and HUp70b in Section \ref{unequal}.  \label{5}}
\begin{tabular}{c c c c}
\hline
& HUp42:  & HUp70a: & HUp70b:\\
& TD-TDE & P\&T-TDE & P\&T-TDE\\
\hline
\small{$\Tilde t_{\rm most_{\rm 1}} (\rm yr)$} & \small{0.0937} & \small{0.0980} & \small{0.1126}\\
\small{$t_{\rm most_{\rm 1}} (\rm yr)$} & \small{0.0845} & \small{0.0760}  & \small{0.1019}\\
\small{$\Tilde t_{\rm most_{\rm 2}} (\rm yr)$} & \small{0.0866} & \small{0.1126} & \small{0.0980}\\
\small{$t_{\rm most_{\rm 2}} (\rm yr)$} & \small{0.1224} &  \small{0.1355} & \small{0.1802}\\
\small{$\Tilde t_{\rm peak_{\rm 1}} (\rm yr)$} & \small{0.1433} & \small{0.1417} & \small{0.1696}\\
\small{$t_{\rm peak_{\rm 1}} (\rm yr)$} & \small{0.1293} &  \small{0.1099} & \small{0.1536}\\
 \small{$\Tilde t_{\rm peak_{\rm 2}} (\rm yr)$} &\small{0.1272} & \small{0.1682} &  \small{0.1408}\\
\small{$t_{\rm peak_{\rm 2}} (\rm yr)$} & \small{0.1797} & \small{0.2025} & \small{0.2589}\\
\small{$\Tilde{\dot M}_{\rm peak_{\rm 1}} (\rm M_{\odot} yr^{-1})$} & \small{0.738} & \small{0.987} & \small{1.622}\\
\small{$\dot M_{\rm peak_{\rm 1}} (\rm M_{\odot} yr^{-1})$} & \small{0.922} &  \small{1.431}  & \small{2.025}\\
\small{$\Tilde{\dot M}_{\rm peak_{\rm 2}} (\rm M_{\odot} yr^{-1})$} & \small{0.574} & \small{1.579} & \small{0.952}\\
\small{$\dot M_{\rm peak_{\rm 2}} (\rm M_{\odot} yr^{-1})$} & \small{0.360} &  \small{0.988}  &  \small{0.442}\\
\small{$\Delta t_{\rm peak} (\rm d)$} & -  & \small{25}  & - \\
\small{$\Delta \dot M_{\rm peak} (\rm M_{\odot} d^{-1})$} &  - &  \small{$\rm 1.5 \times 10^{-3}$} &  - \\
\hline
\end{tabular}
\end{table}

\section{Summary and conclusions}
A stellar tidal disruption occurs when a star passes close enough to the central BH of a galaxy to experience the BH tidal field. The star can be fully or partially torn apart, according to the distance of closest approach (Guillochon \& Ramirez-Ruiz 2013, 2015a). The stellar debris which accretes on to the BH powers a long-lasting single flare (e.g. Rees 1988; Phinney 1989) or even periodic flares, if the star, partially disrupted, keeps on orbiting around the BH (MacLeod et al. 2013). Such events contribute to detect otherwise quiescent BHs of masses complementary to that probed in bright AGN and QSO surveys (Vestergaard \& Osmer 2009). 

Given the high number of field stars in binary systems (Duquennoy \& Mayor 1991b; Fischer \& Marcy 1992), encounters with a galactic central BH can involve stellar binaries instead of single stars. The high central densities and velocity dispersions present in galactic nuclei reduce the number of binaries. Indeed, most binaries are `soft', i.e. the relative velocity of their components is much smaller than the velocities of the field stars. Thus, soft binaries can be separated via close encounters with other stars over the galaxy lifetime (Merritt 2013). However, some binaries are `hard' enough, which also means close enough, to survive encounters with field stars for a longer time. The members of these binaries, under certain conditions, when approaching the central BH can experience total or partial tidal disruption immediately after the tidal binary break-up. From an encounter of this kind, a double-peaked flare is expected to blaze up (Mandel \& Levin 2015). Generally, after binary break-up one star leaves the system while the other binds to the BH (e.g. Antonini et al. 2011). In the case of partial double disruptions, the bound star can be thus repeatedly disrupted, lighting up periodic ($\sim 1 - 10$ yr) single-peaked flares. Hence, we argue that this channel could be one of the most likely mechanisms that allow stars to become bound to central galactic BHs and undergo periodic TDEs, as suggested for IC3599 (Campana et al. 2015).  Periodicity increases the chance of observing and modelling TDE flares, and it could be predicted if a double peak were detected. This is rare but not impossible, given that double TDEs should contribute up to about the $20$ per cent of all TDEs.

This is the first paper that explores the process of double tidal disruption through hydrodynamical simulations, in the aim at detailing the dynamics of the binary-BH interaction (see Figs. \ref{snapshots}, \ref{blue} and \ref{de}) and the shape of the outcoming light curve. 
Based on the results of a set of 14 low-resolution SPH simulations of parabolic equal-mass binary-BH encounters, we set the initial conditions of three high-resolution SPH simulations in order to explore double TDEs of different intensities. For twin stars of equal masses, we found that a knee, rather than a double peak, in the composite light curve is
observed only in the case of grazing double TDEs. Otherwise, flares without knees can be observed, indistinguishably from single-star tidal disruptions (see Fig. \ref{figure1}). 

We also explored the case of unequal-mass binaries experiencing double TDEs, running three additional high-resolution simulations. We found that the most favourable conditions for the visibility of a knee in the composite light curves occur when the difference in mass between the binary components is increased and the star fated to bind to the BH is lighter than the star fated to leave the system (see Fig. \ref{figure5}).
Indeed, the knee becomes more and more defined when the difference in the peak width between the two single-star light curves increases. The less massive star, which is less compact, generates a light curve that is rising and declining on a shorter time-scale. Varying the binary semimajor axis, internal eccentricity and internal orbital plane inclination with respect to the binary CM orbital plane around the BH affects less the shape of the double TDE light curves. These parameters mainly act on the single-star impact parameters, but even if these are different to the maximum degree, they cannot be so much different, otherwise double TDEs are inhibited. 

Starting from the light curve which shows a knee in the case of unequal-mass binaries (Fig. \ref{figure5}, middle panel), we estimated analytically how much the light curve would change when changing the BH mass, $M_{\rm BH}$. We considered the interval between $10^{5}$ and $10^{8} \rm M_{\rm \odot}$ and follow the dependence on $M_{\rm BH}$ of single times and peak accretion rates as reported by Guillochon \& Ramirez-Ruiz (2013, 2015a). We found that $\Delta t_{\rm peak}$ tends to increase whereas $\Delta \dot M_{\rm peak}$ tends to decrease increasing $M_{\rm BH}$ to the point that intermediate values of $M_{\rm BH}$ (i.e. $10^{6}-10^{7} \rm M_{\rm \odot}$) are more favourable to the observation of the knee in the composite light curve.

It is worth noting that relativistic effects should also be taken into account in future studies on double TDEs, especially in the case of deep encounters, given that they could cause deviations of the debris evolution from the one assumed here. Lens-Thirring effects can warp the accretion disc which forms around a spinning BH, powering quasi-periodic oscillations (Franchini, Lodato \& Facchini 2016). 
In-plane relativistic precession leads the stream of debris to self-cross (Shiokawa et al. 2015), speeding up the circularization process (Bonnerot et al. 2016), 
but nodal precession which arises from the BH spin can deflect debris out of its original orbital plane, delaying self-intersection and then circularization, which however depends on the efficiency of radiative cooling (Hayasaki, Stone \& Loeb 2015), and flaring (Guillochon \& Ramirez-Ruiz 2015b).

Up to now, candidate TDE observations have been too widely spaced in time
to allow the notice of a possible knee. The challenge for the future will be to find a way to get more detailed light curves from observations (e.g. Holoien et al. 2016), particularly in the region of peak emission, as well as in the late-time decay. In this way, it will be possible to distinguish between light curves which show or not a knee, opening the opportunity to predict and follow up periodic flares, and to separate TDEs from other phenomena which nowadays could be misinterpreted due to the scarcity of data.
The advent of new telescopes, such as LSST (http://www.lsst.org/lsst), may contribute to such a purpose.

\section*{Acknowledgements} 
We thank the ISCRA staff for allowing us to perform our simulations on the Cineca Supercomputing Cluster GALILEO.
We also thank the anonymous referee for valuable comments on the manuscript and constructive suggestions.

\appendix
\section{Binary star orbits from low-resolution SPH simulations and $N$-body integrator} \label{appA}
This appendix shows the collection of orbits associated with our low-resolution simulations of binary-BH encounters. Tables \ref{app1} and \ref{app2}, respectively, refer to the simulations presented in Tables \ref{1} and \ref{2} in Section \ref{following} and include figures which represent the orbital evolution of the binary components around the BH projected in the ($x,y$) and ($y,z$) planes. Evolutions start at (0,0), (0,0). Units are in $\rm R_{\rm \odot}$. Blue curves reproduce the initial parabolic orbits of the binary CM around the BH, each inferred from the position of the BH, marked in figures with a black dot, and the pericentre radius $r_{\rm p}$. Red and green curves represent the early and late orbital evolution around the BH of each binary component, respectively, inferred from SPH simulations (see also Section \ref{subsection}) and computed through an $N$-body Hermite code (e.g. Hut \& Makino 1995; see Section \ref{following}). The usage of an $N$-body code in drawing advanced orbits enables us to overcome the high computational time required by SPH simulations to track them. 

MGs occur when the binary components progressively reduce their relative separation without being tidally separated, till merging in a single product. This MG product, which corresponds to having the binary components at a fixed minimum distance in our $N$-body simulations, follows the initial parabolic orbit of the binary CM around the BH. Undisturbed binaries (UNs) keep their internal and external orbits unchanged. Double disruptions (D-TDEs) are immediately preceded by binary separation, which can still also occur without stellar disruptions (BBKs). Binary break-up gets one star bound to the BH and leaves the other unbound. The unbound component, if not totally disrupted, may exit the system as hypervelocity star (Hills 1988; Antonini et al. 2011). 

%\newpage
\begin{landscape}
\begin{table}
\caption{Same as Table \ref{1} in Section \ref{following}, also with the orbital evolution of the binary components around the BH for each simulation, projected in the ($x,y$) and ($y,z$) planes. Evolutions start at (0,0), (0,0). Figures are in $\rm R_{\rm \odot}$. The initial binary CM orbits, inferred from the BH position (black dots) and the pericentre radius $r_{\rm p}$, are traced in blue. Red and green curves represent the orbital evolution of the binary components as respectively inferred from SPH and $N$-body simulations. \label{app1}}
\begin{center}
\begin{tabular}{c c c c c c c}
\hline
\footnotesize{$\rm a_{bin}$$\textbackslash $$\rm r_{p}$} &  \footnotesize{$\rm 50.0$} & \footnotesize{$\rm 100.0$} & \footnotesize{$\rm 142.6$} & \footnotesize{$\rm 200.0$} & \footnotesize{$\rm420.0$} & \footnotesize{$\rm 780.0$} \\
$(\rm R_{\rm \odot})$ & & & & & & \\
& & & & & & \\
& \footnotesize{LE1:} & \footnotesize{LE2:} & \footnotesize{LE3:} & \footnotesize{LE4:} & \footnotesize{LE5:} &  \\
\small{$\rm 4.9$} & \footnotesize{TD-} & \footnotesize{ATD-} & \footnotesize{PD-} & \footnotesize{PD-} & \footnotesize{MG} &  \\
& \footnotesize{TDE} &  \footnotesize{TDE} &  \footnotesize{TDE} & \footnotesize{TDE}  & & \\
& & & & & & \\
&  \includegraphics[width=2.7cm, angle=-90]{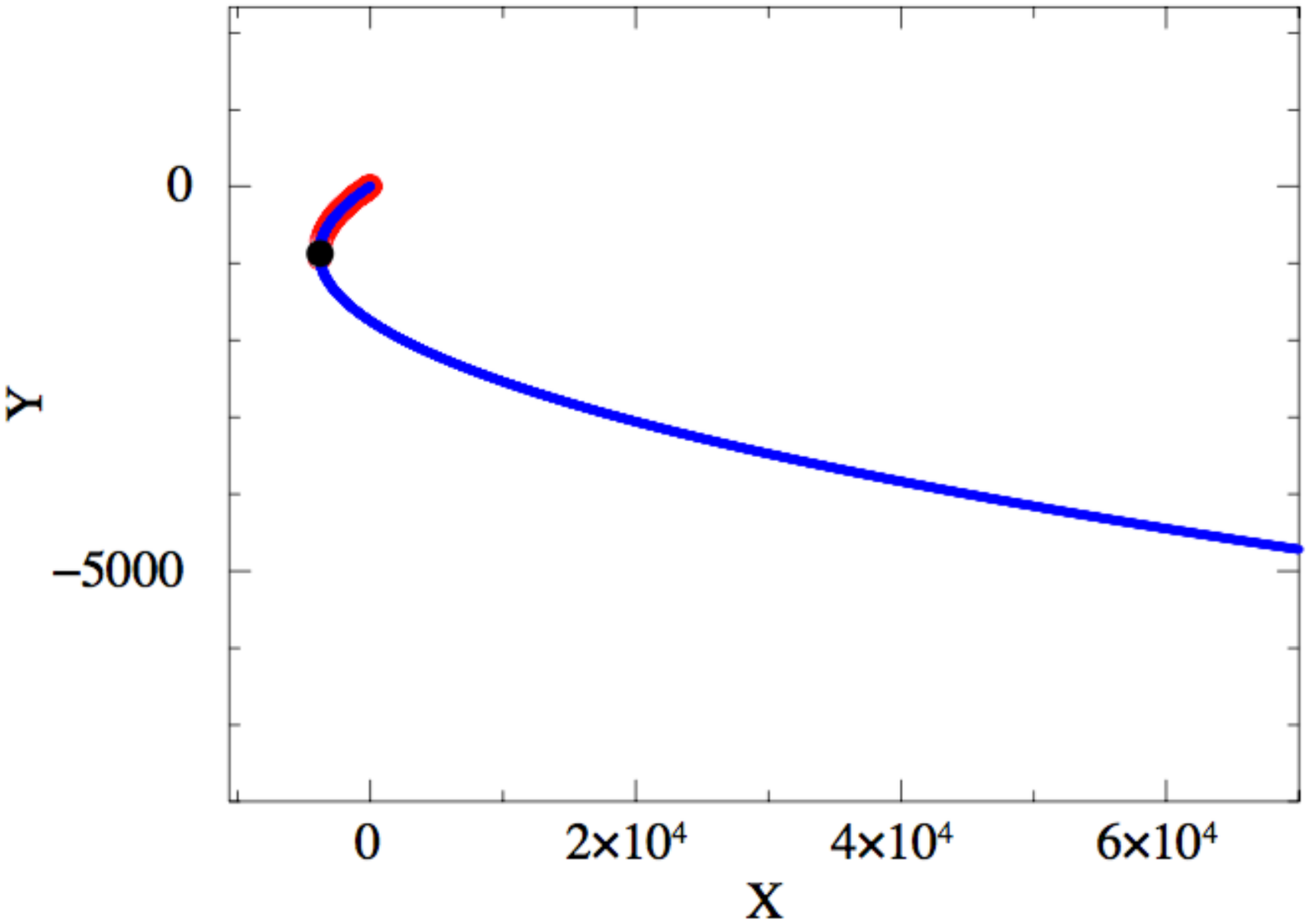} &  \includegraphics[width=2.7cm, angle=-90]{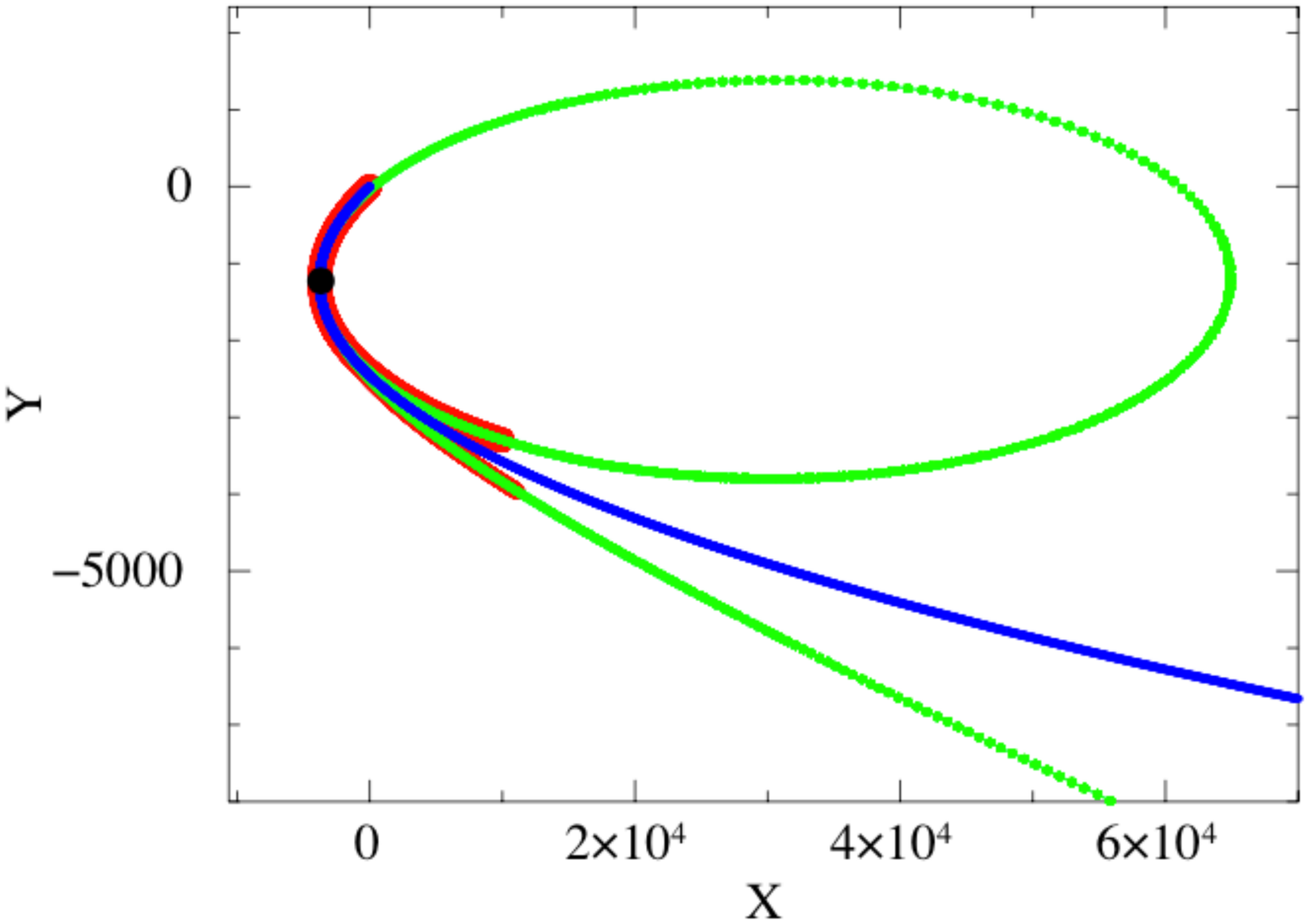} &\includegraphics[width=2.7cm, angle=-90]{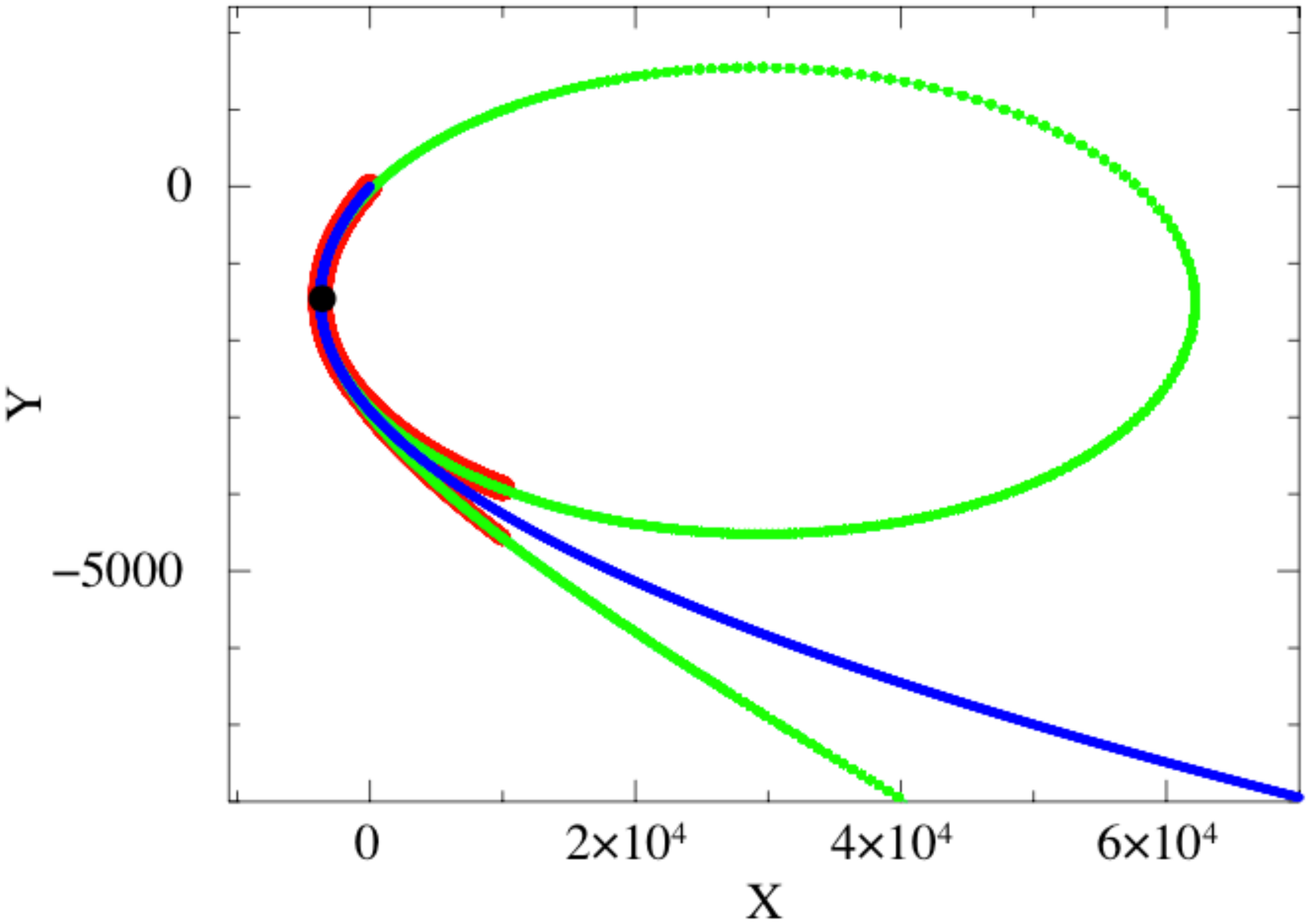}  & \includegraphics[width=2.7cm, angle=-90]{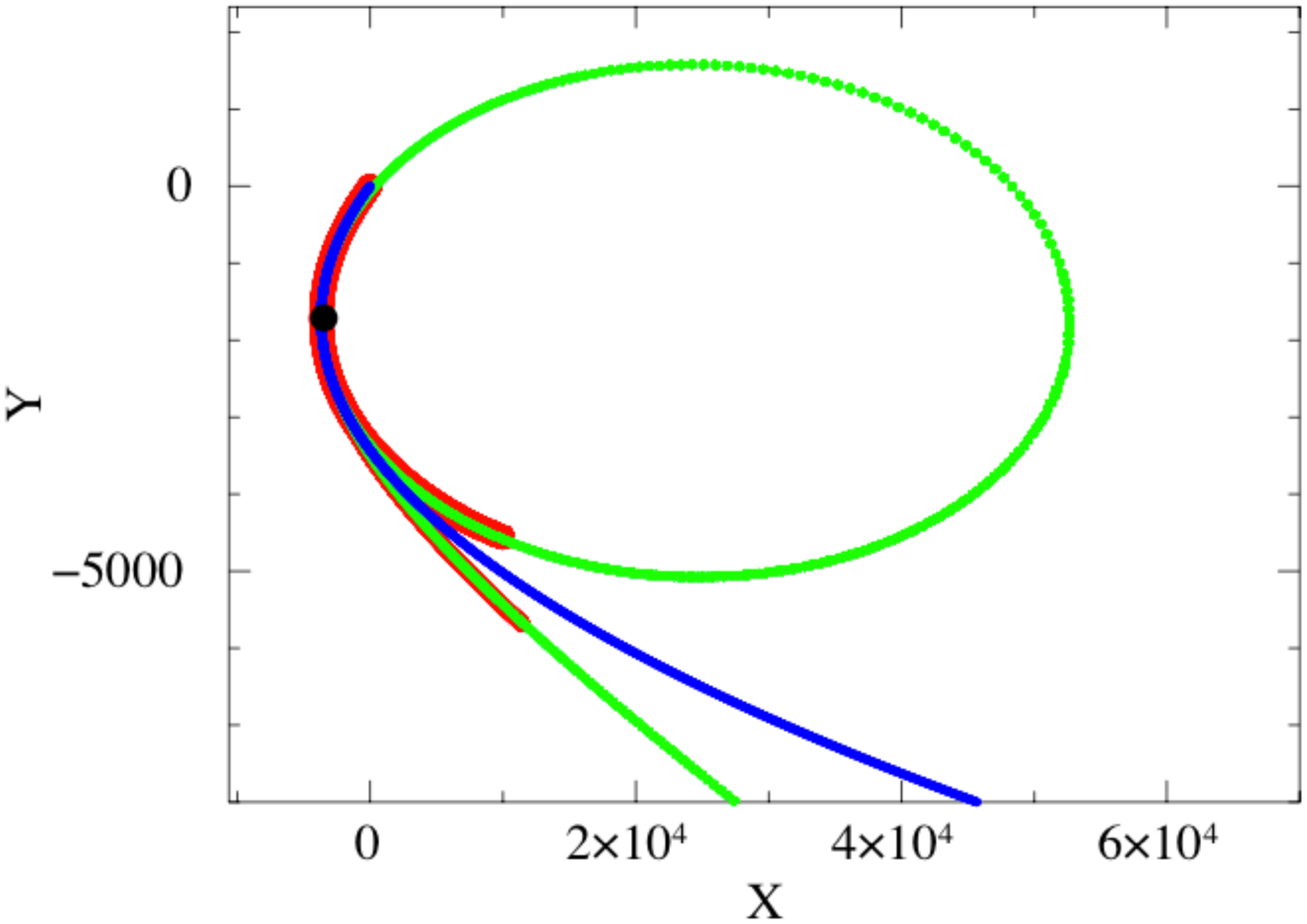}  &\includegraphics[width=2.7cm, angle=-90]{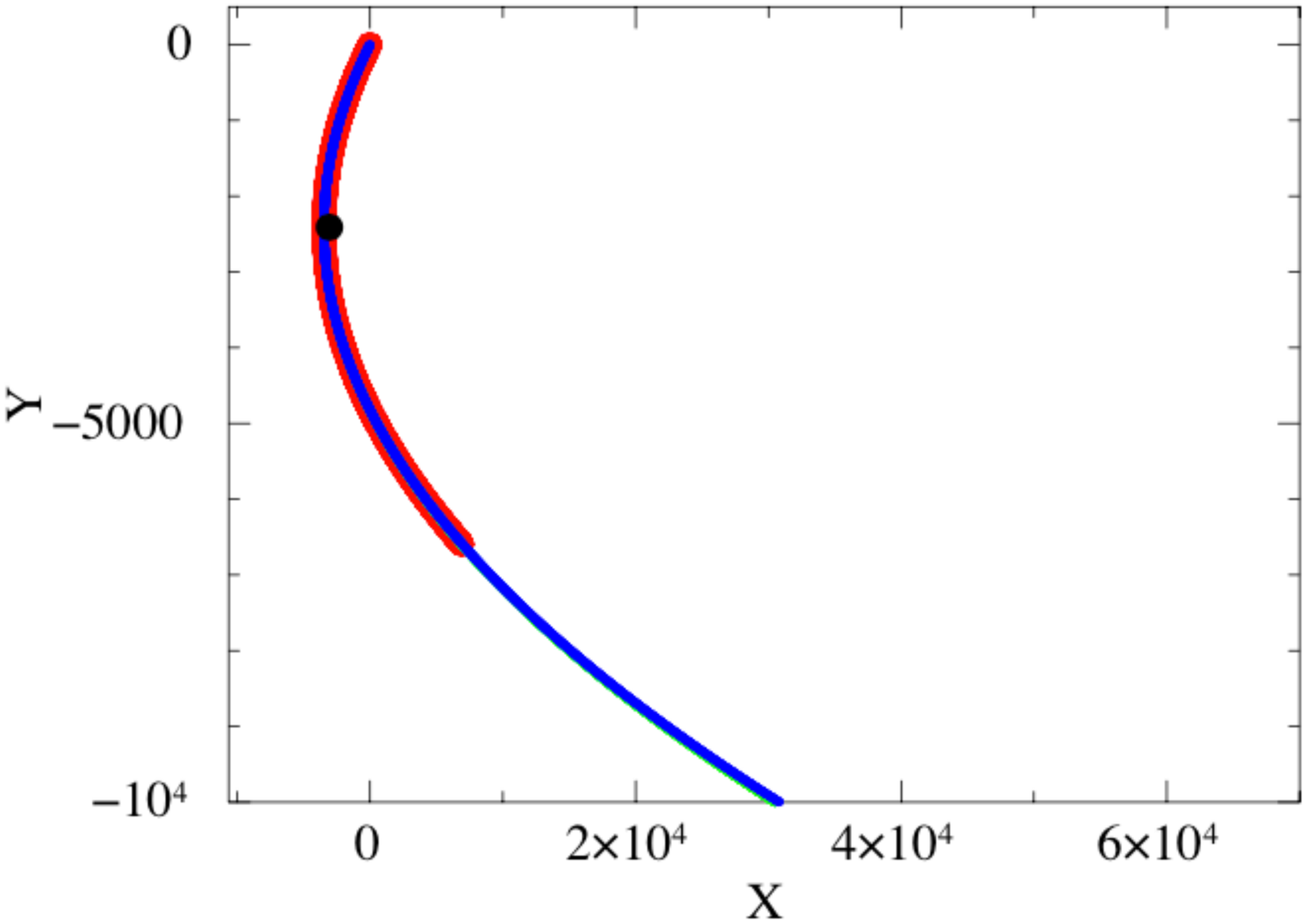}  &  \\
&  \includegraphics[width=2.7cm, angle=-90]{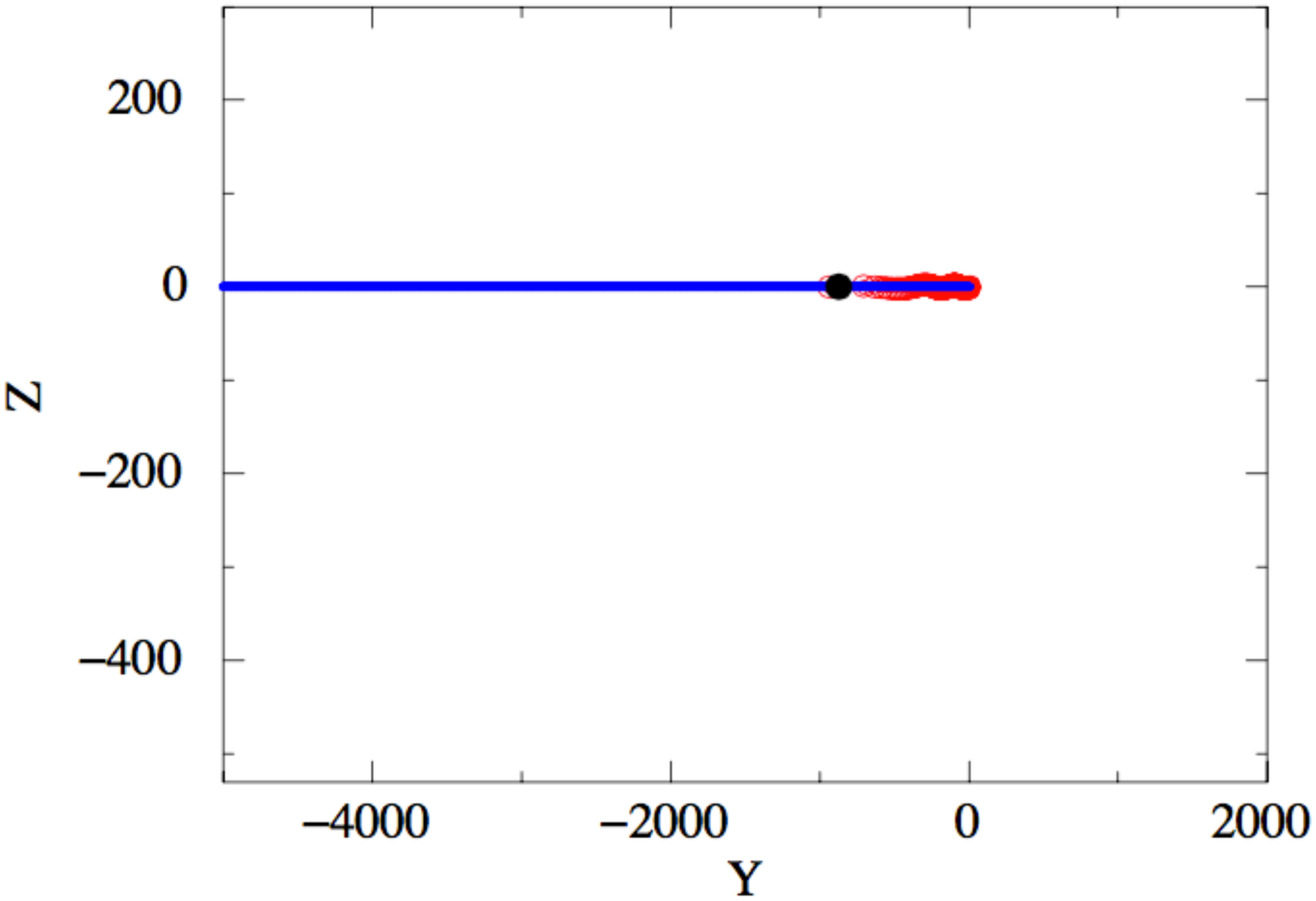}  &   \includegraphics[width=2.7cm, angle=-90]{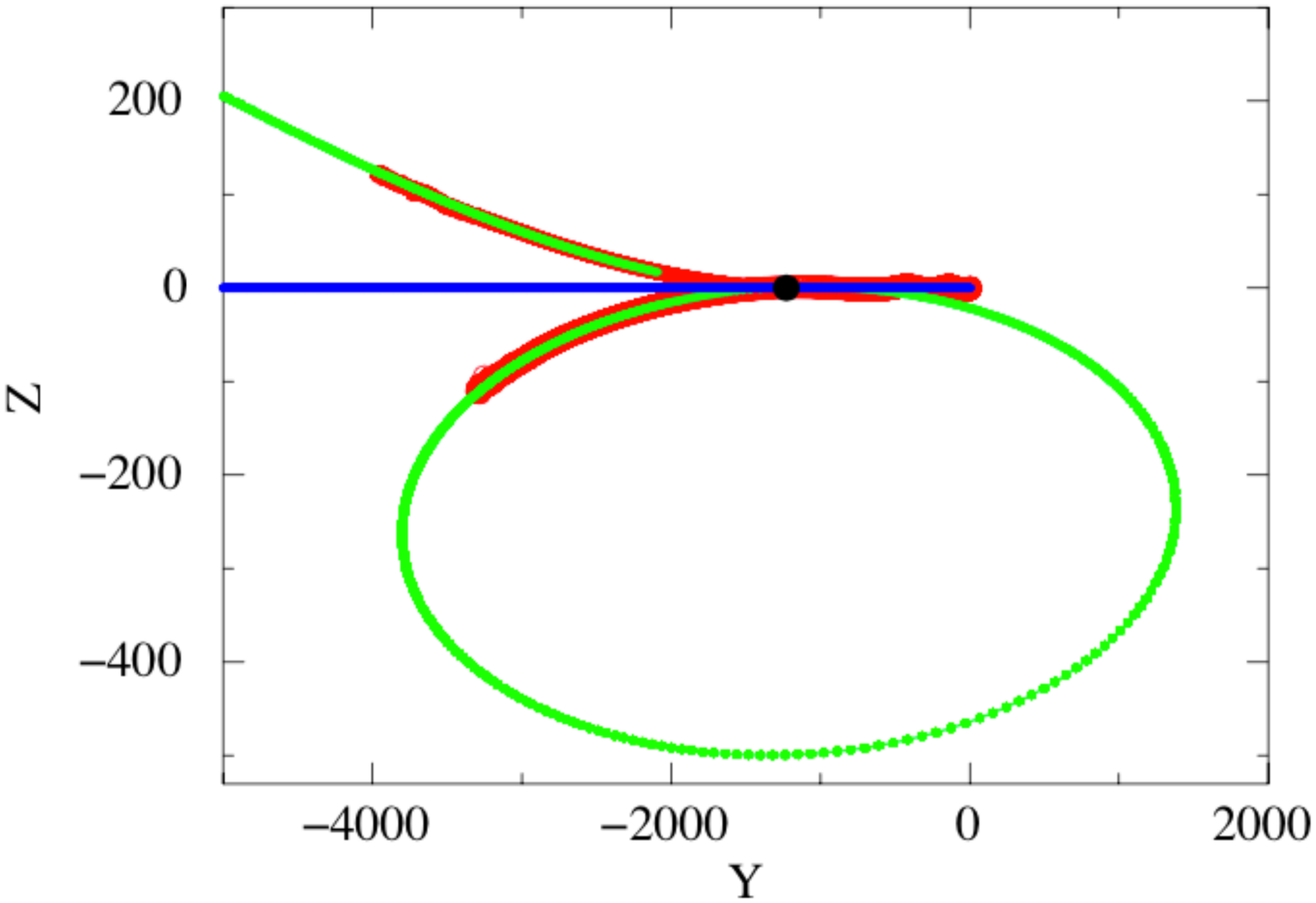} &\includegraphics[width=2.7cm, angle=-90]{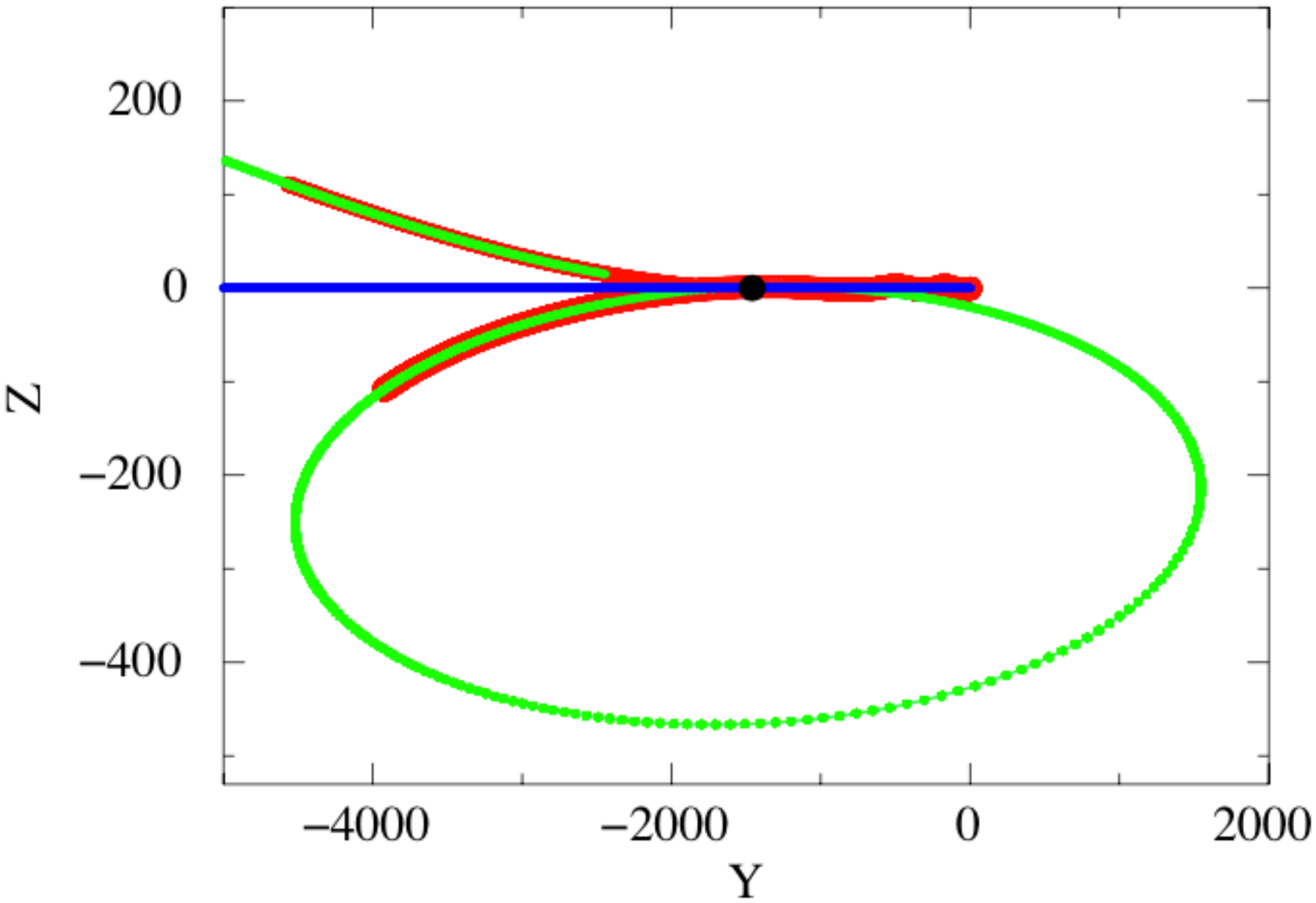}  & \includegraphics[width=2.7cm, angle=-90]{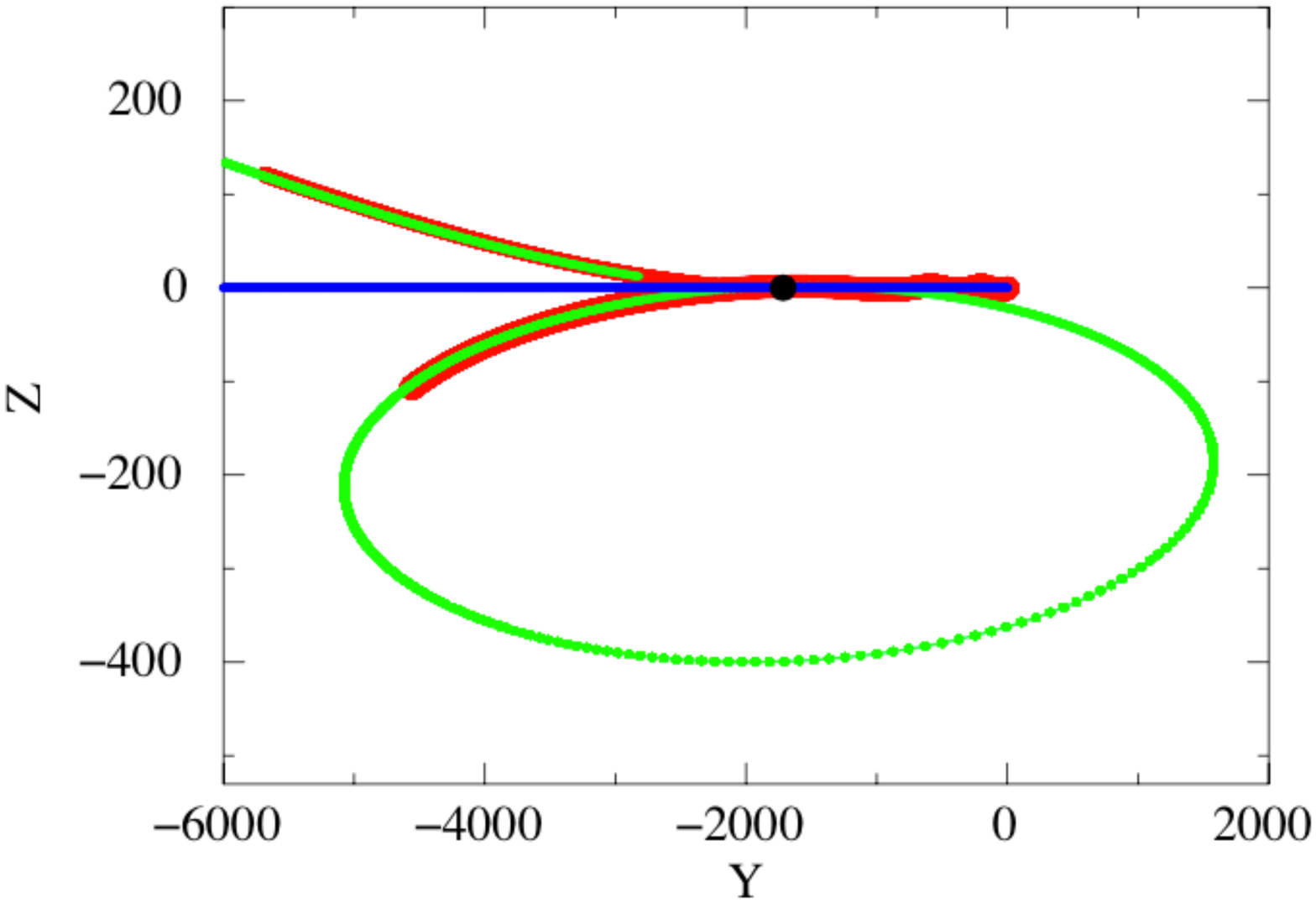}  & \includegraphics[width=2.7cm, angle=-90]{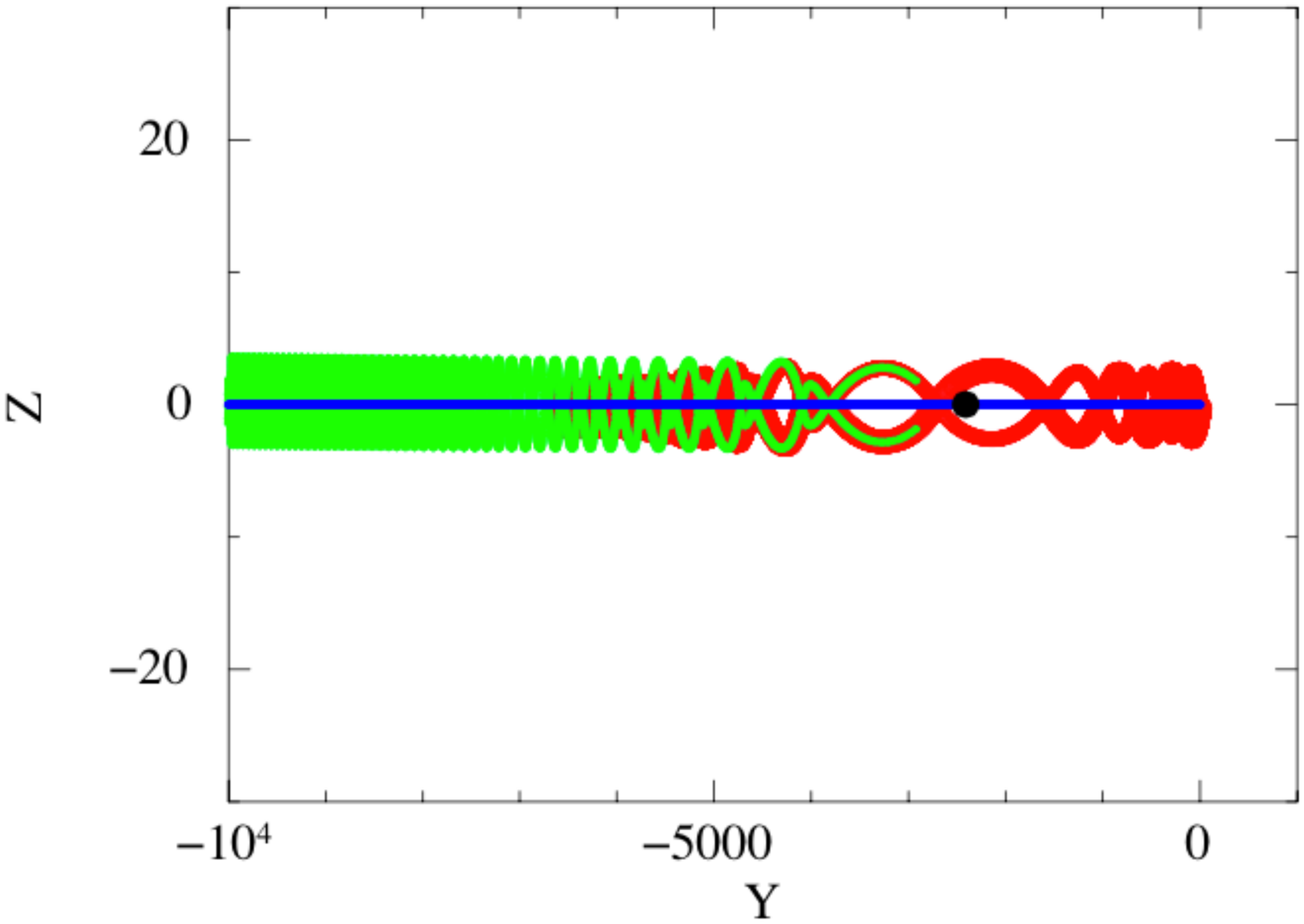}  & \\
& & & & & & \\
& &  \footnotesize{LE9:}  & & \footnotesize{LE10:} & & \footnotesize{LE11:} \\
\small{$\rm 9.8$} & & \footnotesize{ATD-} &  & \footnotesize{PD-} & & \footnotesize{BBK} \\
& & \footnotesize{TDE} & & \footnotesize{TDE} & & \\
& & & & & & \\
&  & \includegraphics[width=2.7cm, angle=-90]{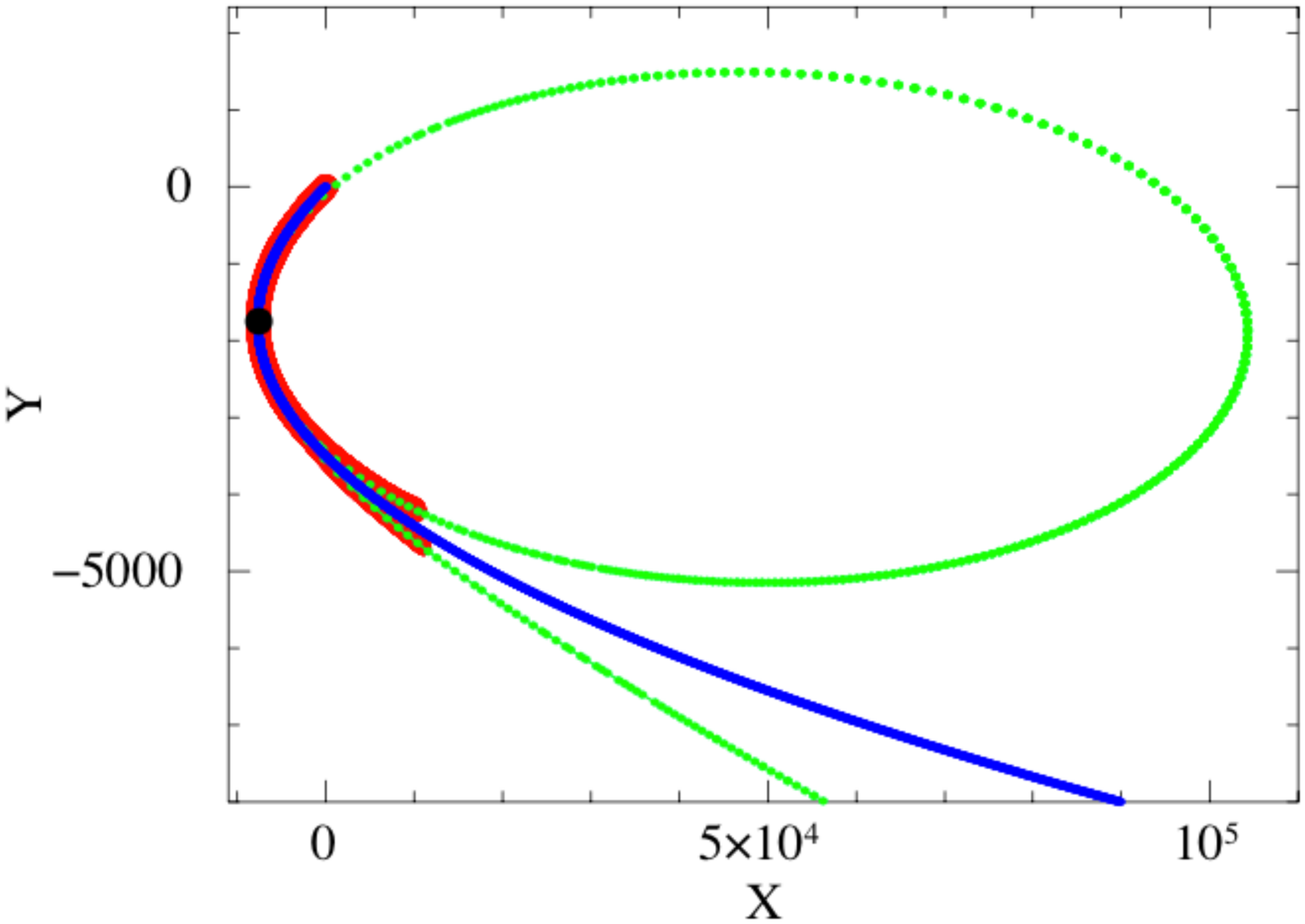}  & & \includegraphics[width=2.7cm, angle=-90]{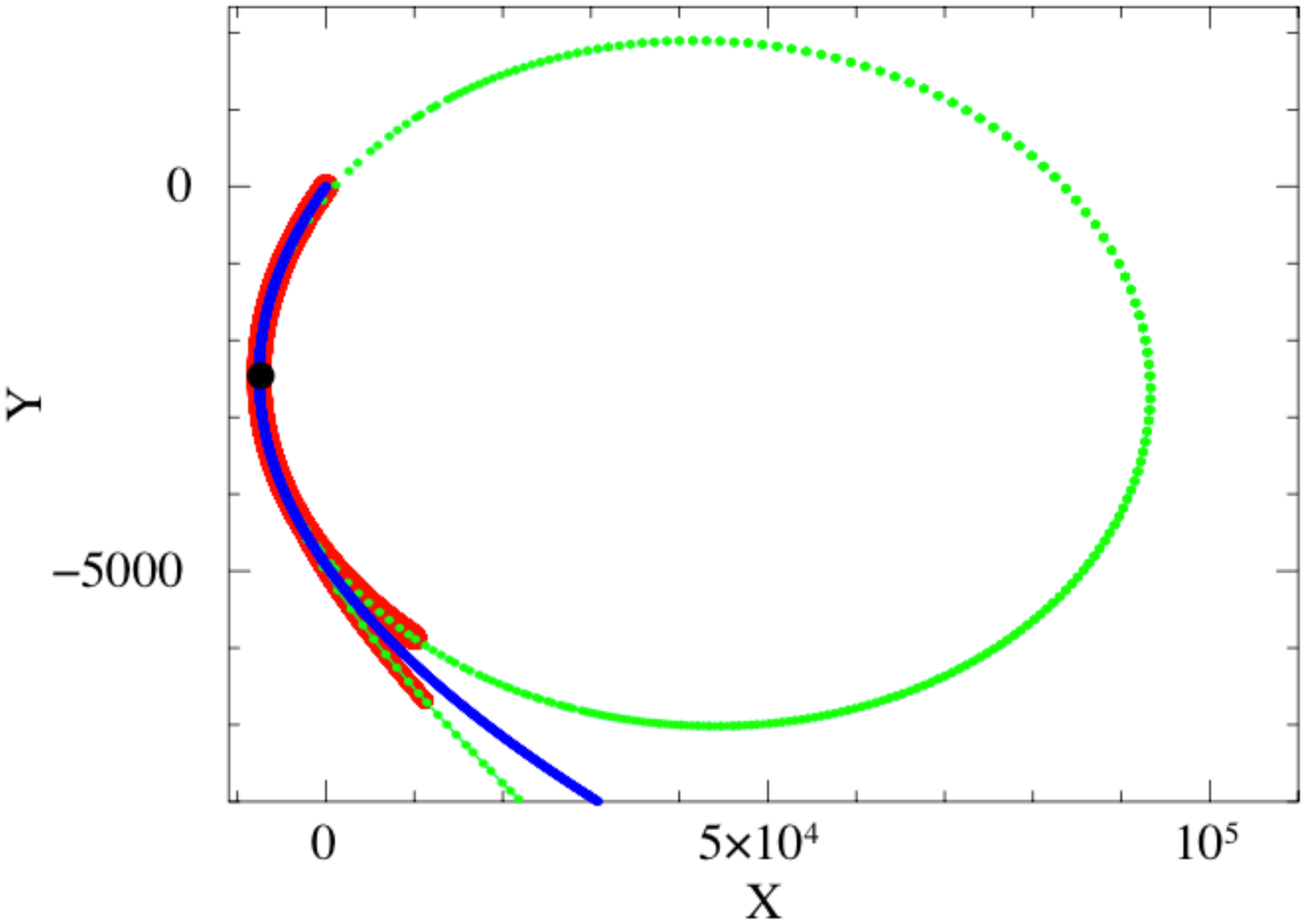} & &  \includegraphics[width=2.7cm, angle=-90]{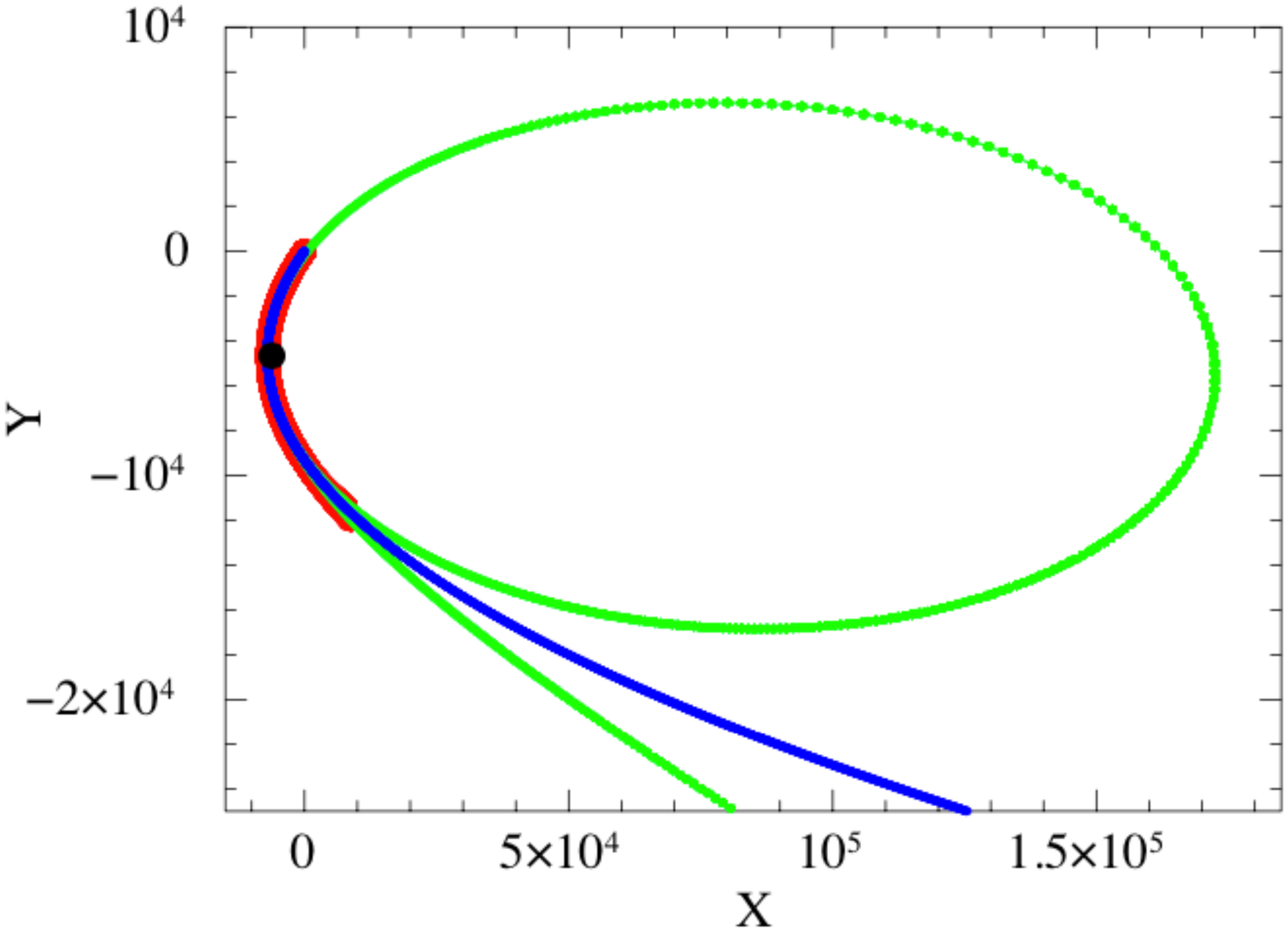} \\
& & \includegraphics[width=2.7cm, angle=-90]{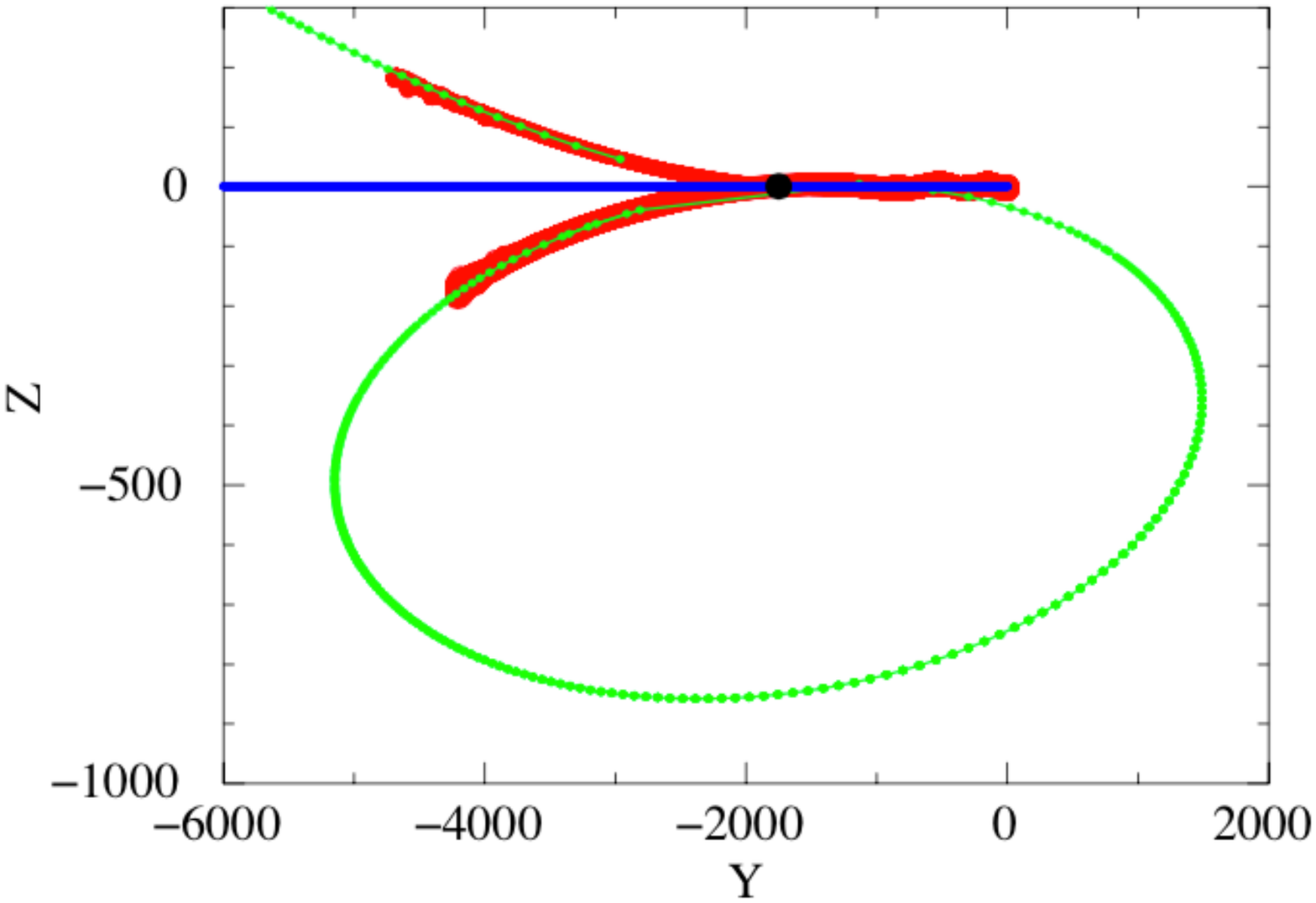}  & & \includegraphics[width=2.7cm, angle=-90]{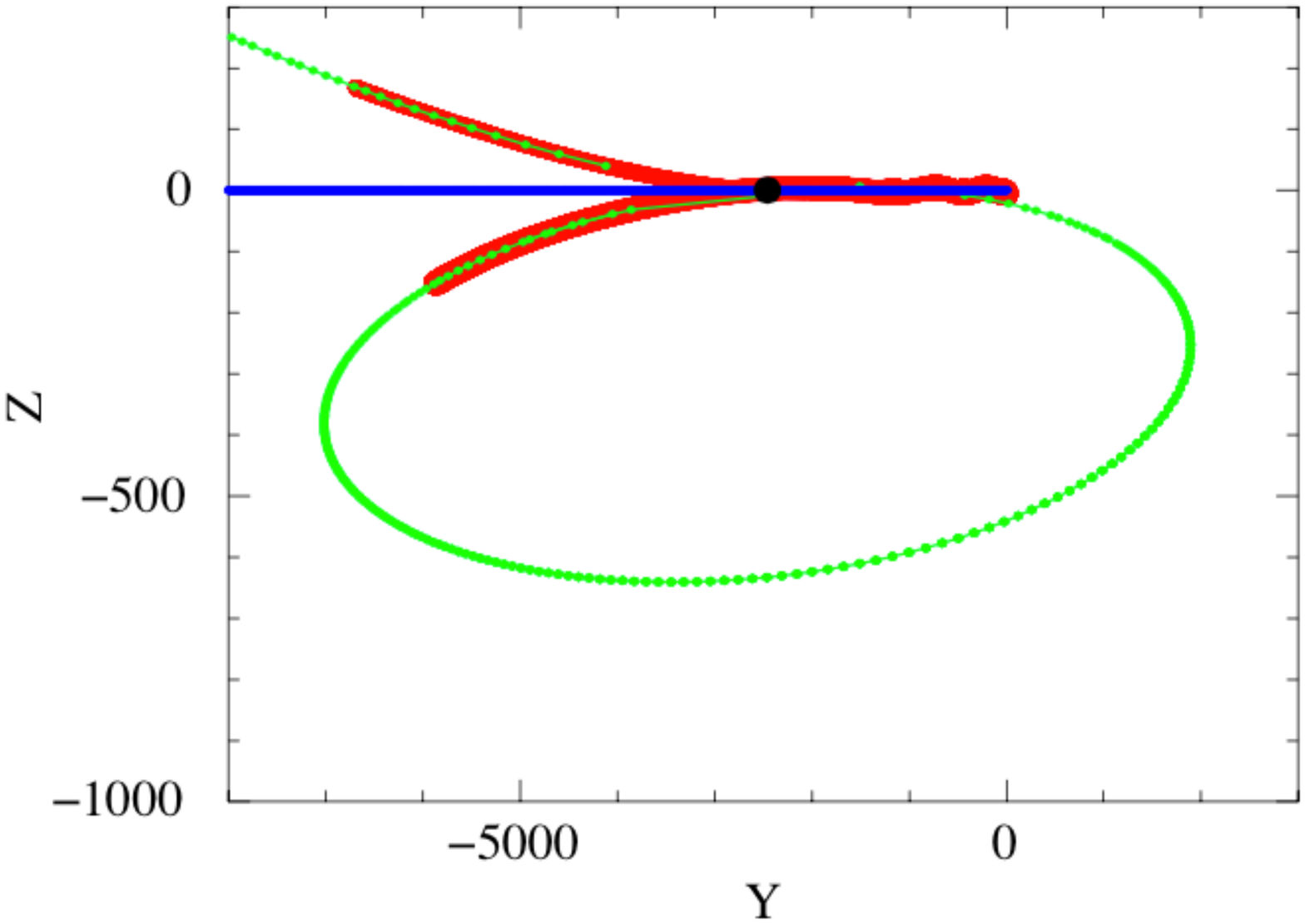} & & \includegraphics[width=2.7cm, angle=-90]{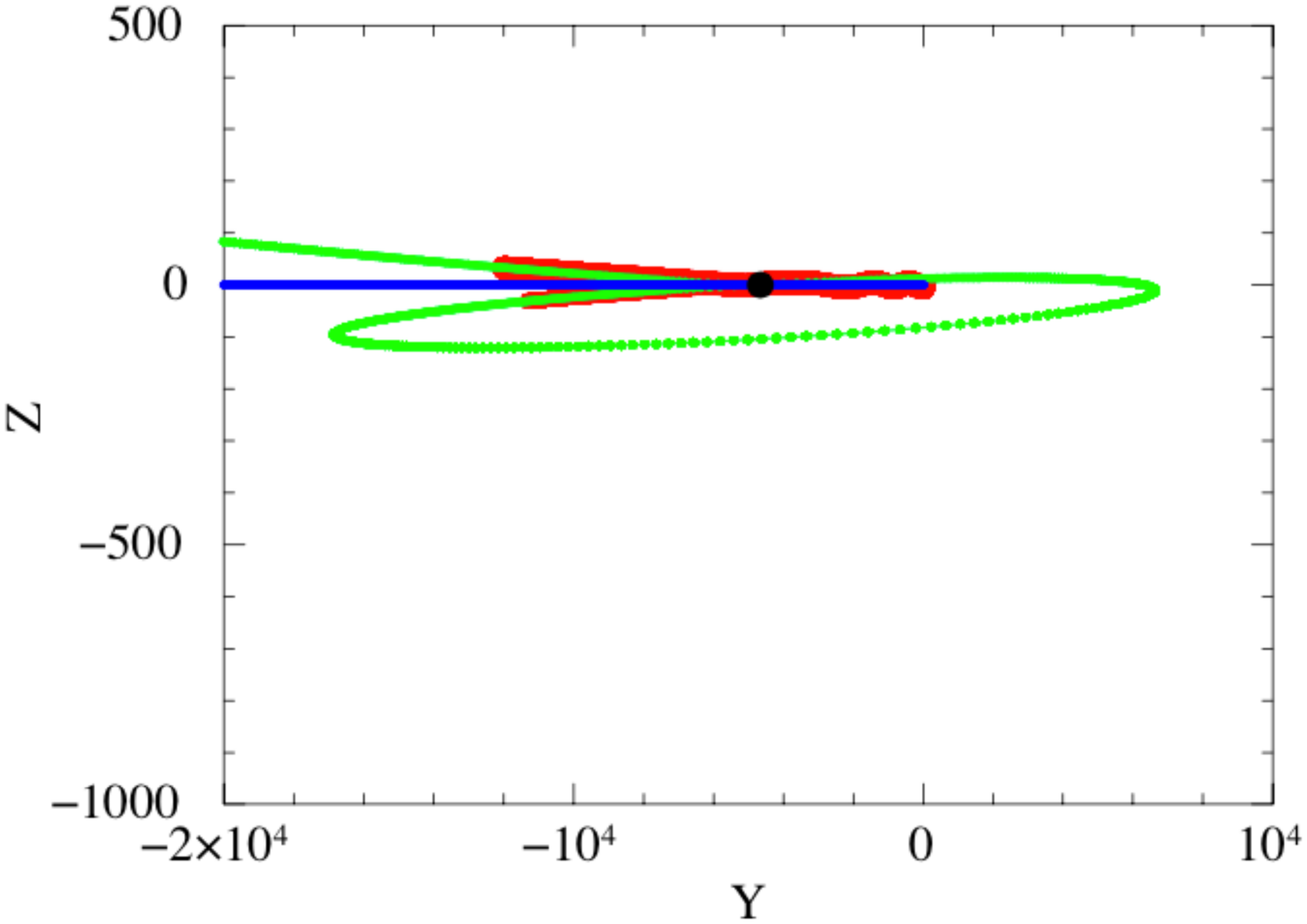}\\
\hline
\end{tabular}
\end{center}
\end{table}
\end{landscape}

\newpage
\begin{landscape}
\begin{table}
\caption{Same as Table \ref{app2}, though following Table \ref{2} in Section \ref{following}. \label{app2}}
\begin{center}
\begin{tabular}{c c c c c c c }
\hline
\footnotesize{$\rm a_{bin}$$\textbackslash$$\rm r_{p}$} &  \footnotesize{$\rm 50.0$} &  \footnotesize{$\rm 100.0$} & \footnotesize{$\rm 142.6$} & \small{$\rm 200.0$} & \footnotesize{$\rm420.0$} & \footnotesize{$\rm 780.0$} \\
$(\rm R_{\rm \odot})$ & & & & & & \\
& & & & & &  \\
& &  \footnotesize{LE6:} &  & \footnotesize{LE7:} & \footnotesize{LE8:} & \\
\small{$\rm 4.9$} & &  \footnotesize{PD-} &  & \footnotesize{MG} & \footnotesize{UN} & \\
& & \footnotesize{TDE}  & & & & \\
& & & & & & \\
& & \includegraphics[width=2.7cm, angle=-90]{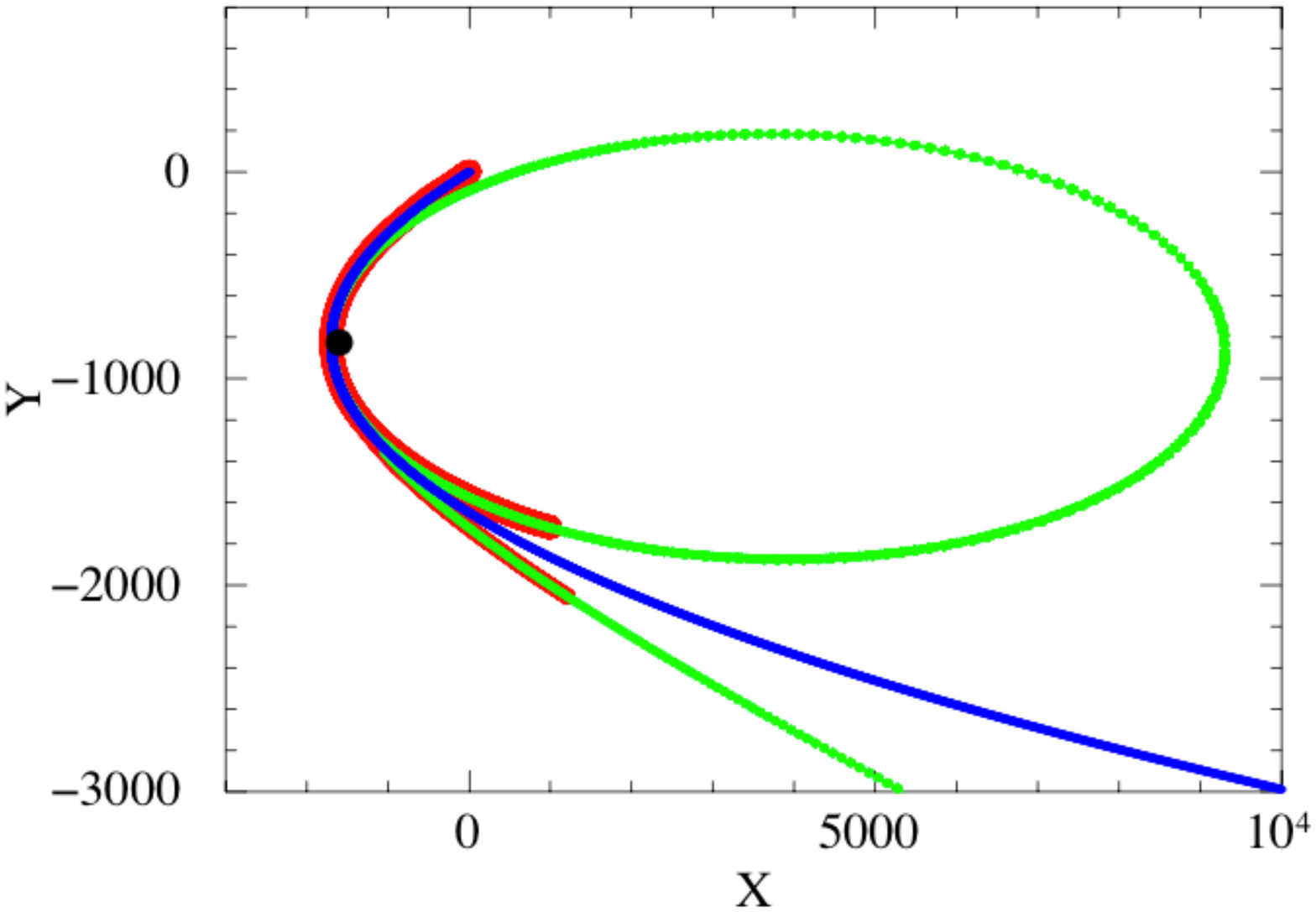} &  & \includegraphics[width=2.7cm, angle=-90]{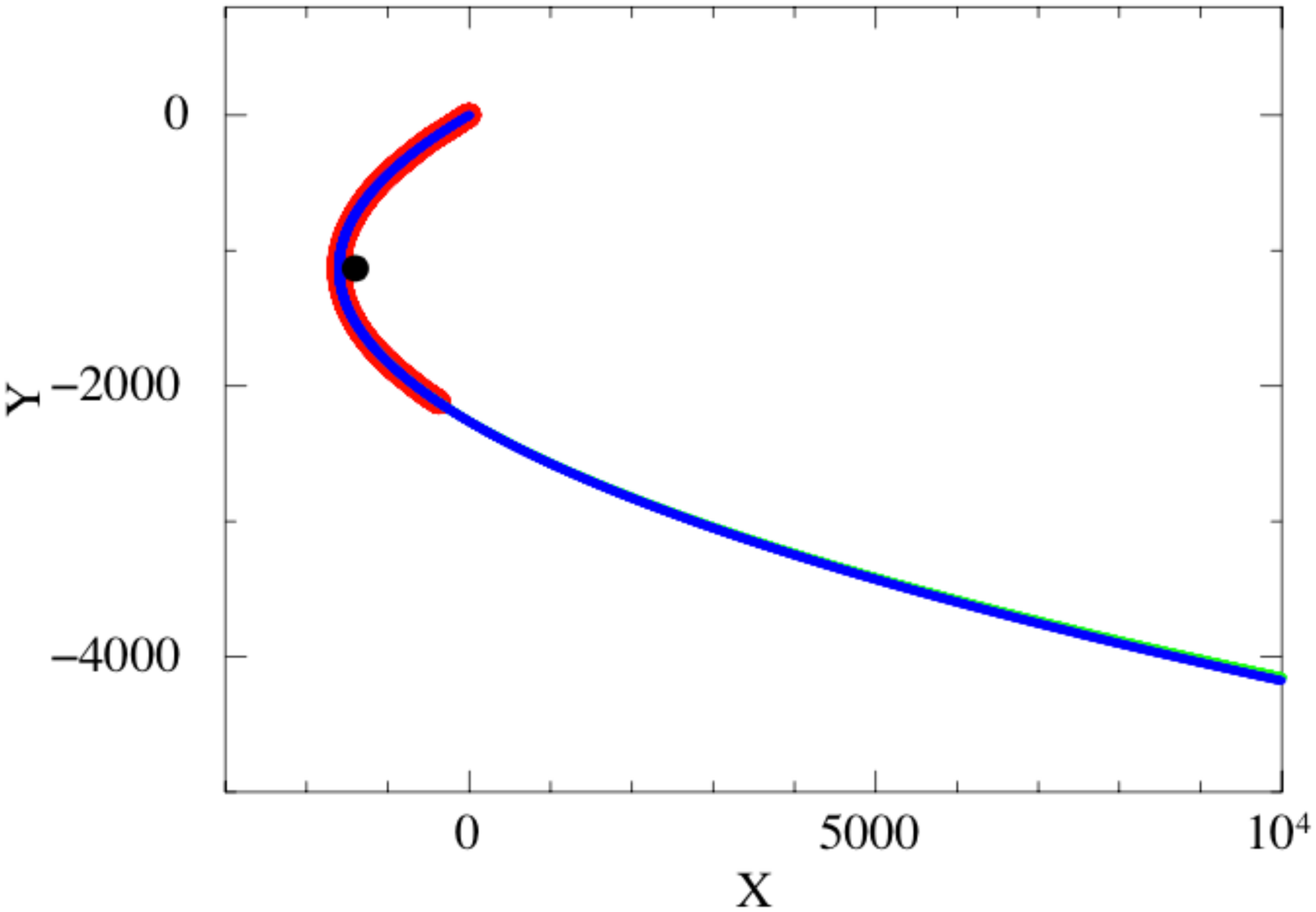} &  \includegraphics[width=2.7cm, angle=-90]{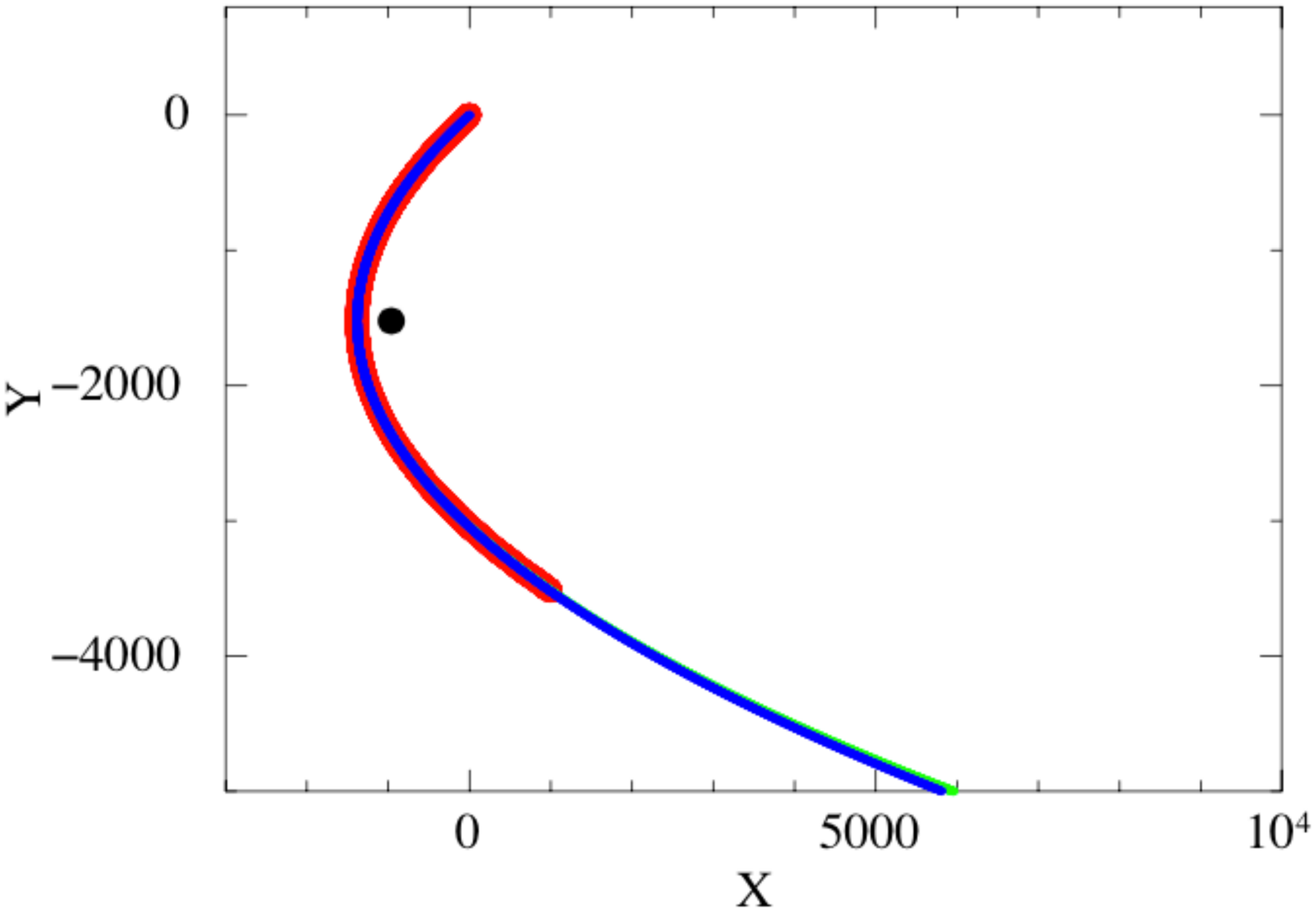}& \\
& &  \includegraphics[width=2.7cm, angle=-90]{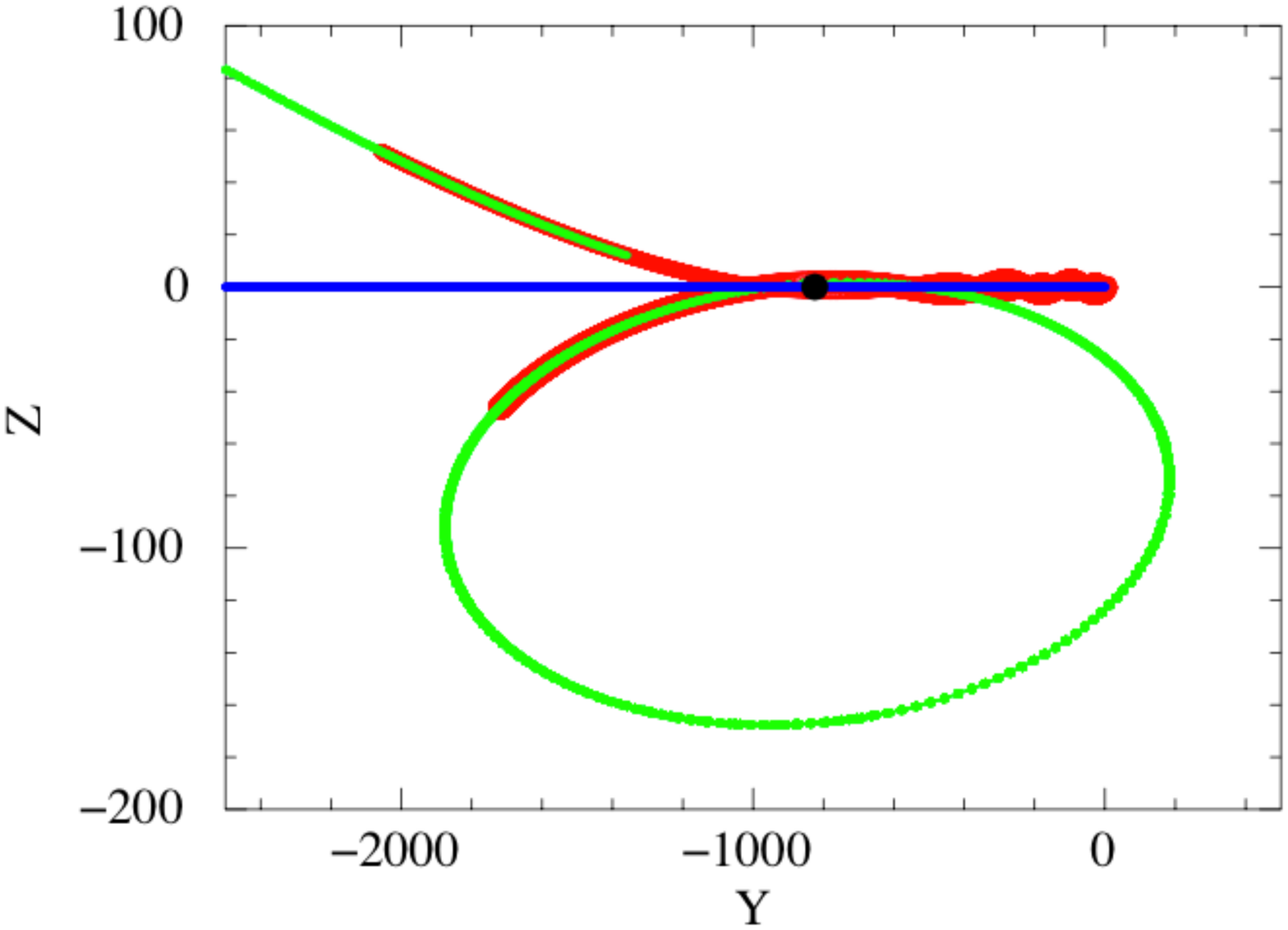}&  & \includegraphics[width=2.7cm, angle=-90]{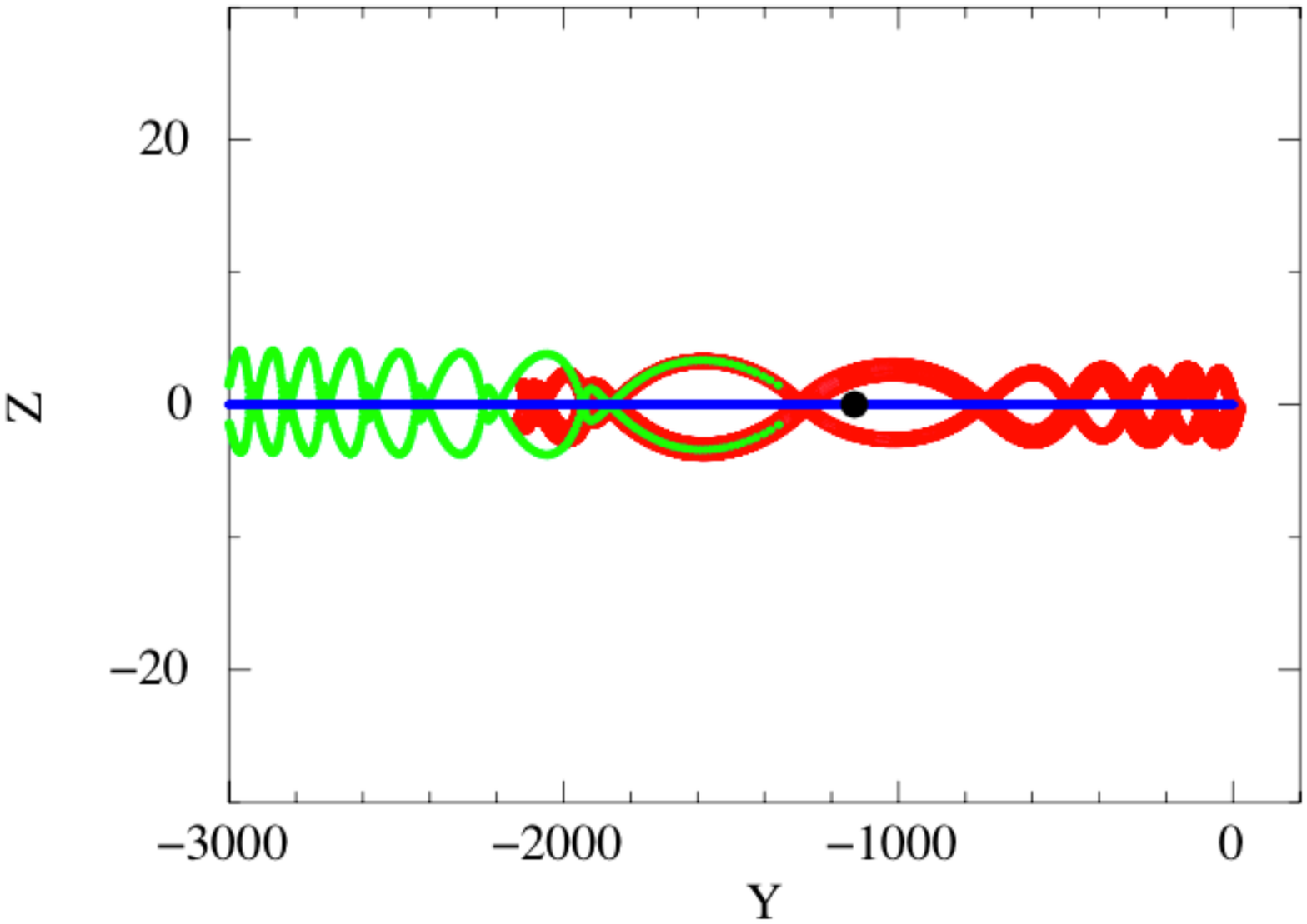}& \includegraphics[width=2.7cm, angle=-90]{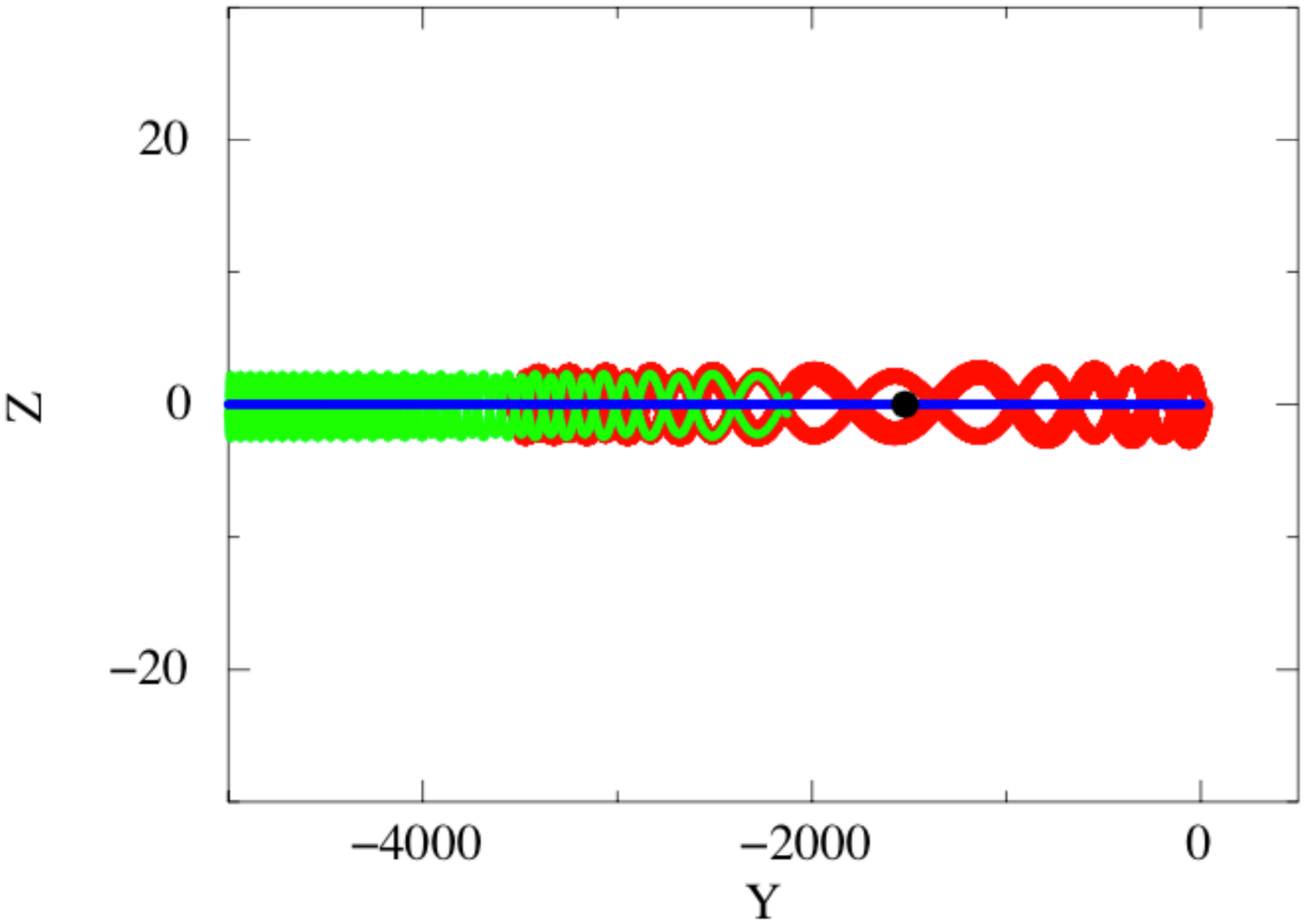}& \\
& & & & & & \\
& &  \footnotesize{LE12:} & & \footnotesize{LE13:} & & \footnotesize{LE14:} \\
\small{$\rm 9.8$} & & \footnotesize{PD-} &  & \footnotesize{BBK} & & \footnotesize{UN} \\
& & \footnotesize{TDE}  & & & & \\
& & & & & & \\
& &   \includegraphics[width=2.7cm, angle=-90]{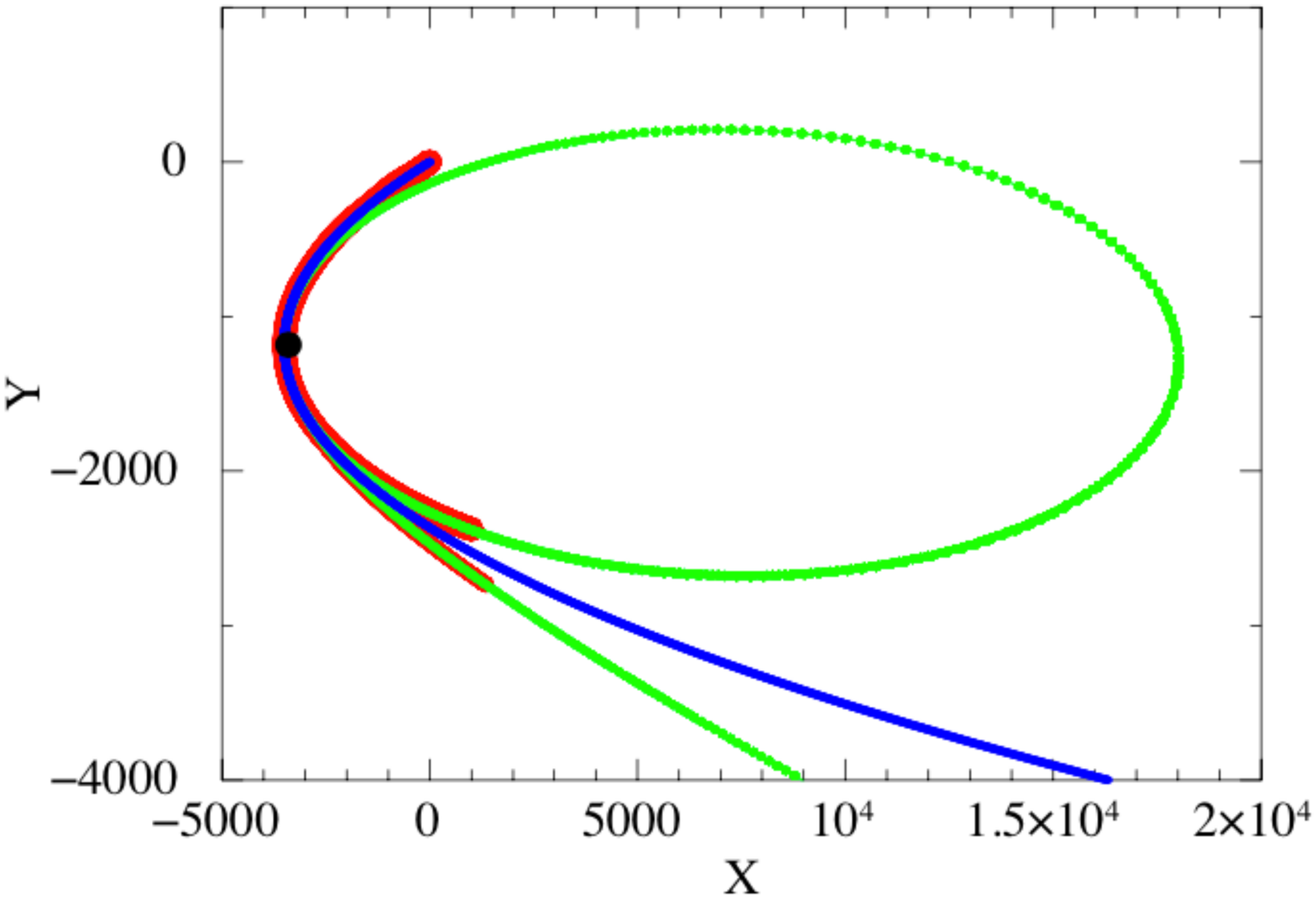}&  &  \includegraphics[width=2.7cm, angle=-90]{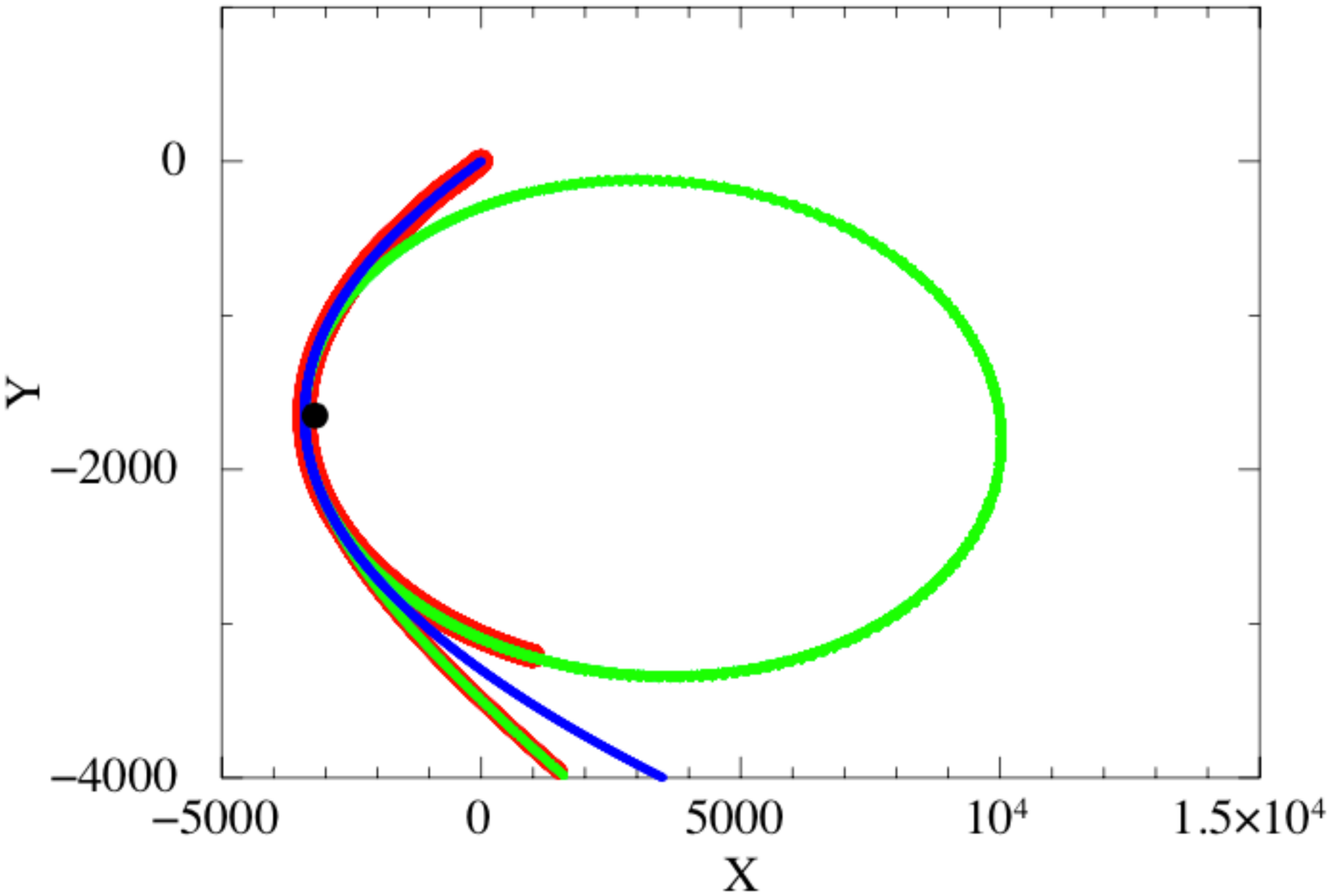} & &\includegraphics[width=2.7cm, angle=-90]{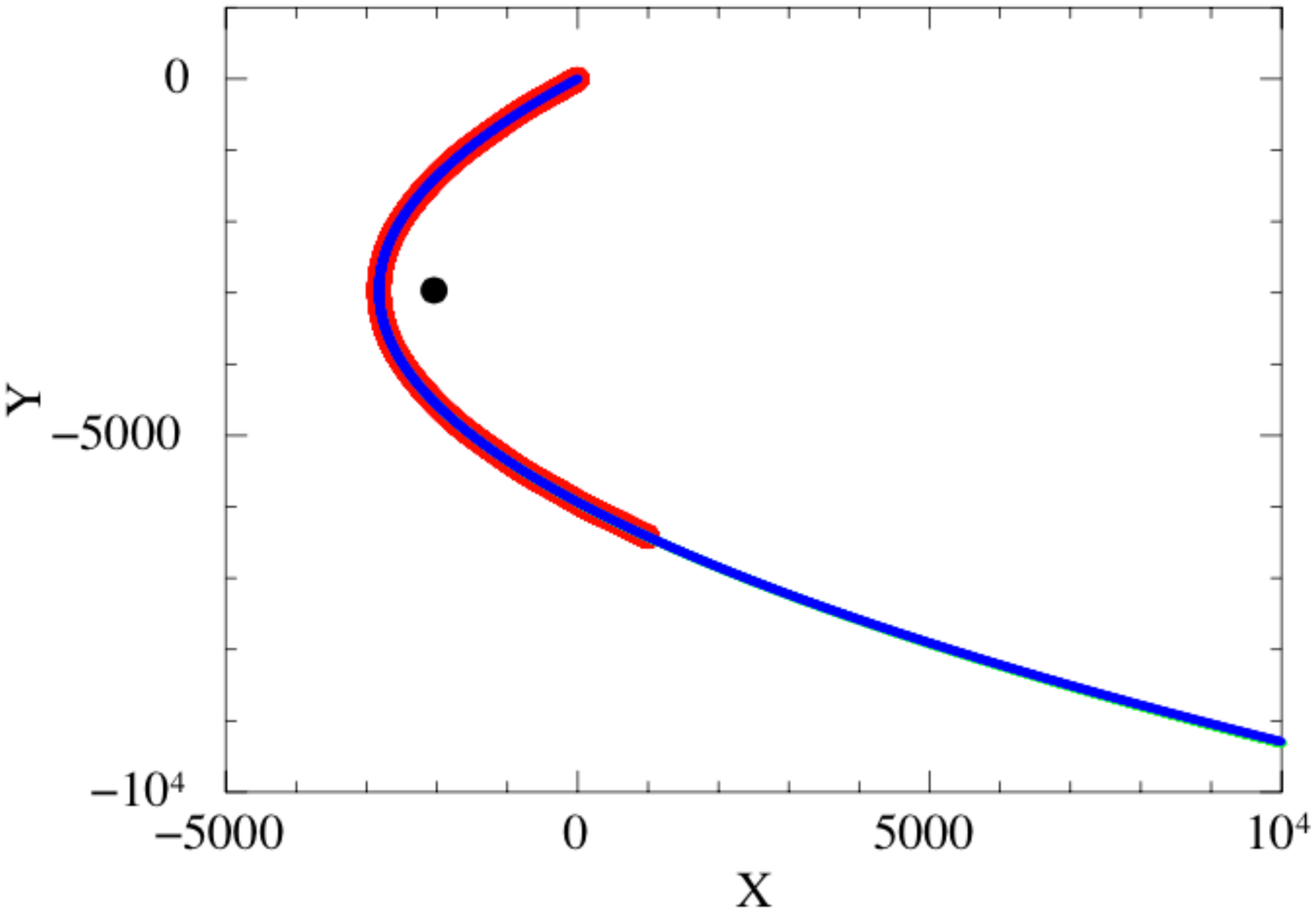}  \\
& & \includegraphics[width=2.7cm, angle=-90]{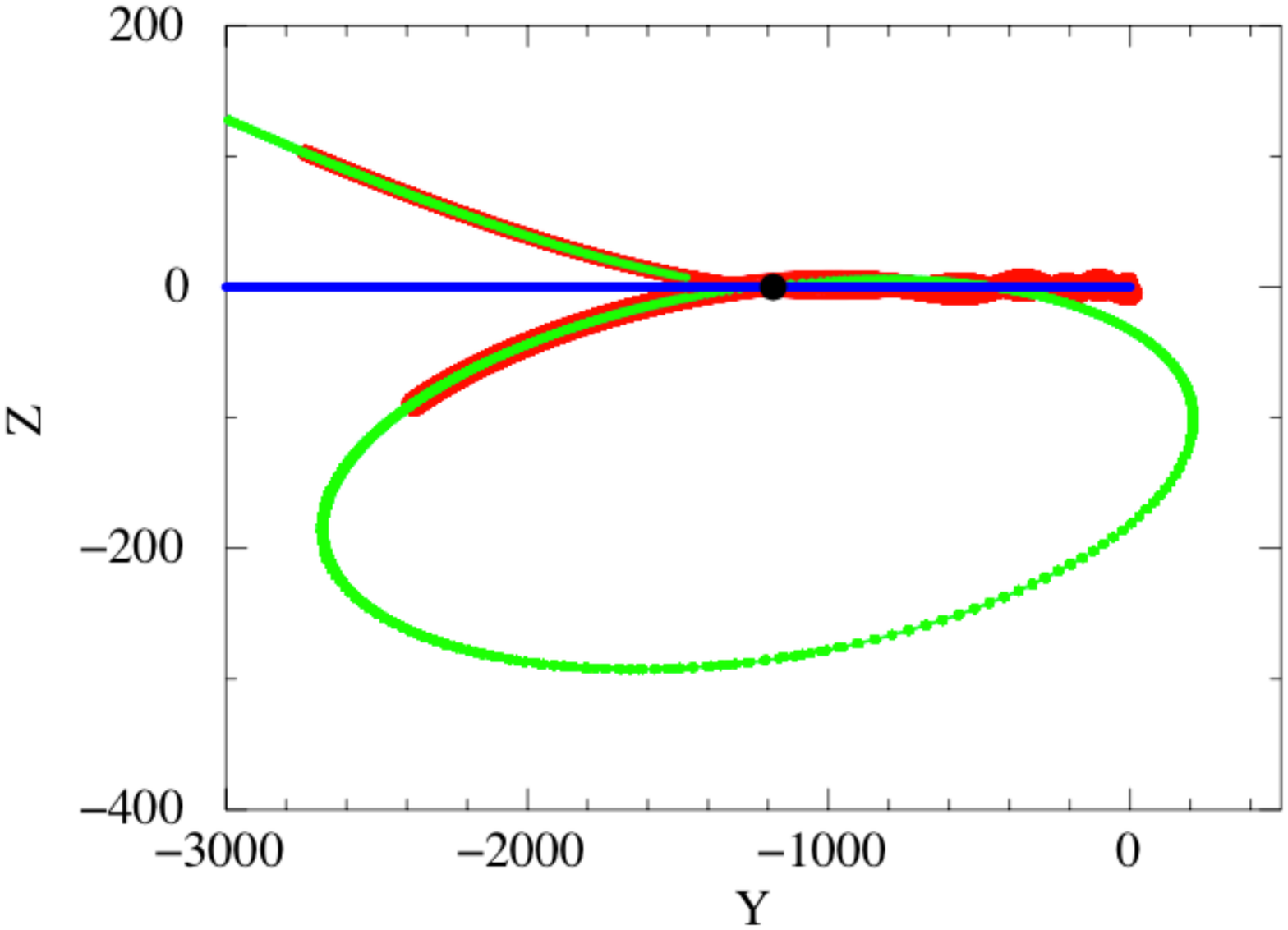} & &  \includegraphics[width=2.7cm, angle=-90]{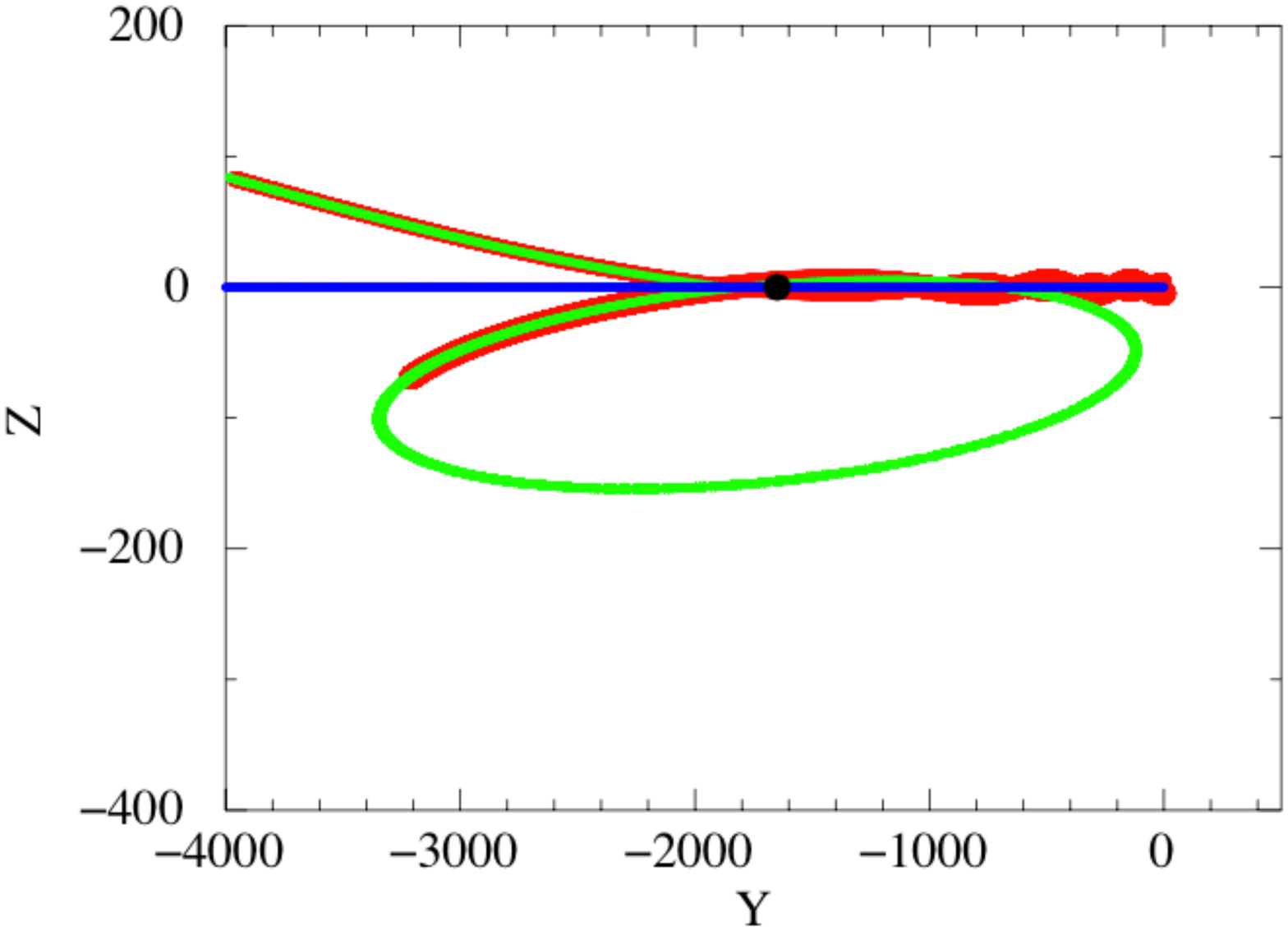}& & \includegraphics[width=2.7cm, angle=-90]{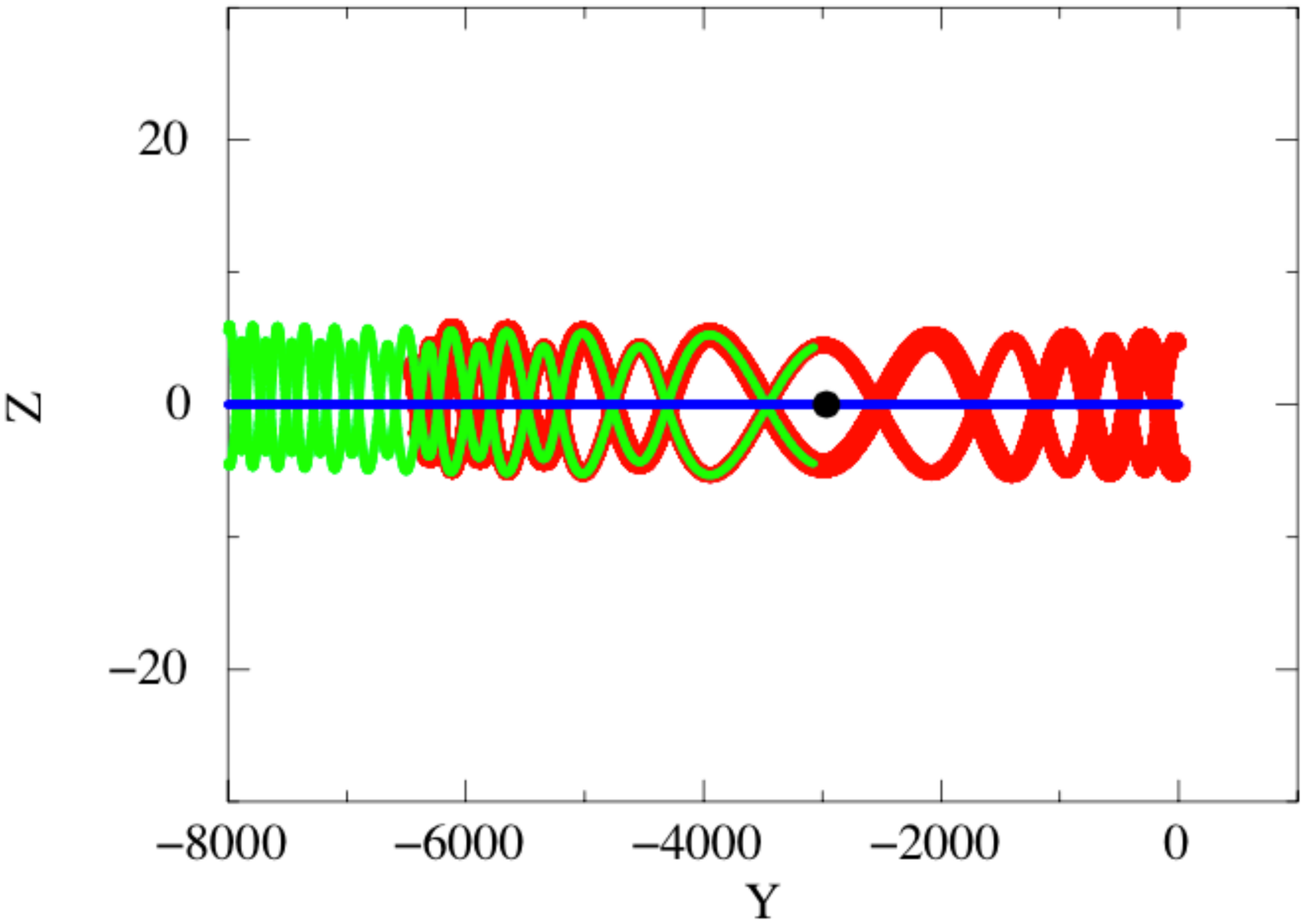}\\
\hline
\end{tabular}
\end{center}
\end{table}
\end{landscape}

\label{lastpage}
\end{document}